\documentclass[12pt]{article}
\usepackage{latexsym, amsmath, epsfig}
\DeclareFontFamily{OT1}{rsfs10}{}
\DeclareFontShape{OT1}{rsfs10}{m}{n}{ <-> rsfs10 }{}
\DeclareMathAlphabet{\mathscript}{OT1}{rsfs10}{m}{n}

\if@twoside\oddsidemargin25mm
\else\oddsidemargin25mm\evensidemargin25mm\marginparwidth25mm\fi
\def\bpl{\Big(}
\def\bpr{\Big)}

\def\brr{\begin{equation}}
\def\err{\end{equation}}
\def\brr{\begin{eqnarray}}
\def\err{\end{eqnarray}}
\def\ba{\left(\begin{array}}
\def\ea{\end{array}\right)}

\newcommand{\dr}{\raise.3ex\hbox{$\stackrel{\leftarrow}{\partial }$}{}}
\newcommand{\dl}{\raise.3ex\hbox{$\stackrel{\rightarrow}{\partial}$}{}}

\newcommand{\ft}[2]{{\textstyle\frac{#1}{#2}}}
\newcommand{\ns}{\normalsize}
\renewcommand{\theequation}{\arabic{section}.\arabic{equation}}

\begin{document}


\begin{titlepage}

\vspace{-3cm}

\title{
   \hfill{\ns CU-TP-986\\}
   \hfill{\ns HU-EP 00/35\\} 
   \hfill{\ns UPR-905T\\} 
   \hfill{\ns hep-th/0010087\\[1cm]}
   {\LARGE An {\it M}-Theory Perspective on Heterotic \\[.1in]
    $K3$ Orbifold Compactifications \\[.5cm]  }}
                       
\author{{\bf
   Michael Faux$^{1}$ \,,\,
   Dieter L{\"u}st$^{2}$ \,
   and Burt A.~Ovrut$^{3}$}\\[5mm]
   {\it $^1$Departments of Mathematics and Physics} \\
   {\it Columbia University} \\
   {\it 2990 Broadway, New York, NY 10027} \\[3mm]
   {\it $^2$Institut f\"ur Physik, Humboldt Universit\"at} \\
   {\it Invalidenstra\ss{}e 110, 10115 Berlin, Germany} \\[3mm]
   {\it $^3$Department of Physics, University of Pennsylvania} \\
   {\it Philadelphia, PA 19104--6396, USA}}
\date{}

\maketitle

\begin{abstract}
\noindent
We analyze the structure of heterotic {\it M}-theory
on $K3$ orbifolds by presenting a comprehensive sequence of 
{\it M}-theoretic models constructed
on the basis of local anomaly cancellation.  This is facilitated 
by extending the technology developed in our previous papers to 
allow one to determine ``twisted" sector states in non-prime orbifolds.
These methods should naturally generalize to  
four-dimensional models, which are of potential phenomenological interest.

\vspace{.3in}
\end{abstract}

\thispagestyle{empty}

\end{titlepage}


\section{Introduction}
Orbifolds describe an interesting category of compactification
schemes likely to provide useful physical models based on {\it M}-theory.
The fundamental example of an {\it M}-theory orbifold,
based on $S^1/{\bf Z}_2$, was presented in
\cite{hw1, hw2}, and examples based on $T^4/{\bf Z}_2$ were
presented in \cite{dasmuk,wittens5}.     
Since these incorporate nonperturbative aspects of related   
models, an algorithm for sifting through the various possibilities
would complement and extend the perturbative string orbifold models classified
more than ten years ago \cite{orbi,imnq}.
Indeed, nonperturbative {\it M}-theory orbifolds
could overcome the 
phenomenological shortcomings endemic in the string 
perturbative case.   

In two previous papers \cite{mlo,phase}, 
we outlined much of the technology needed to 
ascertain microscopic properties of {\it M}-theory orbifolds, emphasizing
the role of local anomaly cancellation in the determination of 
local ``twisted" sector gauge groups and matter content.  In those papers,
as in this present one, we concentrate on the case of 
$S^1/{\bf Z}_2\times K3$ orbifolds\footnote{For a
review on heterotic $K3$ compactifications 
see e.g. \cite{lust}.}, since these encorporate
most of the fundamental issues which attend more general cases.
In \cite{mlo}, we described 
special constraints which arise at fixed-plane intersections,
and delineated the various contributions to both the quantum and inflow
anomalies.  In particular, we introduced the necessity for an
intersection anomaly in orbifolds.  This allows one to determine twisted
gauge groups localized on odd-dimensional orbifold planes, 
which do not themselves support a local anomaly.  
Subsequent papers \cite{bds, ksty} then 
supplied additional and complementary input to this technology.
In \cite{phase}, we also explained how fivebranes may mediate transitions 
linking different ``phases" describing separate solutions to the 
local anomaly cancellation requirements. 
   
In this paper, we extend our previous results and technology 
to include non-prime orbifolds.  
We present a detailed local analysis of anomaly cancellation in all
four $S^1/{\bf Z}_2\times K3$ orbifolds describing 
degenerations of the sort $K3 \to T^4/{\bf Z}_M$.
(These exist for the cases $M$=2,3,4 and 6.)
This rounds out the presentation
in \cite{mlo, phase} and complements the analysis in
\cite{ksty}.  Some of the results in this paper
were previously presented in \cite{ksty}. 
However, our technology differs, and we also have new 
results which more fully unify the overall discussion.
Specifically, we find a systematics associated with the orbifold 
projection of seven-dimensional
gauge groups onto Cartan subgroups.  
This enables a discussion of possible
phase transitions, and indicates smooth interconnections 
linking various possible phases of the moduli space.
 
We expect that intersection anomalies will play
a crucial role in eventual studies of {\it M}-theory phenomenology.
In this paper, we present a comprehensive 
list of models describing portions of the low-energy moduli space 
corresponding to {\it M}-theory compactified on 
$S^1/{\bf Z}_2\times K3$ orbifolds, and we discuss a number of technical and
conceptual issues associated with these.  The technology is developed 
with the expressed intent of enabling a search for 
phenomenologically more interesting
compactifications to four dimensions. 
Our aim is to partially delineate how nonperturbative effects,
particularly those involving fivebrane-mediated phase transitions, can
modify phenomenological predictions based on perturbative orbifolds,
and also to facilitate the search for a more fundamental description 
of {\it M}-theory.

Significant progress along these lines has already been achieved in 
heterotic {\it M}-theory models \cite{losw1}.
These are based on the Ho{\v r}ava-Witten $S^1/{\bf Z}_2$ orbifold
compactified to four dimensions on suitable Calabi-Yau
threefolds $X$.  Clearly, new {\it M}-theory models can be
obtained by replacing these Calabi-Yau manifolds, and indeed
all of $S^1/{\bf Z}_2\times X$, by more general non-perturbative orbifolds.
This paper is a step in that direction.
 
Each of the $S^1/{\bf Z}_2\times T^4/{\bf Z}_M$ orbifolds describes
a brane world involving a pair of parallel ten-planes 
linked by some number of parallel
seven-planes with six-plane intersections. The seven-planes correspond
to the $A_{M-1}$ singularities of the $K3\to T^4/{\bf Z}_M$ orbifold
degenerations\footnote{The four dimensions
transverse to the seven-planes corresond to directions along the $K3$.}.
For reasons motivated in \cite{ksty}, one expects that each
$A_{M-1}$ seven-plane should support an $SU(M)$ gauge theory.
In that paper, consistent models were presented with this feature
for each of $M$=2,3 and 4, but not for the case $M$=6.  For the case
$M$=6, the authors of \cite{ksty} present one model 
(with gauge group $SU(9)\times SU(3)\times SU(2)$)
which requires that the secondary $A_1$ seven-planes support  
$U(1)$ gauge groups instead of $SU(2)$.  
In that case, the $U(1)$ factor is invisible
in the infrared limit, and all anomalies do cancel.  

An observation pointed out in \cite{ksty} is that
whenever infrared-visible $U(1)$ factors appear there exist non-cancelled
anomalies.  They present only one example of this case, which is
an $M$=4 model with gauge group 
$E_6\times SO(10)\times SU(4)\times SU(2)\times U(1)$, which has 
non-cancelled local gravitational, mixed {\it and} gauge anomalies 
in both the abelian factors as well as the nonabelian factor.
In this paper, we present a new class of 
$M$=6 models, with gauge group $E_6\times SU(8)\times SU(2)\times U(1)^{2+b}$,
for $0\le b\le 5$, 
in which all gravitational and mixed anomalies, as well
as all nonabelian gauge anomalies do cancel.   Furthermore, our model
has $SU(N)$ gauge groups on each of the $A_{N-1}$ seven-planes, 
including all secondary planes.
We see the (highly nontrivial) cancellation of 
all gravitational, mixed and nonabelian gauge anomalies 
as an encouraging sign. 
We relegate the resolution of the remaining abelian gauge anomalies
to a future paper.

This paper is structured as follows.  
In Section 2, we review the 
basics of {\it M}-theory orbifold anomaly analyses for the case
of prime orbifolds.  We also introduce
streamlined conventions relative to our previous papers and present
a generalized proof, based on anomaly cancellation, that certain orbifold fixed
planes can sequentially absorb (or emit) fivebranes from (or into) the 
eleven-dimensional bulk while maintaining consistency.  This requires 
that the fivebrane tensor fields involve interesting dynamics which 
allow a localized version of the Green-Schwarz mechanism.
In Section 3, we extend our technology to include nonprime orbifolds. 
In Section 4, we explain how to reduce the anomaly calculation to a set of 
very efficient equations which only involve
rational data parameterizing the orbifold geometry, magnetic
and electric charges, and various group theory factors. 
This enables an algorithmic search for consistent orbifold vertices.
In Section 5, we explain the various consistent models which we have obtained by
applying our technology to the orbifolds mentioned above, and explain how these 
models are interconnected by virtue of fivebrane-mediated phase transitions.
In Section 6, we explain the microscopic details of the models 
described in section 5.  This is done by use of ``vertex diagrams" and 
``ladder diagrams" which conveniently depict the salient geometric properties. 
We also include an appendix which tabulates all of the rational parameters needed
to verify the consistency of each of the vertices presented in the main text.

\setcounter{equation}{0}
\section{Prime Orbifolds}
In this paper, we concentrate on the four 
$S^1/{\bf Z}_2\times T^4/{\bf Z}_M$ orbifold limits of
$S^1/{\bf Z_2}\times K3$, corresponding to $M=2,3,4$ and $6$.
We refer to these succinctly as the ${\bf Z}_2$, ${\bf Z}_3$, 
${\bf Z}_4$ and ${\bf Z}_6$ orbifolds, respectively.
The first two of these are prime orbifolds (since $M$ is a prime integer
in these cases) and the second two are 
non-prime.  Each of these orbifolds has a pair of fixed ten-planes
(associated with the universal ${\bf Z}_2$ factor) and, additionally, 
some number of
parallel fixed seven-planes (associated with the factor ${\bf Z}_M$).
Each of the seven-planes transverally intersects each of the 
ten-planes once, at a particular six-plane.  

For prime orbifolds, the fixed seven-planes are indistinguishable.
For instance, the ${\bf Z}_2$ orbifold has sixteen and the 
${\bf Z}_3$ orbifold has nine of these.  For non-prime orbifolds, 
there are distinguishable sets of fixed seven-planes.
(We describe this in more detail in the following section.)
As a consequence, the analysis of local anomalies in non-prime 
orbifolds is more complicated,
and requires a greater systematics, than in the case of prime orbifolds.
In this section, we recapitulate the systematics involved in 
analyzing prime orbifolds, so that we can readily extend the analysis to
include both prime and non-prime orbifolds in subsequent sections.

Local anomaly matching on the two ten-planes uniquely determines
that these planes each support a ten-dimensional $E_8$ super Yang-Mills
multiplet  \cite{hw2}.
As explained in \cite{phase}, additional, more stringent, constraints 
follow from local anomaly matching at each of the six-planes
describing intersections of the ten-planes and the seven-planes.
Resolving these constraints allows one to determining various 
elements of data.  These include local breaking patterns for the 
$E_8$ factors, a magnetic charge (denoted $g$) associated with each six-plane,
electric parameters (denoted $\eta$ and $\rho$)
associated with seven-dimensional Chern-Simons interactions
\footnote{We refer the reader to \cite{phase} for a discussion of each of 
the parameters $g$, $\eta$ and $\rho$.},  
a {\it seven}-dimensional gauge group $\tilde{\cal G}_7$,
and additional ``twisted" fields.  
The twisted fields can include
seven-dimensional Yang-Mills multiplets and also six-dimensional
$N$=1 vector, hyper and tensor multiplets. 

In general, on a given ten-plane, the associated $E_8$ factor 
is broken to a subgroup $E_8\to {\cal G}\subset E_8$ on the 
embedded intersection six-planes by the nontrivial 
action of ${\bf Z}_M$ on the components of the $E_8$ multiplet.
We assume that ${\cal G}$ is a direct product of simple 
and/or $U(1)$ factors.  We define $p$ as the number of such factors,
so we can write
$E_8\to {\cal G}_1\times \cdots \times {\cal G}_p$.
The field strength associated with ${\cal G}_i$
is described by a matrix-valued two-form $F_i$. 
The twisted gauge group $\tilde{\cal G}_7$ can likewise be broken on the 
intersection six-planes to a subgroup (for instance, its Cartan subgroup) 
by the action of the universal ${\bf Z}_2$ factor. 
Thus, we can write 
$\tilde{\cal G}_7\to \tilde{\cal G}_1\times\cdots\times \tilde{\cal G}_q$, 
where we define $q$ as the number of such factors.
The field strength associated with $\tilde{\cal G}_j$
is described by a  matrix-valued two-form ${\cal F}_j$.

On a given intersection six-plane, the anomaly involves only those gauge
field strengths associated with one of the $E_8$ factors and one of the
$\tilde{\cal G}_7$ factors.  We concentrate
on a particular six-plane and safely suppress
any labelling which distiguishes which six-plane we are considering.
It is convenient
to describe all of the $p+q$ field strengths collectively using a notation
$F_I$, where 
$F_I=\{\,F_1\,,\,...\,,\,F_p\,,\,{\cal F}_1\,,\,...\,,\,{\cal F}_q\,\}$.
In this case, the anomaly on a given six-plane is described by an
eight-form polynomial which we can write as
\brr \widehat{I} =
     A\,{\rm tr}\,R^4
     +B\,(\,{\rm tr}\,R^2\,)^2
     +C^I\,{\rm tr}\,R^2\wedge {\rm tr}\,F_I^2
     +D^{IJ}\,{\rm tr}\,F_I^2\wedge {\rm tr}\,F_J^2
     +E^I\,{\rm tr}\,F_I^4 \,,
\label{ihat}\err
where the various coefficients $A, B, C^I, D^{IJ}$ and $E^I$
are determined by summing each of the various quantum
and inflow anomalies as explained in \cite{phase}.  

If there are no tensor fields in the local twisted spectrum, then the
anomaly (\ref{ihat}) must vanish identically.  In this case, each
of the coefficients $A, B, C^I, D^{IJ}$ and $E^I$ must independently
vanish. If there is one twisted tensor field, then a local Green-Schwarz
mechanism can cancel the anomaly provided $\widehat{I}$ 
factorizes as a product of two identical four forms.  
In this case, the requirement that $\widehat{I}$ be a complete square
follows from supersymmetry (since tensor multiplets contain
anti-self dual tensor fields). In the general case, if there are
$n_T$  local tensor multiplets, then anomaly matching requires
that $\widehat{I}$ be a sum of $n_T$ complete squares.
In any case, cancellation or factorization of $\widehat{I}$ 
each requires that $A=0$ and $E^I=0$, since these correspond to 
non-factorizable terms in the anomaly.

Henceforth, we assume that $A=0$ and $E^I=0$, as required
by the preceeding discussion. We analyze the requirement $A=0$ in more
detail in the next section, and discuss the $E^I=0$ requirement
during the course of this section.    
The remaining coefficients are given by
\footnote{Note that we have scaled $\eta$ relative to the convention 
used in \cite{phase}, such that 
$Y_4=\frac{1}{4\pi}\,(-\frac{\eta}{2\,f}\,{\rm tr}\,R^2
+\rho\,{\rm tr}\,{\cal F}^2\,)$.
In this way, $\eta$ invariably turns out to be unity.}
\brr B &=& \frac{1-\eta}{2f}-\frac{n_T}{8}
     \nonumber\\[.1in]
     C^I\,{\rm tr}\,F_I^2 &=&
     (\,\frac{g}{2}+\frac{\eta}{f}\,)\,{\rm tr}\,F^2
     +\frac{1}{6}\,{\rm trace}'\,F^2
     +\frac{1}{6}\,{\rm trace}\,{\cal F}^2
     +\rho\,{\rm tr}\,{\cal F}_7^2
     \nonumber\\[.1in]
     D^{IJ}\,{\rm tr}\,F_I^2\wedge {\rm tr}\,F_J^2
     &=& -\frac{g}{2}\,(\,{\rm tr}\,F^2\,)^2
     -\frac{2}{3}\,{\rm trace}'\,F^4
     -2\,\rho\,{\rm tr}\,F^2\wedge {\rm tr}\,{\cal F}_7^2
     -\frac{2}{3}\,{\rm trace}\,{\cal F}^4 \,.
     \nonumber\\   
\label{bcd}\err
Various mnemonics are involved in (\ref{bcd}). Traces denoted
``${\rm tr}$" describe group theoretic reduction of
the $E_8$ trace into traces over the various subgroup factors $F_i$,
defined above. Traces denoted
``${\rm trace}$" are quantum traces, which include an extra minus sign 
when hypermultiplets are involved.  Traces denoted 
``{\rm trace}\,$'$\,"
are {\it weighted} quantum traces which also involve divisors related
to orbifold-plane multiplicity.  For the prime orbifolds, this is
defined simply
\brr {\rm trace}'F^{r}=\frac{1}{f}\,\sum_{i=1}^p\,
     \sum_{\cal R}\,(-1)^{2J}\,
     {\rm tr}_{\cal R}\,F_i^r \,,
\label{wqt}\err
where $f$ is the number of seven-planes for the orbifold in question,
$r$ is any arbitrary power, 
and $J$ is the highest spin in the multiplet contributing the 
anomalous coupling.  (Thus, for hypermultiplets $J=1/2$ and for
vector multiplets $J=1$.)  Terms in (\ref{bcd}) involving $F$ describe 
anomalies arising from ten-dimensional fields.  Terms involving
${\cal F}$ describe anomalies due to additional six or seven dimensional
twisted fields \footnote{Since the twisted fields generically transform under
both ${\cal G}$ and $\tilde{\cal G}_7$, the terms in (\ref{bcd}) mnemonically 
involving ${\cal F}$ will, in fact, include contributions from 
both $F_i$ and ${\cal F}_i$.}.
Finally, terms involving ${\cal F}_7$ are inflow 
contributions involving only the seven-dimensional twisted group
$\tilde{\cal G}_7$.  Note that the cancellation of $E^I$ is implicit 
in the last equation of (\ref{bcd}), since the right hand side
can only factorize if this condition is satisfied. 

If $n_T=0$, then we require that $B=C^I=D^{IJ}=0$.
If $n_T=1$ then, according to the discussion in the preceeding
paragraph, we have a weaker requirement that the 
anomaly factorize as a complete square.
This is only possible if 
\brr 4\,B\,D^{IJ}=C^I\,C^J \,.
\label{bdc}\err
Since this requirement includes the stronger case where all coefficients
vanish, equation (\ref{bdc}) describes a generic constraint 
valid when $n_T\le 1$.

\vspace{.1in}
\noindent
{\it Horizontal Hierarchy}:\\[.1in]
Note that if we locally increment the hypermultiplet multiplicity
and the tensor multiplet multiplicity each by one
gauge singlet, and simultaneously increment the magnetic charge by one such
that $n_H\to n_H+1$, $n_T\to n_T+1$ and $g\to g+1$, then
equation (\ref{bcd}) tells us that
\brr B &\longrightarrow& B-\ft18
     \nonumber\\[.1in]
     C^I\,{\rm tr}\,F_I^2 &\longrightarrow&
     C^I\,{\rm tr}\,F_I^2+\ft12\,{\rm tr}\,F^2
     \nonumber\\[.1in]
     D^{IJ}\,{\rm tr}\,F_I^2\wedge {\rm tr}\,F_J^2
     &\longrightarrow&
     D^{IJ}\,{\rm tr}\,F_I^2\wedge {\rm tr}\,F_J^2
     -\ft12\,(\,{\rm tr}\,F^2\,)^2\,.
\label{shift}\err
Furthermore, $A=0$ and $E^I=0$ are invariant under this shift as well.
Therefore, if we substitute the modified values of the coefficients back into
(\ref{ihat}), we determine 
\brr \widehat{I} &\longrightarrow&
     \widehat{I}-\ft18\,(\,{\rm tr}\,R^2\,)^2
     +\ft12\,{\rm tr}\,R^2\wedge {\rm tr}\,F^2
     -\ft12\,(\,{\rm tr}\,F^2\,)^2
     \nonumber\\[.1in]
     &=& \widehat{I}
     -\ft18\,(\,{\rm tr}\,R^2-2\,{\rm tr}\,F^2\,)^2\,.
\err
So we observe that the net anomaly changes by the addition of one
complete square.  But, since the local twisted sector changes
by the addition of one tensor multiplet, the net anomaly can be
cancelled by a local Green-Schwarz mechanism mediated by the
anti self-dual tensor field in this multiplet.  We interpret this
process as describing the absorption of a fivebrane onto the 
relevant intersection six-plane.
\begin{figure}
\begin{center}
\includegraphics[width=5in,angle=0]{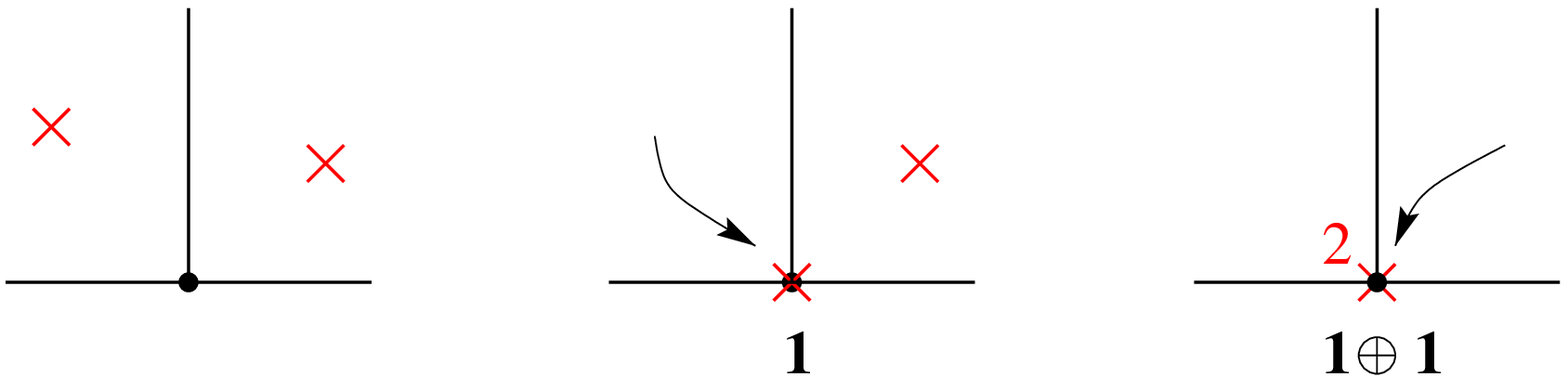}\\[.2in]
\parbox{6in}{Figure 1: Any number of fivebranes can sequentially attach
to a consistent intersection six-plane while retaining anomaly cancellation.
Gauge-singlet twisted hypermultiplets are indicated by labels
beneath the vertex.}
\end{center}
\end{figure}

This analysis generalizes to include the absorption of any integer number 
$M$ of fivebranes.  In this general case, we have $n_H\to n_H+M$,
$n_T\to n_T+M$ and $g\to g+M$, and we determine
\brr \widehat{I} &\longrightarrow&
     \widehat{I}-\frac{M}{8}\,(\,{\rm tr}\,R^2-2\,{\rm tr}\,F^2\,)^2
     \nonumber\\[.1in]
     &=& \widehat{I}
     -\sum_{i=1}^M\,
     \ft18\,(\,{\rm tr}\,R^2-2\,{\rm tr}\,F^2\,)^2\,.
\err
So, regardless of how many fivebranes we sequentially pile onto the 
six-plane, we can maintain anomaly freedom at the vertex provided each 
successive tensor multiplet independently mediates a Green-Schwarz mechanism.
The process of two fivebranes successively absorbed onto a particular
intersection six-plane is depicted in Figure 1, which employs the 
sort of ``vertex-diagrams" which were introduced 
and explained in \cite{phase}.

We aim to classify sets of consistent local vertices.
For reasons also explained in \cite{phase},   
these are conveniently assembled into arrays such that each column contains
vertices with a specific number $n_T$ of twisted tensors, and each row
contains vertices with a specific value of the difference $g-n_T$,
where $g$ is the magnetic charge.
The consistent sequential absorption of fivebranes therefore implies
a horizontal motion in such an array of vertices, and we refer to the
set of solutions in a given row of this array as 
describing a horizontal hierarchy.

\setcounter{equation}{0}
\section{Non-Prime Orbifolds}
Non-prime orbifolds have more than one type of fixed plane.  
Hyperplanes fixed under the full discrete group we call
{\it primary}.  {\it Secondary}
fixed planes are invariant under some subgroup of ${\bf Z}_M$
but comprise nontrivial multiplets under ${\bf Z}_M$.
For instance, the ${\bf Z}_4$ orbifold has four primary fixed planes 
invariant under ${\bf Z}_4$ and six additional secondary planes, each describing
a ${\bf Z}_4$ doublet of planes invariant under ${\bf Z}_2\subset {\bf Z}_4$.
Finally, the ${\bf Z}_6$ orbifold has one primary fixed plane 
invariant under
${\bf Z}_6$, four ${\bf Z}_3\subset {\bf Z}_6$ secondary planes assembled 
as ${\bf Z}_6$ doublets and five more ${\bf Z}_2\subset {\bf Z}_6$
secondary planes assembled as ${\bf Z}_6$ triplets.

For the case of prime orbifolds, it is straightforward to determine
the untwisted contribution to the local anomaly\footnote{By ``untwisted"  
we refer to the local contribution due to nonlocalized fields, such as
the bulk supergravity fields and (for the case of the
six-dimensional anomalies) the ten-dimensional fields.}.  This is
obtained by determining the {\it total} untwisted anomaly via
index theorems and then distributing
this result uniformly over the indistinguishable fixed planes.
For the case of nonprime orbifolds, the untwisted contribution is less
straightforward to determine due to the distinguishability
of the fixed planes.  As explained in \cite{ksty}, it is nevertheless
clear how to resolve this problem.  In a nonprime orbifold, the 
secondary planes are each identical to a primary fixed plane in 
a corresponding prime orbifold.  
(For instance, in the case of the ${\bf Z}_6$ orbifold
the secondary ${\bf Z}_3\subset {\bf Z}_6$ fixed-planes are locally
identical to the primary fixed planes in the ${\bf Z}_3$
orbifold.)  But, since we know the untwisted anomaly for the prime 
orbifolds, we can readily determine the 
total anomaly due to all secondary fixed-planes in the nonprime orbifolds.
To determine
the untwisted contribution to the local anomaly on the primary
fixed planes of a nonprime orbifold, we first determine the total 
anomaly via index theorems, and then subtract off the net contribution 
from all of the secondary fixed-planes.  What remains is then uniformly
distributed over the set of primary fixed-planes, which are 
indistinguishable. 

It is straightforward to systemetize the process described
in the previous paragraph.  However, the analysis is
facilitated if we first define a few useful conventions.
We refer to a primary fixed-plane as a ``1-type" fixed-plane
and a secondary fixed-plane as ``2-type".  If there are two 
distingishable types of secondary fixed-planes (as in the case
of the ${\bf Z}_6$ orbifold), then one
of these is denoted ``2-type" and the other is denoted ``3-type",
with the distintion left as a choice of convention.  And so forth.
We then define $N_n$ as the number of $n$-type fixed-planes 
in the orbifold in question and $B_n$ as the number of
primary-fixed planes for a prime orbifold of $n$-type.  Finally,
$h_{(n)}$ is the number of (untwisted) hypermultiplets derived
from the bulk supergravity in a prime orbifold of $n$-type
\footnote{Note that in addition to the untwisted hypermultiplets
there is always one untwisted tensor multiplet asociated with the
bulk supergravity.}.
For each type of fixed plane, we can then define an ``effective" number of
$n$-type fixed planes as
\brr f_{(n)}=\frac{N_1}{N_n}\,B_n \,.
\err
Each of the parameters defined so far plays a special role in the
determination of the local anomalies.
Also useful are the ``effective" fixed plane multiplicity and the
``net effective hypermultiplet ratio", respectively defined by
\brr f &\equiv& (\,\frac{1}{f_{(1)}}
     -\sum_{n\ne 1}\frac{1}{f_{(n)}}\,)^{-1}
     \nonumber\\[.1in]
     {\cal H} &=& \frac{h_{(1)}}{f_{(1)}}
     -\sum_{n\ne 1}\,\frac{h_{(n)}}{f_{(n)}} \,.
\label{fh}\err
Furthermore, the ${\bf Z}_M$ projection can act to leave different
subgroups ${\cal G}^{(n)}$ invariant at the different $n$-type
fixed planes, owing to the fact that on secondary planes only a subgroup
of ${\bf Z}_M$ induces the projection on the $E_8$ root lattice.
The relevant projection is partially accounted for by the
multiplicities of vector and hyper multiplets which remain from the
$E_8$ fields on the fixed six-planes of $n$-type.  These are
defined as $V^{(n)}$ and $H^{(n)}$, respectively.

Cancellation of the ${\rm tr}\,R^4$ anomaly locally, on a given
intersection six-plane, requires
\brr n_H-n_V=30g-29n_T+P+Q
\label{nhnv}\err
where $P$ is the supergravity contribution and $Q$ is the 
contribution deriving from the ten-dimensional $E_8$ fields.
These are given by
\brr P &=& \frac{122}{f}-\frac{\cal H}{2}
     \nonumber\\[.1in]
     Q &=& \frac{1}{f_{(n)}}\,(\,V^{(1)}-H^{(1)}\,)
     -\sum_{n\ne 1}\frac{1}{f_{(n)}}\,(\,V^{(n)}-H^{(n)}\,) \,.
\label{pq}\err
Note that equation (\ref{nhnv}) is the requirement that $A=0$, as described
in the previous section.

For each of the four $S^1/{\bf Z}_2\times T^4/{\bf Z}_M$ orbifolds,
the parameters $N_n$, $B_n$, $h_{(n)}$ and
$f_{(n)}$ for each type of fixed plane, and also the
values of $f$, ${\cal H}$ and $P$ which can be derived from these using
(\ref{fh}) and (\ref{pq}), are listed in Table 1.
From the data exhibited in Table 1, we can roughly characterize the 
orbifold geometry in each case by the pictures shown in Figure 2.

\begin{figure}
\begin{center}
${\bf Z}_2$ Orbifold:\\[.1in]
\begin{tabular}{|cc|cc|cc|}
\hline
&&&&&\\[-5mm]
\hspace{.1in}$n$\hspace{.05in} & 
\hspace{.05in}$P_n$ \hspace{.1in} & 
\hspace{.1in}$N_n$ \hspace{.05in} & 
\hspace{.05in}$B_n$ \hspace{.1in} & 
\hspace{.1in}$h_{(n)}$ \hspace{.05in} & 
\hspace{.05in}$f_{(n)}$ \hspace{.1in} \\[.1in]
\hline
\hline
&&&&&\\[-5mm]
1 & ${\bf Z}_2$ & 16 & 16 & 4 & 16 \\[.1in]
\hline
\end{tabular}\\[1mm]
\begin{tabular}{ccc}
\hspace{.1in}$f=16$ \hspace{.05in} & 
\hspace{.05in}${\cal H}=\frac{1}{4}$ \hspace{.1in} &
\hspace{.05in}$P=\frac{15}{2}$ \hspace{.05in} 
\end{tabular}\\[.3in]
${\bf Z}_3$ Orbifold:\\[.1in]
\begin{tabular}{|cc|cc|cc|}
\hline
&&&&&\\[-5mm]
\hspace{.1in}$n$\hspace{.05in} & 
\hspace{.05in}$P_n$ \hspace{.1in} & 
\hspace{.1in}$N_n$ \hspace{.05in} & 
\hspace{.05in}$B_n$ \hspace{.1in} & 
\hspace{.1in}$h_{(n)}$ \hspace{.05in} & 
\hspace{.05in}$f_{(n)}$ \hspace{.1in} \\[.1in]
\hline
\hline
&&&&&\\[-5mm]
1 & ${\bf Z}_3$ & 9 & 9 & 2 & 9 \\[.1in]
\hline
\end{tabular}\\[.1in]
\begin{tabular}{ccc}
\hspace{.1in}$f=9$ \hspace{.05in} & 
\hspace{.05in}${\cal H}=\frac{2}{9}$ \hspace{.1in} &
\hspace{.05in}$P=\frac{121}{9}$ \hspace{.05in}
\end{tabular}\\[.3in]
${\bf Z}_4$ Orbifold:\\[.1in]
\begin{tabular}{|cc|cc|cc|}
\hline
&&&&&\\[-5mm]
\hspace{.1in}$n$\hspace{.05in} & 
\hspace{.05in}$P_n$ \hspace{.1in} & 
\hspace{.1in}$N_n$ \hspace{.05in} & 
\hspace{.05in}$B_n$ \hspace{.1in} & 
\hspace{.1in}$h_{(n)}$ \hspace{.05in} & 
\hspace{.05in}$f_{(n)}$ \hspace{.1in} \\[.1in]
\hline
\hline
&&&&&\\[-5mm]
1 & ${\bf Z}_4$ & 4 & 4 & 2 & 4 \\[.1in]
2 & ${\bf Z}_2$ & 6 & 16 & 4 & $\ft{32}{3}$ \\[.1in]
\hline
\end{tabular}\\[.1in]
\begin{tabular}{ccc}
\hspace{.1in}$f=\frac{32}{5}$ \hspace{.05in} & 
\hspace{.05in}${\cal H}=\frac{1}{8}$ \hspace{.1in} &
\hspace{.05in}$P=19$ \hspace{.05in}
\end{tabular}\\[.3in]
${\bf Z}_6$ Orbifold:\\[.1in]
\begin{tabular}{|cc|cc|cc|}
\hline
&&&&&\\[-5mm]
\hspace{.1in}$n$\hspace{.05in} & 
\hspace{.05in}$P_n$ \hspace{.1in} & 
\hspace{.1in}$N_n$ \hspace{.05in} & 
\hspace{.05in}$B_n$ \hspace{.1in} & 
\hspace{.1in}$h_{(n)}$ \hspace{.05in} & 
\hspace{.05in}$f_{(n)}$ \hspace{.1in} \\[.1in]
\hline
\hline
&&&&&\\[-5mm]
1 & ${\bf Z}_6$ & 1 & 1 & 2 & 1 \\[.1in]
2 & ${\bf Z}_3$ & 4 & 9 & 2 & $\ft94$ \\[.1in]
3 & ${\bf Z}_2$ & 5 & 16 & 4 & $\ft{16}{5}$ \\[.1in]
\hline
\end{tabular}\\[.1in]
\begin{tabular}{ccc}
\hspace{.1in}$f=\frac{144}{35}$ \hspace{.05in} & 
\hspace{.05in}${\cal H}=-\frac{5}{36}$ \hspace{.1in} &
\hspace{.05in}$P=\frac{535}{18}$ \hspace{.05in}
\end{tabular}\\[.5in]
\parbox{6in}{Table 1: Some of the orbifold data associated with 
the four $S^1/{\bf Z}_2\times K3$ orbifolds.}
\end{center}
\end{figure}

\begin{figure}
\begin{center}
\includegraphics[width=1.5in,angle=0]{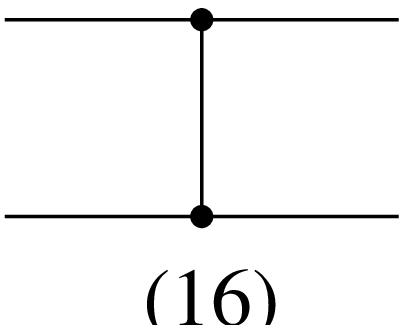}\\[.6in]
\includegraphics[width=1.5in,angle=0]{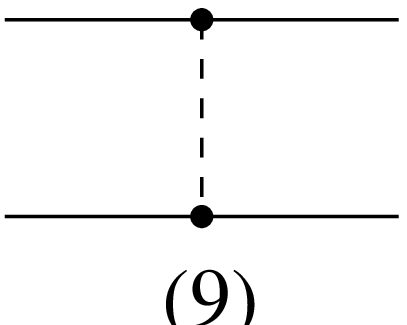}\\[.6in]
\includegraphics[width=2.5in,angle=0]{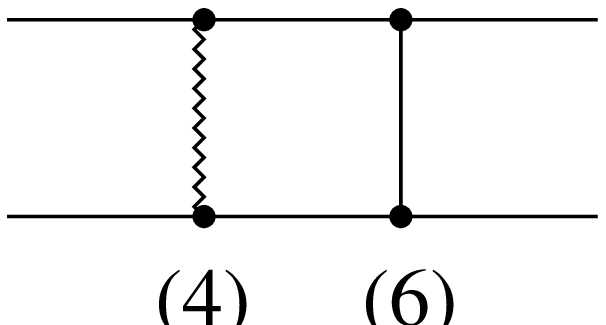}\\[.6in]
\includegraphics[width=3.5in,angle=0]{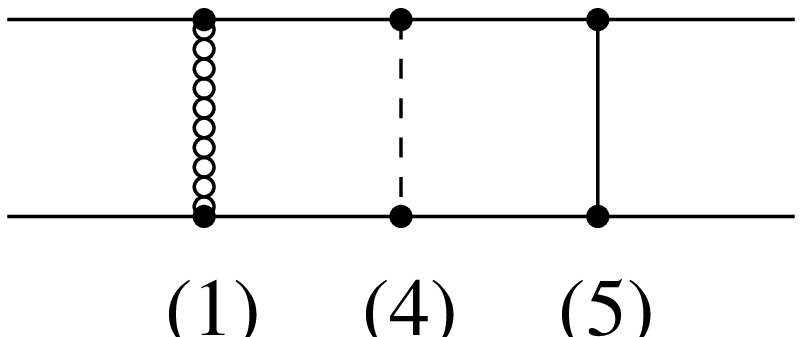}\\[.6in]
\parbox{6in}{Figure 2: Orbifold
fixed-plane geometry for the four $S^1/{\bf Z}_2\times T^4/{\bf Z}_M$
orbifolds.  Top to bottom, we depict the cases $M$=2, 3, 4
and 6.  Horizontal lines are ten-planes, vertical lines are seven-planes,
and the intersections are six-planes.
Solid, dotted, jagged and bubbly lines respectively denote 
${\bf Z}_2$, ${\bf Z}_3$, ${\bf Z}_4$ and ${\bf Z}_6$ fixed planes.
Multiplicities are shown by the numbers in parentheses.}
\end{center}
\end{figure}
    
In the general case, the anomaly is described by
equation (\ref{bcd}).  However, in the case of nonprime 
orbifolds the weighted quantum
trace should include the relevant secondary-plane subractions.
As a result, we modify equation (\ref{wqt}), replacing it
with the following expression,
 \brr {\rm trace}'\,F^r \equiv \sum_{i=1}^p\sum_{\cal R}\,\bpl
     (-1)^{2J}\,\frac{1}{f_{(1)}}\,{\rm tr}_{\,{\cal R}_{(1)}}\,F_i^r
     -\sum_{n\ne 1}\,\frac{1}{f_{(n)}}\,(-1)^{2J}\,
     \,{\rm tr}_{\,{\cal R}_{(n)}}\,F_i^r 
     \,\bpr\,.
\label{wqtnp}\err
Note that for the case of prime orbifolds, $f_{(1)}=f$ is the only
fixed-plane multiplicity parameter, so that in that case 
(\ref{wqtnp}) is identical to (\ref{wqt}).  Thus, equation (\ref{wqtnp})
incorporates the case of prime orbifolds.
 
The horizontal hierarchy is also valid in
the case of non-prime orbifolds.  This is because the shift (\ref{shift})
in the anomaly only involves the first terms in (\ref{bcd}),
which, in turn,
only depend on local physics at the prime orbifold point (i.e. they do not
involve the secondary subtractions).  Thus, in the general case, 
available fivebranes can move to and consistently wrap any
fixed six-plane in the orbifold.
As a corollary, it is possible for six-planes to emit these fivebranes
back into the bulk.  Alternatively, it is possible that a wrapped fivebrane
(which is more properly identified as a torsion-free sheaf 
\cite{opp}) can smoothly connect on moduli space to a small gauge instanton.
In this case, the $E_8$ factor is broken to a 
subgroup, the twisted tensor field is replaced
with various hypermultiplet moduli, and the magnetic charge increases
by one unit.  Fractional transitions are also possible, in which
the fivebrane data is distributed over some number of
six-planes.

Thus, it is sensible to seek consistent
orbifold vertices which are linked in moduli space to 
vertices without twisted tensor fields (i.e. $n_T=0$).  
In this case, all data is resolved by imposing that
the anomaly vanish identically. The most basic of these solutions
harbors no gauge instanton.
In that case, the local magnetic charge assumes a minimal 
value which, for primary orbifold points, is given by $g_{\rm min}=-12/f$.
More interesting solutions have 
nontrivial breaking of the $E_8$ factor. In these cases
$g>-12/f$, and the difference $g-g_{\rm min}$ describes 
a topological charge of the small instanton (or toron)
residing on the fixed-plane.
 
\setcounter{equation}{0}
\section{Local Anomaly Cancellation}
In order to determine the hierarchically-linked sector of
moduli space defined in the preceeding paragraph, 
for each orbifold compactification under consideration, we should first
determine anomaly-free vertices with $n_T=0$.  Specifically, 
we seek sets of magnetic and electric charges, as well
specific $E_8$ breaking patterns and twisted hypermultiplet
representation content, which ensures that $A=0$ 
(i.e. equation (\ref{nhnv})) and also ensures that each of the 
coefficients $B$, $C^I$ and $D^{IJ}$ defined
in equation (\ref{bcd}) vanishes identically.
To systemetize such a search, equation (\ref{bcd}) can 
be more conveniently re-expressed soley in terms of various multiplicities
which appear in the untwisted and twisted spectrum,
and in terms of rational indices which characterize 
representations.  In this way, the anomaly cancellation requirement
reduces to a set of (overconstrained) algebraic relations amongst 
various rational and integer parameters.  To determine these
relations,  we require a few more conventional definitions.

Each representation of each group factor appears with a multiplicity.
We define $n_1(\,{\cal R}\,)$ as the number of vector 
multiplets minus the number of hypermultiplets transforming in 
the representation ${\cal R}$ of the group ${\cal G}_I$ in the spectrum 
corresponding to the primary orbifold points. 
(The relevant group factor ${\cal G}_I$ is invariably clear from the context.)
Similarly, there are representation multiplicities associated with 
each secondary orbifold point.   These are defined in analogy to 
$n_1({\cal R})$.  Thus, $n_n({\cal R})$ is the number of vector
multiplets minus the number of hypermultiplets transforming in
the representation ${\cal R}$ in the spectrum corresponding to 
$n$-type fixed planes.
Finally, the parameter $\tilde{n}({\cal R})$ 
describes the representation content of the twisted spectrum
in analogy to the definitions made above.
We also need to decompose the $E_8$ factors in terms
of the gauge group ${\cal G}\subset E_8$ which remains unbroken on the
primary fixed points, using the relevant branching rule \cite{slansky}.
This is needed to determine the inflow anomaly.
We therefore define $n_0({\cal R})$ as the 
(nonnegative) multiplicity parameter associated 
with this decomposition.

It proves useful to define an ``effective multiplicity parameter" 
for each representation which appears in the untwisted spectrum,
\brr {\rm z}(\,{\cal R}\,) &\equiv&
     \frac{1}{f_{(1)}}\,n_1({\cal R})
     -\sum_{n\ne 1}\,\frac{1}{f_{(n)}}\,n_n({\cal R}) \,.
\err
This definition usefully accounts for both the secondary subtractions
as well as the appropriate distribution factors.  It is 
also useful to define a number of ``anomaly weight factors"
\brr X_I &=& \ft{1}{30}\,\sum_{\cal R}\,n_0(\,{\cal R}\,)\,
     {\cal I}_2(\,{\cal R}\,)
     \nonumber\\[.1in]
     Z_I^{(4)} &=& \sum_{\cal R}\,\bpl\,{\rm z}(\,{\cal R}\,)
     +\tilde{n}(\,{\cal R}\,)\,\bpr\,{\cal I}_4(\,{\cal R}\,)
     \nonumber\\[.1in]
     Z_I^{(2)} &=& \sum_{\cal R}\,\bpl\,{\rm z}(\,{\cal R}\,)
     +\tilde{n}(\,{\cal R}\,)\,\bpr\,{\cal I}_2(\,{\cal R}\,)
     \nonumber\\[.1in]
     Z_I^{(2,2)} &=& \sum_{\cal R}\,\bpl\,{\rm z}(\,{\cal R}\,)
     +\tilde{n}(\,{\cal R}\,)\,\bpr\,{\cal I}_{2,2}(\,{\cal R}\,)
     \nonumber\\[.1in]
     Z_{IJ}^{(X)} &=& \sum_{{\cal R}_I\times{\cal R}_J} 
     \bpl\,{\rm z}(\,{\cal R}_I\times {\cal R}_J\,)
     +\tilde{n}(\,{\cal R}_I\times {\cal R}_J\,)\,
     \bpr\,{\cal I}_2(\,{\cal R}_I\,)\,{\cal I}_2(\,{\cal R}_J\,) \,,
\label{awf}\err
where the rational numbers
${\cal I}_2(\,{\cal R})$, ${\cal I}_{2,2}(\,{\cal R}\,)$
and ${\cal I}_4(\,{\cal R}\,)$ are representation indices for the
representation ${\cal R}$, which are defined in \cite{phase}.
The weight factors (\ref{awf}) encorporate the twisted multiplicities
$\tilde{n}({\cal R})$ in a way which facilitates an algorithmic 
search for consistent vertices.

In terms of the weight factors given in
(\ref{awf}), we can more efficiently describe the anomaly
coefficients defined in (\ref{bcd}), expressing them soley in terms 
of the multiplicities,  representation indices, and orbifold plane
multiplicities.  We find
\footnote{Symmetrization has {\it weight one}, so that
$\rho_{(I}\,X_{J)}=\ft12(\rho_I\,X_J+\rho_J\,X_I)$.}
\brr B &=& \frac{1-\eta}{2\,f}
     -\frac{n_T}{8}
     \nonumber\\[.1in]
     C^I &=& (\,\frac{g}{2}+\frac{\eta}{f}\,)\,X_I
     +\ft16\,Z_I^{(2)}
     +\rho_I
     \nonumber\\[.1in]
     D^{II} &=& 
     -\ft12\,g\,X_I^2
     -\ft23\,Z_I^{(2,2)}
     -2\,\rho_I\,X_I
     \nonumber\\[.1in]
     D^{IJ} &=&
     -\ft12\,g\,X_I\,X_J
     -2\,Z_{IJ}^{(X)}
     -\rho_{(I}\,X_{J)} \hspace{.3in} ,\,I\ne J \,,
\label{tidy}\err
where $\rho_I=\rho$ if ${\cal G}_I={\cal G}_7$ and $\rho_I=0$ otherwise.  Equation (\ref{tidy}) describes the anomaly completely in terms of 
rational data.

Cancellation of all anomalies is necessary when $n_T=0$.  
In this case, the requirement $B=0$ implies $\eta=1$,
as is readily seen from equation (\ref{tidy}).
We henceforth assume that $\eta=1$.  In this case,   
the complete set of
anomaly cancellation requirements are given by (\ref{nhnv}), which
ensures cancellation of the ${\rm tr}\,R^4$ anomaly, 
and also the following
\brr Z_I^{(4)} &=& 0
     \nonumber\\[.1in]
     Z_I^{(2)} &=& -3\,(\,g+\frac{2}{f}\,)\,X_I
     -6\,\rho_I
     \nonumber\\[.1in]
     Z_I^{(2,2)} &=& -\ft34\,g\,X_I^2
     -3\,\rho_I\,X_I
     \nonumber\\[.1in]
     Z_{IJ}^{(X)} &=& -\ft14\,g\,X_I\,X_J
     -X_{(I}\,\rho_{J)} \hspace{.4in},\,I\ne J \,,
\label{zzzz}\err
which correspond, respectively, to the cancellation of ${\rm tr}\,F_I^4$,
${\rm tr}\,R^2\wedge {\rm tr}\,F_I^2$,
$(\,{\rm tr}\,F_I^2\,)^2$,
and ${\rm tr}\,F_I^2\wedge {\rm tr}\,F_J^2$ anomalies.  

Equation (\ref{zzzz}) is of central importance in this paper.
The primary objective in our analysis, which is summarized in the next
section, is to discover and classify consistent local vertices
by solving equation (\ref{zzzz}).  
This means identifying 
permitted breaking patterns of the ten-dimensional $E_8$ gauge
group,  any extra seven-dimensional twisted gauge
groups, and any extra twisted matter.  In addition, this
means identifying the appropriate magnetic charge $g$ associated
with the vertex in question, and also to identify the 
remaining electric parameter $\rho$ associated with the intersecting
seven-plane. 

\setcounter{equation}{0}
\section{Consistent Models}
In this section, we analyze each of the four 
$S^1/{\bf Z}_2\times T^4/{\bf Z}_M$ orbifolds and establish a 
classification scheme for the solutions to the anomaly matching problem.
In each case we obtain a set of
vertices by finding solutions to the constraints
listed in Section 4.  The vertices are represented by specialized
``vertex diagrams" which are exhibited in the following section.
The technical aspects of the anomaly analysis are
summarized in the tables in the appendix.
The interested reader is encouraged to work through the appendix
to verify the anomaly matching for each of the cited vertices.
In each case, we use the elemental vertices as building blocks
which we assemble into a variety of 
global ``models", represented by ladder diagrams, 
which are also explained in \cite{phase}.
There is an extra constraint imposed on the global construction. 
Namely, the net magnetic charge, obtained by summing the
values of $g$ on each of the six-planes and total number of (unit-charged)
fivebranes living in the bulk, must vanish.

We must also specify the manner in which the universal ${\bf Z}_2$
projection acts on the seven-dimensional fields.  As explained
in \cite{phase}, there are three possibilities.  
The seven-dimensional fields necessarily comprise an adjoint 
Yang-Mills supermultiplet, which decomposes in six dimensions as
one vector multiplet and one hypermultiplet, each transforming
as the adjoint of $\tilde{\cal G}_7$.  The universal ${\bf Z}_2$ acts
to project out some of this six-dimensional field content.  
The first possibility is that this projects out the hypermultiplet half, 
leaving a six-dimensional adjoint vector multiplet. The second possibility 
is the opposite of this, whereby the vector multiplet is projected out,
leaving a hypermultiplet.  This second case is permitted only if 
$\tilde{\cal G}_7$
is identified with one gauge factor in the 
unbroken subgroup ${\cal G}\subset E_8$
supplied from ten-dimensions.  (In this way, a gauge connection is supplied
from ten-dimensions, allowing us to project out the seven-dimensional
vector and still retain six-dimensional gauge invariance.)  On our vertex diagrams, we 
respectively denote the first two possibilities by a 
$V$ or an $H$ label on a downward arrow next to the vertical line 
which depicts the seven-plane.  The third possibility is that
the universal ${\bf Z}_2$ acts to break $\tilde{\cal G}_7$ to a
subgroup ${\cal H}\subset \tilde{\cal G}_7$.   On our vertex diagrams,
this case is denoted by an ${\cal H}$ next to the
downward arrow.  For our purposes, we only consider the special case 
where ${\cal H}$ is the Cartan subgroup of $\tilde{\cal G}_7$.  
To illustrate our vertex-diagram
conventions, we depict the three possibilities in Figure 3 for the 
choice $\tilde{\cal G}_7=SU(N)$.
\begin{figure}
\begin{center}
\includegraphics[width=6in,angle=0]{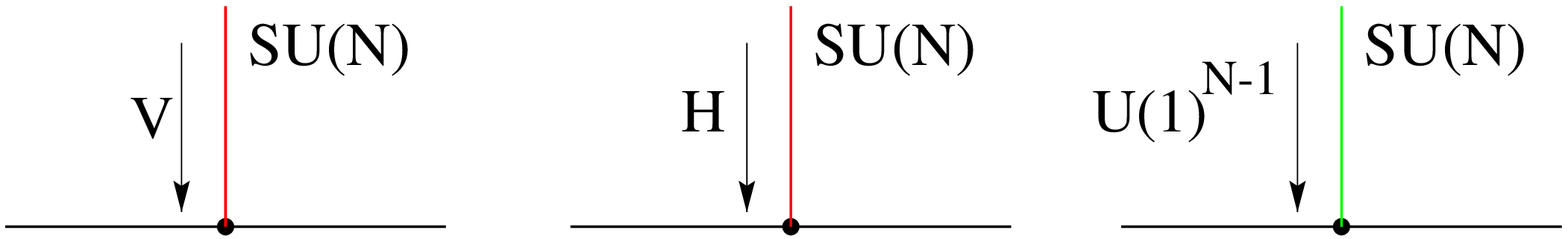}\\[.3in]
\parbox{6in}{Figure 3: Three possibilities indicating different 
projections induced by the action of universal ${\bf Z}_2$ on
$\tilde{\cal G}_7=SU(N)$.  Respectively, the three diagrams
indicate a projection onto an $SU(N)$ adjoint vector multiplet,  
an $SU(N)$ adjoint hypermultiplet, and the Cartan subgroup.
In the latter case, the surviving vector and hypermultiplet
content is explained in the text.}
\end{center}
\end{figure}
 
For the cases where $\tilde{\cal G}_7$ is broken to its Cartan
subgroup, we need to understand how the projection acts on the 
hypermultiplets.  We specialize the discussion to those cases
where $\tilde{\cal G}_7=SU(N)$, so that the Cartan subgroup is
$U(1)^{N-1}$.  Clearly, we have $N-1$ surviving vector multiplets
in this case.  But there remains a certain freedom in the way that
${\bf Z}_2$ can act on the hypermultiplet half of the seven-dimensional
fields.  Characteristically, we use anomaly cancellation to
resolve this question.  What we typically find is a need
for retaining precisely $2(N-1)$ hypermultiplets out of the original
$N^2-1$. This can be realized if ${\bf Z}_2$ acts as a modified 
Weyl reflection on the $SU(N)$ root lattice.  Under the modification,
each Weyl-invariant root except for the one defining the Weyl
reflection itself receives a parity flip. (Thus, we retain only one
of the Weyl-invarant roots, while the remaining roots are halved.)
Although this prescription is ad-hoc, it does illustrate
how the needed accounting can be accomodated at the level of discrete
projections.  Obviously, a more fundamental description of {\it M}-theory
is needed to provide a better explanation.  But for the time being, 
anomaly matching is the best criteria which we have available.

Another issue concerns the electric parameter $\rho$.  As it turns out,
for any consistent primary ${\bf Z}_M$ orbifold vertex, this parameter 
is unity whenever the seven-dimensional gauge group is $SU(M)$, provided
this group is unbroken by the universal ${\bf Z}_2$ projection.
However, when this group is broken to its Cartan subgroup $U(1)^{M-1}$ 
we require $\rho=0$.  Furthermore, 
in addition to the ``conventional" vertices which involve
$\tilde{\cal G}_7=SU(M)$ 
(either broken or unbroken by the universal ${\bf Z}_2$),
there exist additional ``unconventional" vertices, in which
the seven-dimensional gauge group is $U(1)^{M-1}$ to begin with.
In these cases, we also require $\rho=0$.  As a rule of thumb,
whenever a vertex supports an unbroken $SU(M)$ from the seven-plane
then $\rho=1$.  Otherwise $\rho=0$.  As explained in \cite{phase},
the possibility of two different values for $\rho$ poses an interesting
riddle if we want to fully understand transitions which connect
a ``$\rho\!=\!0$" phase to a ``$\rho\!=\!1$" phase.  We have more
to say about this riddle below.

We exhibit the microscopic details of the models in 
Section 6 and in the appendix. 
Here we offer a more superficial description of our results, 
and also explain how
these relate to the results in \cite{ksty}.  A ``model" is obtained by 
assembling a collection of consistent vertices in such a manner that 
the global constraints are also satisfied.  There are two global 
constraints.  The first is that the sum of all magnetic charges, including
a unit positive charge for each bulk fivebrane, must vanish.  This 
tells us that the sum of all of the independent orbifold vertex charges
within a given orbifold 
must be nonpositive.
In this way, the magnetic balance can be achieved, if necessary, by
including fivebranes in the bulk.  The second global constraint is that
the electric parameters $\eta$ and $\rho$ associated with vertices linked
by a particular seven-plane must be identical. This is because these parameters
ostensibly derive from seven-dimensional interactions common to 
each of the two vertices in question.
 In this paper, we
also impose a third global constraint: that there exists a sensible
small-radius limit in which the untwisted states associated with 
individual vertices can be collectively resolved into a coherent low-energy untwisted 
spectrum.  This implies that the untwisted spectrum associated to 
each six-plane intersection within a given ten-plane must combine
into representations of a particular low-energy gauge group
${\cal G}_{\rm eff}\subset E_8$ 
\footnote{This precludes other potentially interesting
situations which might describe orbifolds without sufficient moduli 
for the compact dimensions to be ``shrunk" to a point, thereby 
describing universes with interesting ``large" extra dimensions.}.

The interesting models which satisfy our criteria are indicated 
in Table 2.  In that table we indicate, for each case, the effective gauge group
obtained when all of the compact dimensions are shrunk to a point.
In this way, we describe effective six-dimensional field theories
with $N=1$ supersymmetry, with the indicated gauge groups, and 
with the indicated multiplicities $N_H$, $N_V$ and $N_T$, respectively
describing the total nunber of hyper, vector and tensor multiplets,
including both twisted and untwisted contributions.  More complete
descriptions of each of these models are given below. 
It may prove useful to examine the figures in Section 6 while 
reading this section.
 Some of these models, specifically Models
1, 3 with $a=0$, 5 and 9 with $b=0$ are not new to this paper, but
were previously presented in \cite{ksty}.  This includes one model for
each of the four K3 orbifolds and one special orbifold, namely Model 3,
with the unconventional seven-dimensional $U(1)$ gauge group.  
The new models which we present in this paper fall into two categories.  The first
category includes ``fully nonperturbative" models which include
extra tensor multiplets and a completely unbroken $E_8\times E_8$ gauge group.
For the cases of the ${\bf Z}_2$, ${\bf Z}_3$ and ${\bf Z}_4$ orbifolds,
these new models are smoothly connected to the aforementioned models via
phase transitions involving small instanton/fivebrane phase transitions.
The second category of new models is given by Model 10, which is a new
``perturbative" solution which permits us to 
discuss some interesting features of the ${\bf Z}_6$ orbifold.

\begin{figure}
\begin{center}
\begin{tabular}{|c||c|c|ccc|}
\hline
&&&&&\\[-5mm]
Orbifold & Model & Gauge Group & $N_H$ & $N_V$ & $N_T$ \\[.1in]
\hline
&&&&&\\[-5mm]
${\bf Z}_2$ & 1 & $E_7\times SO(16)\times SU(2)$ & 500 & 256 & 1 \\[.1in]
& 2 & $E_8\times E_8\times U(1)^{16}$ & 60 & 512 & 25 \\[.1in]
& 3 & $E_8\times E_7\times SU(2)\times U(1)^{a}$ & 628+16a & 384+16a & 1 \\[.1in]
& 4 & $E_8\times E_8\times U(1)^{a}$ & 44+16a & 496+16a & 25 \\[.1in]
\hline
&&&&&\\[-5mm]
${\bf Z}_3$ & 5 & $E_6\times SU(9)\times SU(3)$ & 410 & 166 & 1 \\[.1in]
& 6 & $E_8\times E_8\times U(1)^{16}$ & 62 & 514 & 25 \\[.1in]
\hline
&&&&&\\[-5mm]
${\bf Z}_4$ 
& 7 & $SO(10)\times SU(8)\times SU(4)\times SU(2)$ & 370 & 126 & 1 \\[.1in]
& 8 & $E_8\times E_8\times U(1)^{16}$ & 62 & 514 & 25 \\[.1in]
\hline
&&&&&\\[-5mm]
${\bf Z}_6$ 
& 9 & $SU(9)\times SU(6)\times SU(3)\times SU(2)\times U(1)^b$ 
& 359+16b & 115+16b & 1 \\[.1in]
& 10 & $E_6\times SU(8)\times SU(2)\times U(1)^{2+b}$ & 379+16b & 135+16b & 1 \\[.1in]
& 11 & $E_8\times E_8\times U(1)^{16}$ & 62 & 514 & 25 \\[.1in]
\hline
\end{tabular} \\[.3in]
\parbox{6in}{Table 2: A tabulation of the orbifold models discussed 
in the text. The unspecified parameters $a$ and $b$ can take any integer
value such that 
$0\le a\le 16$ and $0\le b\le 5$.} 
\end{center}
\end{figure}

Note that Model 2 and Model 4 with $a=16$ have the same gauge group, 
but have a different
microscopic description.    
In Model 2, the seven-dimensional gauge
group at each of the sixteen seven-planes 
is $SU(2)$, which is broken to $U(1)$
by the universal ${\bf Z}_2$ projection. 
(This is depicted in Figure 6.) 
In Model 4, on the other hand, the seven-dimensional gauge groups
are $U(1)$ to begin with. (This is shown in Figure 8.)
As explained at length in \cite{phase},    
we envision a phase transition mediated
by fivebranes which links Model 1 with Model 2, and another phase
transition mediated by fivebranes which links Model 3 with Model 4.
Model 1 and Model 2 then live on what we would call the ``conventional"
branch of moduli space, and Models 3 and 4 live on an ``unconventional" branch.    
Models 3 and 4 are included in our survey because they correspond to 
the ${\bf Z}_2$ secondary plane used in Model 9.

For the ${\bf Z}_3$ and the ${\bf Z}_4$ orbifolds, we exhibit only the
conventional solutions.  In each case, there is a ``perturbative"
solution (with only one tensor multiplet) and a ``nonperturbative"
solution (with 25 tensor multiplets).   In each case, the perturbative and the
nonperturbative solutions are arguably smoothly connected on moduli space
due to fivebrane-mediated phase transitions.

As we mentioned above, each of the ``perturbative" models 
  except for Model 10 
were described in \cite{ksty}.  For the cases of the ${\bf Z}_2$, ${\bf Z}_3$ 
and ${\bf Z}_4$ orbifolds, these have fully conventional descriptions
in the sense that each fixed seven-plane, both primary and secondary,
supports a conventional $SU(N)$ gauge group which is unbroken by the 
universal ${\bf Z}_2$ projection.  This is not so for the case of the ${\bf Z}_6$
orbifold, however.  
For instance, Model 9 requires that the ${\bf Z}_2$ secondary plane
be unconventional.  Specifically, in the microscopic description
(shown in Figure 17), we see that 
each of the five ${\bf Z}_2$ secondary planes is similar to 
the ${\bf Z}_2$ primary
planes in Model 3. This has the unconventional seven-dimensional $U(1)$ gauge group 
rather than the conventional $SU(2)$. 

There is a model similar to Model 9, which we could call Model $9'$,
with the same gauge group
as Model 9 but with a different microscopic description.
In Model $9'$, the ${\bf Z}_2$ secondary planes do support the
conventional $SU(2)$ gauge group on the 
${\bf Z}_2$ secondary seven-planes, but these $SU(2)$ factors are
projected to the $U(1)$ Cartan subgroup on each of two six-plane
intersections. Model $9'$ is more directly analogous to Model 10.  
Model 9, however, more accurately describes the situation presented
in \cite{ksty}, since that paper does not address the possibility of
unconventional projections.  
The difference between Model 9 and Model $9'$ indicates
a generic phenomenon: anomaly cancellation may indicate a need for
{\it either} an unconventional seven-dimensional gauge group {\it or} 
a conventional
gauge group with an unconventional projection. But in some cases, the local
anomaly matching alone may not be sufficient to remove the ambiguity. 
      
Model 10 is interesting because it provides
a second ``perturbative" ${\bf Z}_6$ solution.  Notably, this
case also requires either unconventional vertices
or conventional vertices with unconventional projections.  
But in Model 10 it is the
primary ${\bf Z}_6$ vertices rather than the secondary 
${\bf Z}_2$ vertices which are distinguished:  
 either the primary seven-plane supports the unconventional $U(1)^5$
gauge group or it supports the conventional $SU(6)$ gauge group
which is projected to the $U(1)^5$ Cartan subgroup at the 
 six-plane intersections.  In the microscopic description of Model 10 
(shown in Figure 18) we exhibit the first of these possibilities.
(The second possibility could be called Model $10'$, and would be more directly
analogous to Model 9.)
Models $9'$ and 10 are smoothly connected 
on moduli space to Model 11 via phase transitions involving small 
instanton/fivebrane transitions similar to those discussed in \cite{phase}.

\setcounter{equation}{0} 
\section{Microscopic Details}
In this section, we present a more detailed analysis of
the models described above.  We do this by exhibiting a sequence 
of figures and tables which conveniently display the microscopic
structure of each model.   Specifically, 
for each of the four $K3$ orbifolds we list a set of consistent vertices
followed by a set of ladder diagrams which pictorially illustrate the
microscopic structure of each of the models described in the previous section.
We first give a description of how to interpret the figures associated with
the ${\bf Z}_2$ orbifold.  
This makes the remaining figures self-explanatory.  
  
A set of consistent vertices for the ${\bf Z}_2$ orbifold 
are shown in Figure 4a. The left-hand column in that figure depicts
the ``basic" vertices, without local tensor fields (i.e. $n_T=0$).  
The vertices in the right hand column have $n_T=1$, and
are obtained from those in the left by the addition
of a fivebrane.  (Additional columns are generated by 
adding additional fivebranes.)  The upper left hand vertex is
the most basic of all, since this represents a situation which 
includes neither fivebranes nor small gauge instantons.  As explained above, 
such a vertex has the smallest possible magnetic charge for this
orbifold.  Proceeding down the left-hand column are
solutions in which the ten-dimensional $E_8$ gauge group is
broken to subgroups, indicating the presence of small gauge instantons.
In these cases, the magnetic charge is necessarily greater than
the minimum value of $g_{\rm min}\!=\!-12/f\!=\!-3/4$, and the difference
$g\!-\!g_{\rm min}\!=\!g\!+\!3/4$ describes a topological charge 
of the instanton.  In \cite{phase} we postulated that vertices
degenerate in magnetic charge may be linked by phase transitions.
Thus, the upper right-hand vertex in Figure 4a could mutate
into the lower left-hand vertex (since these each have
magnetic charge $g=1/4$) and vice-versa.  Each of the vertices
in Figure 4a are ``conventional" because they
have seven-dimensional gauge group $SU(2)$.

For the case of vertices with unbroken $E_8$ it is, in fact,
impossible to decide on the basis of anomaly cancellation
alone whether the seven-dimensional gauge group is
actually $SU(2)$ broken to $U(1)$ by the
projection, or whether the seven-dimensional gauge group is
$U(1)$ itself.
Nevertheless, the ``conventional" choice of $SU(2)$ is suggested  
by the other vertices in the second and third row
of Figure 4a, which {\it are} required by anomaly cancellation
to support $SU(2)$ on the seven-planes.  A commonality in this respect
is seemingly important if the afformentioned phase transitions
are to be possible. 
 
In Figure 4b, we list additional ``unconventional" vertices for the 
${\bf Z}_2$ orbifold, which 
have seven-dimensional $U(1)$ gauge group rather than $SU(2)$.   
Once again, the left hand column includes
``basic" vertices, which have $n_T=0$. The vertices in the 
right hand column each have $n_T=1$, and 
are obtained from those in the left by the addition of a
fivebrane.  Notice that there are two vertices in Figure 4b with 
the minimal charge $-3/4$ and unbroken $E_8$:
one with a ``vector" projection
and the other with a ``hyper" projection. 
This degeneracy reflects
a limitation in the anomaly cancellation criteria.
This is because the fermions in the $U(1)$ multiplet are gauge singlets.
As a result, the only constraint available to us 
comes from gravitational anomalies, summarized by the
requirement $A=0$. For the case with $n_T=0$ and unbroken $E_8$,  
this requires $n_H\!-\!n_V=\ft12$,
where $n_H$ and $n_V$ are the effective multiplicities of twisted
hyper and vector multiplets.  As explained in
\cite{mlo,phase}, these include a factor of one-half for seven-dimensional
fields.  Thus, there are three possibilities: the conventional one
and the two unconventional choices with opposite projections.  For the case 
with hyper projection we have $n_H\!-\!n_V=1/2-0=1/2$. For the
case with vector projection we need one additional (six-dimensional) 
twisted singlet hypermultiplet, so that $n_H-n_V=1-1/2=1/2$. The 
single hypermultiplet is indicatd by the ${\bf 1}$ on the 
upper left-hand vertex in Figure 4b.

In Figures 5 and 6 we assemble the conventional ${\bf Z}_2$ vertices shown
in Figure 4a into global models along the lines described above.
In Figures 7 and 8 we similarly assemble the unconventional ${\bf Z}_2$
vertices shown in Figure 4b into global models.
In each case, the two
ten-planes are represented by parallel horizontal lines and the 
sixteen seven-planes are collectively depicted by one vertical line.  
(Since the sixteen seven-planes are indistinguishable,
we only depict one of these in the diagrams.
A complete global picture of the relevant model is therefore obtained by
visualizing sixteen copies of the associated figure.)   
Below each figure is a table in which all of the low energy  
hypermultiplets are listed in terms of their representation content.
 
There are alternatives to 
Models 3 and 4 obtained by making a change in the 
choice of $U(1)$ projection, replacing the V projection
with an $H$ projection.  This is due to the
degeneracy, described above, which is indicated in Figure 4b.  
By changing one $V$ projection to an $H$ projection, the effect is 
to subtract one $U(1)$ factor from the effective gauge group,
and to remove one vector and one hypermultiplet from the spectrum.
The versions which we have illustrated in Models 3 and 4 have the
maximal number of $U(1)$ factors. By altering the projections on some
subset of the fixed-seven planes, we can construct any alternative to
Models 3 and 4 with abelian factor $U(1)^a$ where $0\le a\le 16$.
Similar comments apply to Models 9 and 10.

 On the basis of this discussion, the figures associated with the 
${\bf Z}_3$, ${\bf Z}_4$ and ${\bf Z}_6$ orbifolds are sufficiently
similar that we do not need to present separate detailed description 
of each case; the figures speak for themselves.  The consistency of the
vertices presented in each case can be verified by working through the
numbers found in the tables in the appendix.  Table 3 lists all
of the figures which follow in this section. 

\begin{figure}
\begin{center}
\begin{tabular}{|c|c|c|}
\hline
Orbifold & Figure & Description \\[.1in]
\hline
\hline
${\bf Z}_2$ & 4a & Conventional ${\bf Z}_2$ vertices \\[.1in]
& 4b & Unconventional ${\bf Z}_2$ vertices \\[.1in]
& 5 & Model 1 \\[.1in]
& 6 & Model 2 \\[.1in]
& 7 & Model 3 \\[.1in]
& 8 & Model 4 \\[.1in]
\hline 
${\bf Z}_3$ & 9 & Conventional ${\bf Z}_3$ vertices \\[.1in]
& 10 & Model 5 \\[.1in]
& 11 & Model 6 \\[.1in]
\hline
${\bf Z}_4$ & 12 & Conventional ${\bf Z}_4$ vertices \\[.1in]
& 13 & Model 7 \\[.1in]
& 14 & Model 8 \\[.1in]
\hline
${\bf Z}_6$ & 15 & Conventional ${\bf Z}_6$ vertices \\[.1in]
& 16 & Model 9 \\[.1in]
& 17 & Model 10 \\[.1in]
& 18 & Model 11 \\[.1in]
\hline
\end{tabular} \\[.3in]
\parbox{2in}{Table 3: A Table of Figures}
\end{center}
\end{figure}  
 
\begin{figure}
\begin{center}
\includegraphics[width=6in,angle=0]{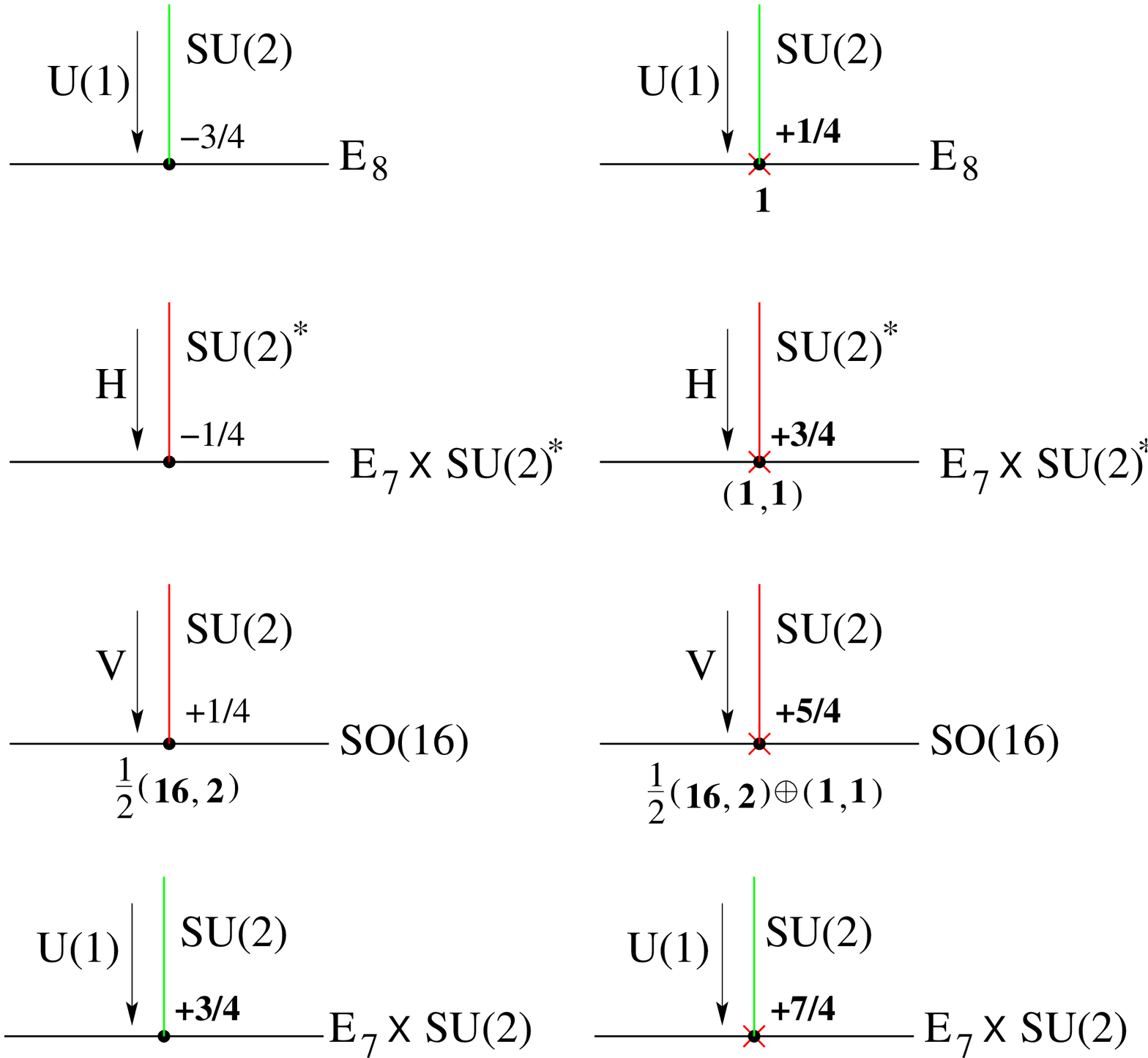}\\[.4in]
\parbox{6in}{Figure 4a: Conventional primary vertices  
in the ${\bf Z}_2$ orbifold.  These vertices have $SU(2)$ gauge
group on the seven-planes.  The asterix indicates identification
of factors.}
\end{center}
\end{figure}

\begin{figure}
\begin{center}
\includegraphics[width=6in,angle=0]{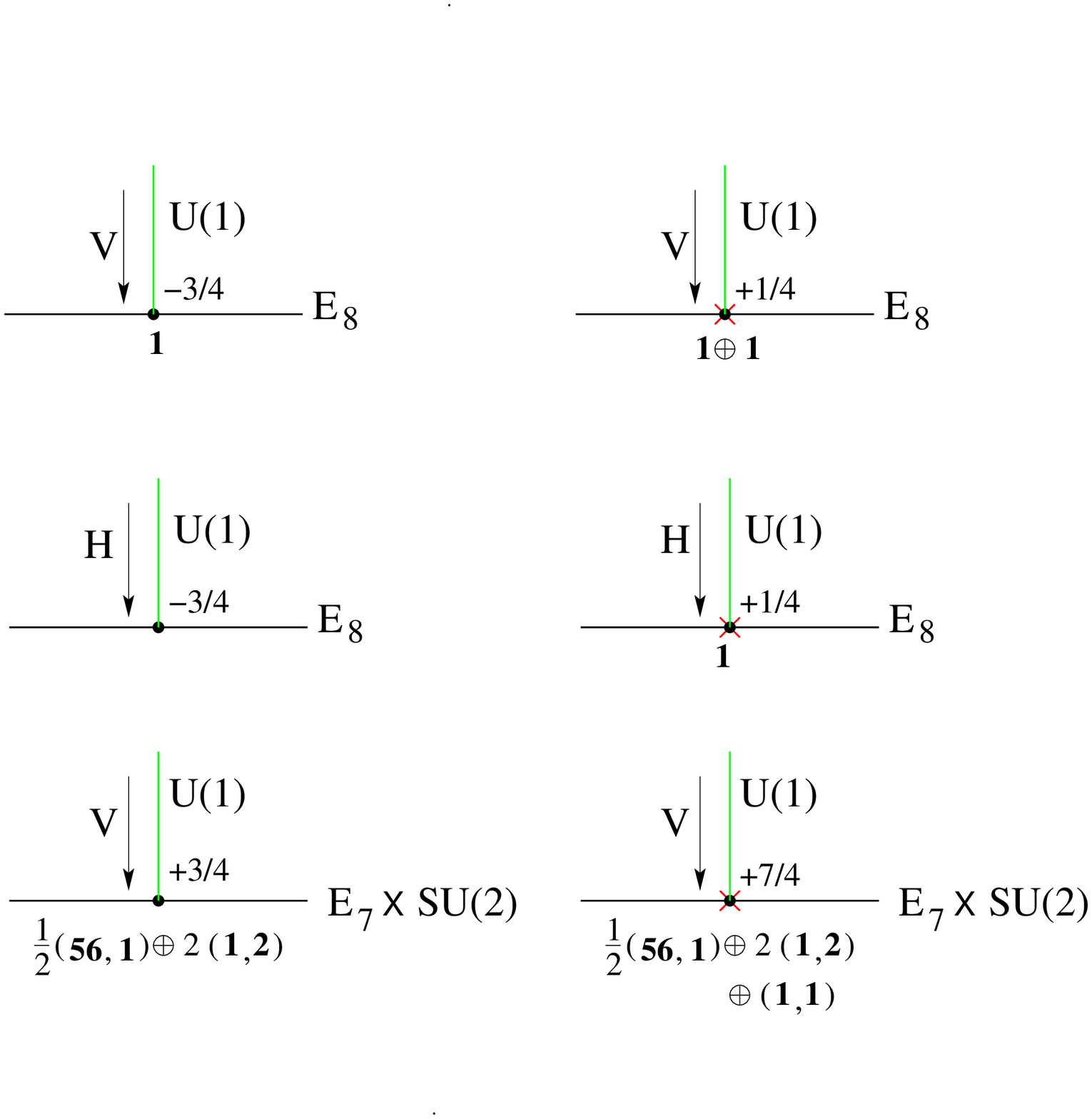}\\[.4in]
\parbox{6in}{Figure 4b:  Unconventional primary vertices  
in the ${\bf Z}_2$ orbifold.  These vertices have $U(1)$ gauge
group on the seven-planes.}
\end{center}
\end{figure}

\begin{figure}
\begin{center}
\includegraphics[height=2in,angle=0]{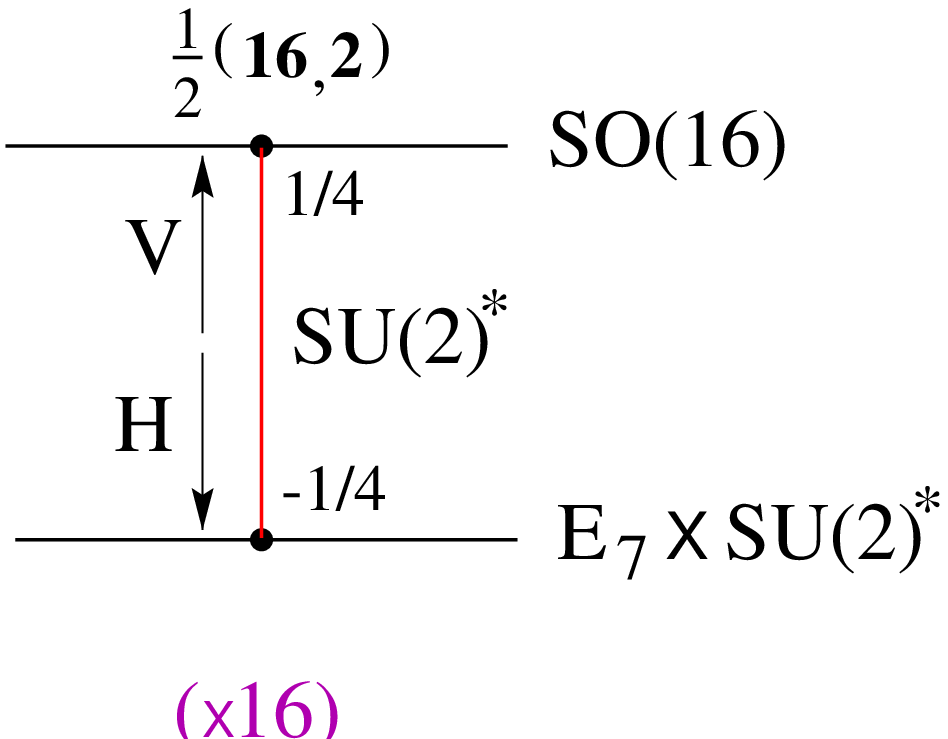} \\[.5in]
\parbox{5in}{Figure 5: Global picture of Model 1, which has
gauge group $E_7\times SO(16)\times SU(2)$.  This model has
one tensor multiplet, 256 Vector multiplets transforming as
$({\bf 133},{\bf 1},{\bf 1})
\oplus ({\bf 1},{\bf 120},{\bf 1})
\oplus ({\bf 1},{\bf 1},{\bf 3})$ and 500 hypermultiplets
transforming as shown in Table 4.}\\[.5in]
\begin{tabular}{|c||c||c||c|}
\hline
\multicolumn{4}{|c|}{} \\[-5mm]
\multicolumn{4}{|c|}{500 Hypers} \\[.1in]
\hline
\hline
&  Sugra & $4({\bf 1},{\bf 1},{\bf 1})$ & 4 \\[.1in]
Untwisted & $M_{10}^{\rm top}$      
& $({\bf 1},{\bf 128},{\bf 1})$ & 128 \\[.1in]
& $M_{10}^{\rm bottom}$   & 
$({\bf 56},{\bf 1},{\bf 2})$ & 112 \\[.1in]
\hline
& $M_6^{\rm top}$ & $16\times \ft12({\bf 1},{\bf 16},{\bf 2})$ & 256 \\[.1in]
${\rm Twisted}_1$ & $M_6^{\rm bottom}$ &  &  \\[.1in]
& $M_7$ &  &  \\[.1in]
\hline
\end{tabular}\\[.3in]
\parbox{4.5in}{Table 4: The hypermultiplet content of Model 1, in terms of
$E_7\times SO(16)\times SU(2)$ representations.}
\end{center}
\end{figure}
  
\begin{figure}
\begin{center}
\includegraphics[width=2in,angle=0]{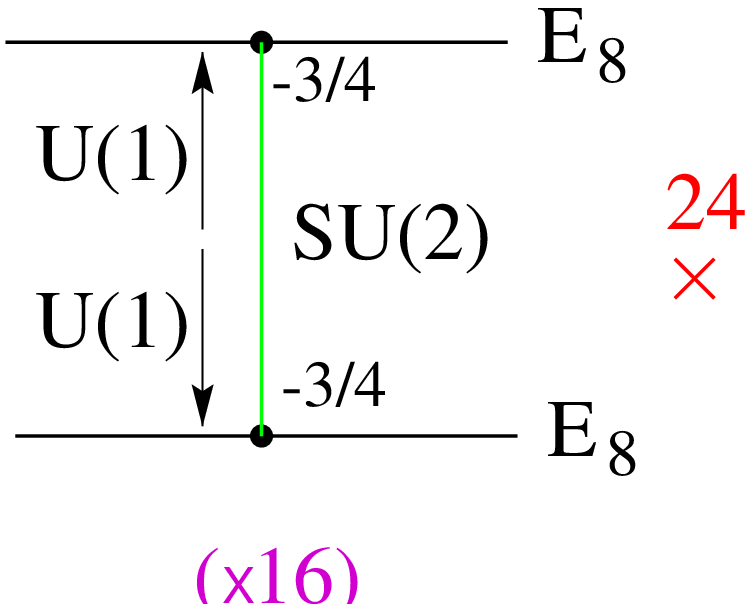} \\[.5in]
\parbox{5in}{Figure 6: Global picture of Model 2, which has
gauge group $E_8\times E_8\times U(1)^{16}$ and 24 fivebranes.
This model has 25 tensor multiplets (including 24 from the fivebranes),
512 vector multiplets transforming as
$({\bf 248},{\bf 1})\oplus ({\bf 1},{\bf 248})
\oplus 16\,({\bf 1},{\bf 1})$ and 60 hypermultiplets
transforming as shown in Table 5.}\\[.5in]
\begin{tabular}{|c||c||c||c|}
\hline
\multicolumn{4}{|c|}{} \\[-5mm]
\multicolumn{4}{|c|}{60 Hypers} \\[.1in]
\hline
\hline
&&&\\[-5mm]
&  Sugra & $4({\bf 1},{\bf 1})$ & 4 \\[.1in]
Untwisted & $M_{10}^{\rm top}$      
&   &   \\[.1in]
& $M_{10}^{\rm bottom}$   & 
  &   \\[.1in]
\hline
&&&\\[-5mm]
& $M_6^{\rm top}$ &   &   \\[.1in]
${\rm Twisted}_1$ & $M_6^{\rm bottom}$ &  &  \\[.1in]
& $M_7$ & $16\times 2\,({\bf 1},{\bf 1})$ & 32 \\[.1in]
\hline
&&&\\[-5mm]
Fivebranes & & $24\,({\bf 1},{\bf 1})$ & 24 \\[.1in]
\hline
\end{tabular}\\[.3in]
\parbox{4.5in}{Table 5: The hypermultiplet content of Model 2,
in terms of $E_8\times E_8$ representations.}
\end{center}
\end{figure}
 
\begin{figure}
\begin{center}
\includegraphics[width=3in,angle=0]{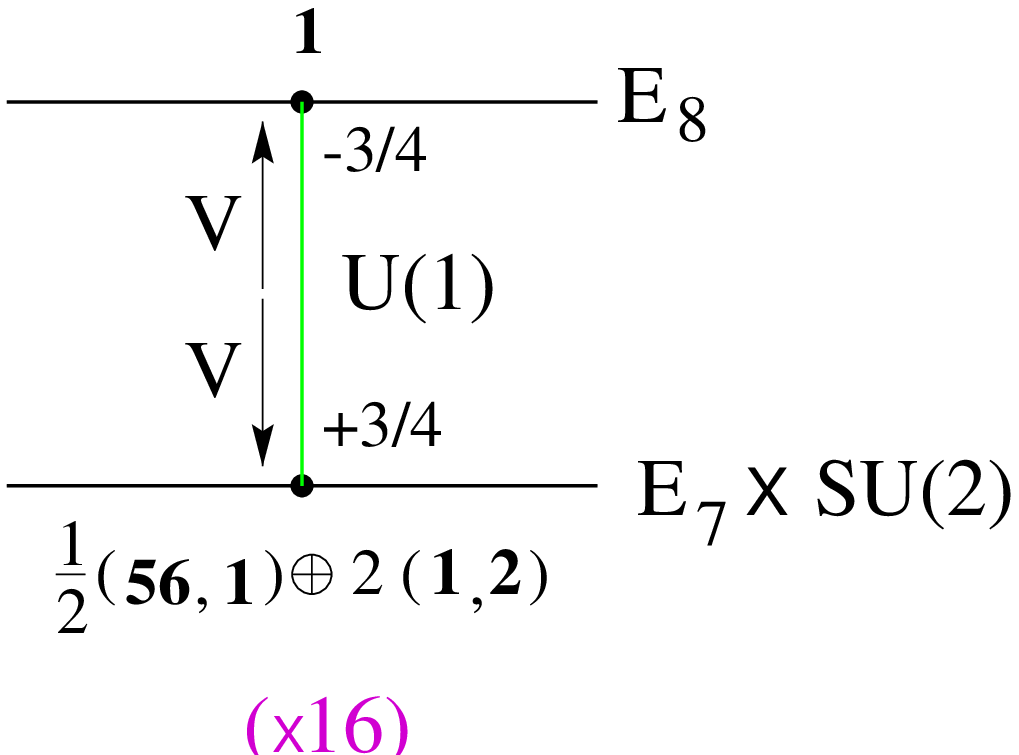} \\[.5in]
\parbox{5in}{Figure 7: Global picture of Model 3, which has
gauge group $E_8\times E_7\times SU(2)\times U(1)^{16}$.
This model has one untwisted tensor multiplet,
400 vector multiplets transforming as 
$({\bf 248},{\bf 1},{\bf 1})\oplus ({\bf 1},{\bf 133},{\bf 1})
\oplus ({\bf 1},{\bf 1},{\bf 3})\oplus 16\,({\bf 1},{\bf 1},{\bf 1})$,
and 644 hypermultiplets transforming as shown in Table 6.}\\[.5in]
\begin{tabular}{|c||c||c||c|}
\hline
\multicolumn{4}{|c|}{} \\[-5mm]
\multicolumn{4}{|c|}{644 Hypers} \\[.1in]
\hline
\hline
&&&\\[-5mm]
&  Sugra & $4({\bf 1},{\bf 1})$ & 4 \\[.1in]
Untwisted & $M_{10}^{\rm top}$      
&   &   \\[.1in]
& $M_{10}^{\rm bottom}$   & $({\bf 1},{\bf 56},{\bf 2})$
  & 112  \\[.1in]
\hline
&&&\\[-5mm]
& $M_6^{\rm top}$ & $16\times ({\bf 1},{\bf 1},{\bf 1})$ & 16  \\[.1in]
${\rm Twisted}_1$ & $M_6^{\rm bottom}$ & 
$16\times [\ft12\,({\bf 1},{\bf 56},{\bf 1})
\oplus 2\,({\bf 1},{\bf 1},{\bf 2})\,]$ & 512 \\[.1in]
& $M_7$ &  &  \\[.1in]
\hline
\end{tabular} \\[.3in]
\parbox{4.5in}{Table 6: The hypermultiplet content of Model 3 in terms of
$E_8\times E_7\times SU(2)$ representations.}
\end{center}
\end{figure}
 
\begin{figure}
\begin{center}
\includegraphics[height=2in,angle=0]{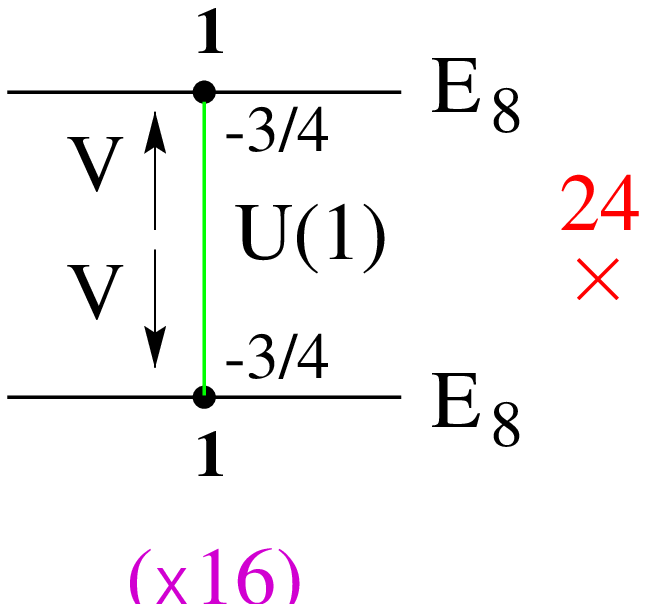} \\[.2in]
\parbox{5in}{Figure 8: Global picture of Model 4, which has 
gauge group $E_8\times E_8\times U(1)^{16}$.
This model has 25 tensor multiplets (including 24 from the fivebranes) 
and 512 vector multiplets transforming as 
$({\bf 248},{\bf 1})\oplus ({\bf 1},{\bf 248})
\oplus 16\,({\bf 1},{\bf 1})$,
and 60 hypermultiplets transforming as shown in Table 7.}\\[.5in]
\begin{tabular}{|c||c||c||c|}
\hline
\multicolumn{4}{|c|}{} \\[-5mm]
\multicolumn{4}{|c|}{60 Hypers} \\[.1in]
\hline
\hline
&&&\\[-5mm]
&  Sugra & $4({\bf 1},{\bf 1})$ & 4 \\[.1in]
Untwisted & $M_{10}^{\rm top}$      
&   &   \\[.1in]
& $M_{10}^{\rm bottom}$   &  
  &   \\[.1in]
\hline
&&&\\[-5mm]
& $M_6^{\rm top}$ & $16\times ({\bf 1},{\bf 1})$ & 16  \\[.1in]
${\rm Twisted}_1$ & $M_6^{\rm bottom}$ & 
$16\times ({\bf 1},{\bf 1})$  & 16 \\[.1in]
& $M_7$ &  &  \\[.1in]
\hline
&&&\\[-5mm]
Fivebranes & & $24\times ({\bf 1},{\bf 1})$ & 24   \\[.1in]
\hline
\end{tabular}\\[.3in]
\parbox{4.5in}{Table 7: The hypermultiplet content of Model 4 in terms of
$E_8\times E_8$ representations.}
\end{center}
\end{figure}

\begin{figure}
\begin{center}
\includegraphics[width=6in,angle=0]{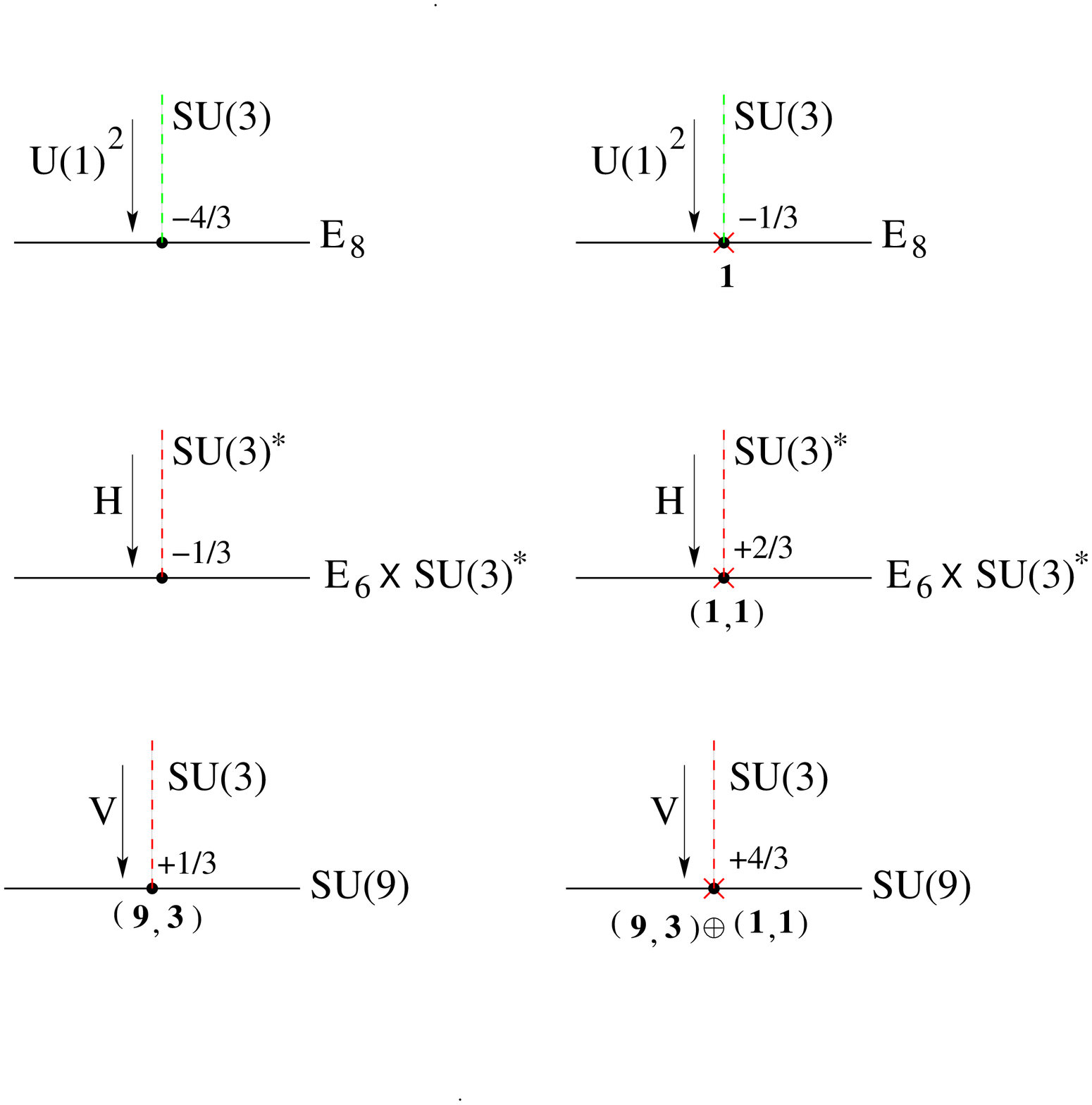}\\[.3in]
\parbox{5in}{Figure 9: Consistent primary vertices in the 
${\bf Z}_3$ orbifold}
\end{center}
\end{figure}

\begin{figure}
\begin{center}
\includegraphics[height=2in,angle=0]{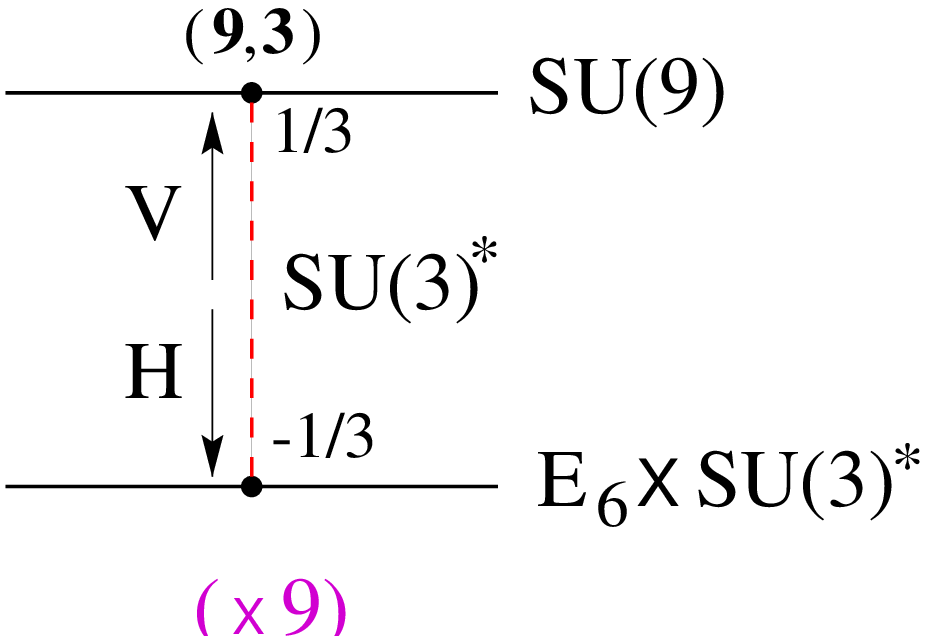} \\[.2in]
\parbox{5in}{Figure 10: Global picture of Model 5, which has
$6D$ gauge group $E_6\times SU(9)\times SU(3)$.
This model has
one untwisted tensor multiplet, 
166 vector multiplets transforming as  
$({\bf 78},{\bf 1},{\bf 1})\oplus ({\bf 1},{\bf 80},{\bf 1})
\oplus ({\bf 1},{\bf 1},{\bf 8})$,
and 410 hypermultiplets transforming as shown in Table 8.}\\[.5in]
\begin{tabular}{|c||c||c||c|}
\hline
\multicolumn{4}{|c|}{} \\[-5mm]
\multicolumn{4}{|c|}{410 Hypers} \\[.1in]
\hline
\hline
&&&\\[-5mm]
&  Sugra & $2({\bf 1},{\bf 1},{\bf 1})$ & 2 \\[.1in]
Untwisted & $M_{10}^{\rm top}$      
& $({\bf 1},{\bf 84},{\bf 1})$ & 84  \\[.1in]
& $M_{10}^{\rm bottom}$   & 
$({\bf 27},{\bf 1},{\bf 3})$ & 81 \\[.1in]
\hline
& top & $9\times ({\bf 1},{\bf 9},{\bf 3})$ & 243 \\[.1in]
${\rm Twisted}_1$ & bottom &  &  \\[.1in]
& $M_7$ &  &  \\[.1in]
\hline
\end{tabular} \\[.3in]
\parbox{4.5in}{Table 8: The hypermultiplet content of Model 5 in terms of
$E_6\times SU(9)\times SU(3)$ representations.}
\end{center}
\end{figure}

\begin{figure}
\begin{center}
\includegraphics[width=2.3in,angle=0]{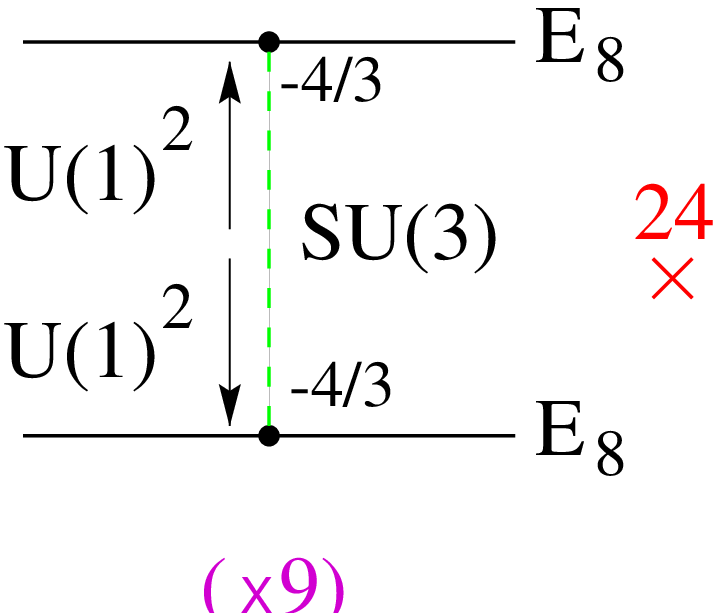} \\[.5in]
\parbox{5in}{Figure 11: Global picture of Model 6, which has gauge group
$E_8\times E_8\times U(1)^{18}$ and 24 fivebranes.
This model has
25 tensor multiplets (including 24 from the fivebranes) and
514 vector multiplets transforming as 
$({\bf 248},{\bf 1})\oplus ({\bf 1},{\bf 248})
\oplus 18\,({\bf 1},{\bf 1})$,
and 62 hypermultiplets transforming as shown in Table 9.}\\[.3in]
\begin{tabular}{|c||c||c||c|}
\hline
\multicolumn{4}{|c|}{} \\[-5mm]
\multicolumn{4}{|c|}{62 Hypers} \\[.1in]
\hline
\hline
&&&\\[-5mm]
&  Sugra & $2({\bf 1},{\bf 1},{\bf 1})$ & 2 \\[.1in]
Untwisted & $M_{10}^{\rm top}$      
&   &    \\[.1in]
& $M_{10}^{\rm bottom}$   & 
  &   \\[.1in]
\hline
&&&\\[-5mm]
& top &   &   \\[.1in]
${\rm Twisted}_1$ & bottom &  &  \\[.1in]
& $M_7$ & $9\times 4\,({\bf 1},{\bf 1},{\bf 1})$ & 36 \\[.1in]
\hline
&&&\\[-5mm]
Fivebranes & & $24\,({\bf 1},{\bf 1},{\bf 1})$ & 24 \\[.1in]
\hline
\end{tabular}\\[.3in]
{Table 9: The hypermultiplet content of Model 6 in terms of 
$E_8\times E_8$ representations.}
\end{center}
\end{figure}

\begin{figure}
\begin{center}
\includegraphics[width=6in,angle=0]{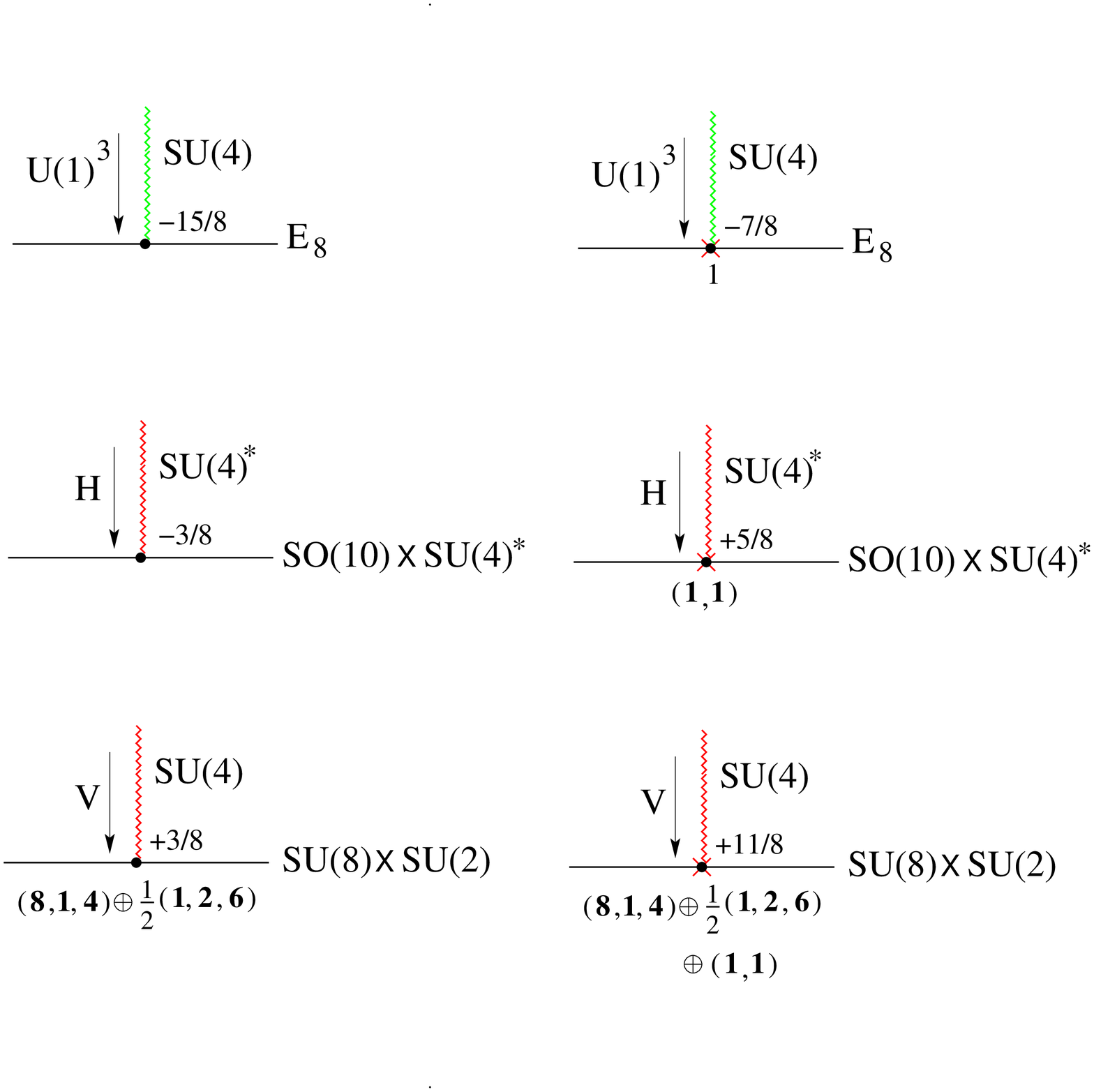}\\[.3in]
\parbox{5in}{Figure 12: Consistent primary vertices in the 
${\bf Z}_4$ orbifold}
\end{center}
\end{figure}
 
\begin{figure}
\begin{center}
\includegraphics[height=2.6in,angle=0]{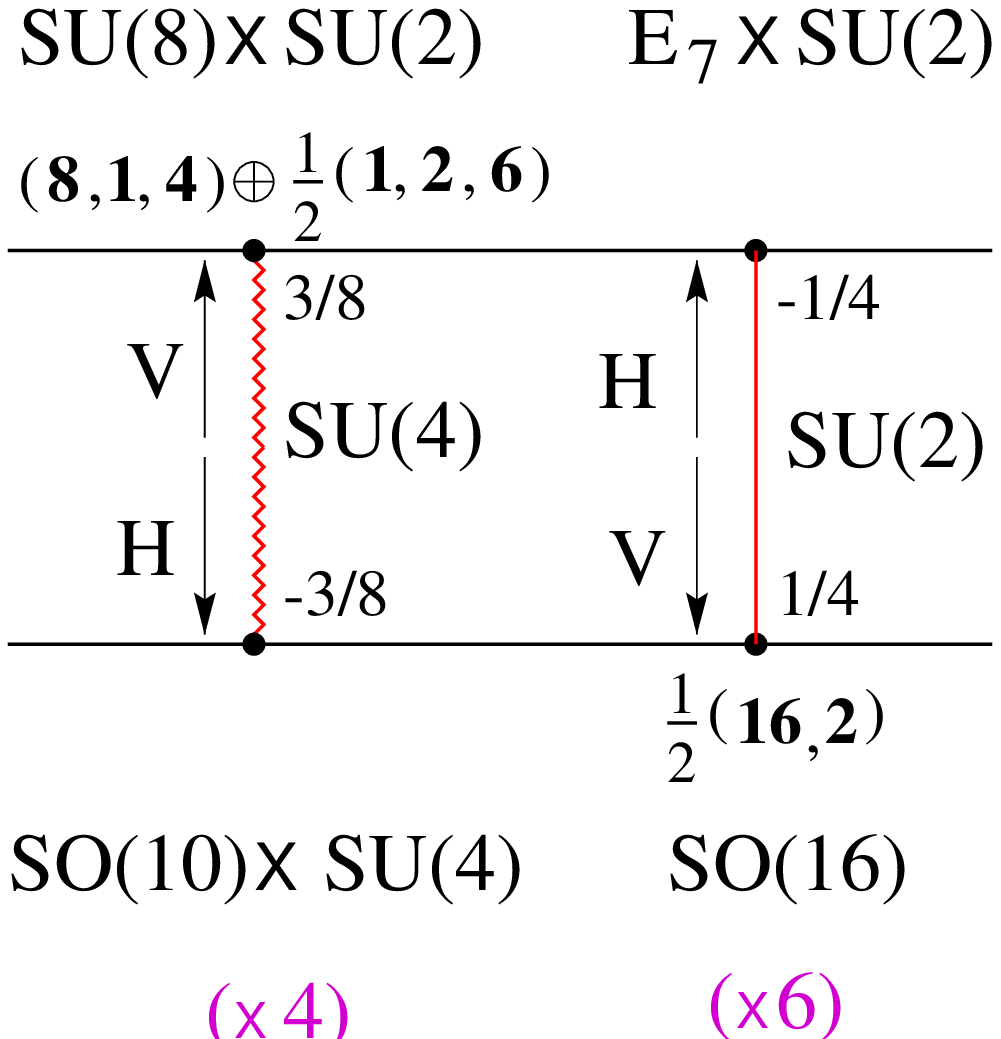} \\[.3in]
\parbox{5in}{Figure 13: Global picture of Model 7, which has gauge group
$SO(10)\times SU(8)\times SU(4)\times SU(2)$.
This model has
one untwisted tensor multiplet, 
126 vector multiplets transforming as  
$({\bf 45},{\bf 1},{\bf 1},{\bf 1})\oplus ({\bf 1},{\bf 63},{\bf 1},{\bf 1})
\oplus ({\bf 1},{\bf 1},{\bf 15},{\bf 1})\oplus 
({\bf 1},{\bf 1},{\bf 1},{\bf 3})$,
and 370 hypermultiplets transforming as shown in Table 10.} \\[.5in]
\begin{tabular}{|c||c||c||c|}
\hline
\multicolumn{4}{|c|}{} \\[-5mm]
\multicolumn{4}{|c|}{370 Hypers} \\[.1in]
\hline
\hline
&&&\\[-5mm]
&  Sugra & $2({\bf 1},{\bf 1},{\bf 1},{\bf 1})$ & 2 \\[.1in]
Untwisted & $M_{10}^{\rm top}$      
& $({\bf 1},{\bf 28},{\bf 1},{\bf 2})$ & 56 \\[.1in]
& $M_{10}^{\rm bottom}$   & 
$({\bf 16},{\bf 1},{\bf 4},{\bf 1})$ & 64 \\[.1in]
\hline
& top & $4\times[({\bf 1},{\bf 8},{\bf 1},{\bf 4})
        \oplus\ft12({\bf 1},{\bf 1},{\bf 2},{\bf 6})]$ & 152 \\[.1in]
${\rm Twisted}_1$ & bottom &  &  \\[.1in]
& $M_7$ &  &  \\[.1in]
\hline
& top &  & \\[.1in]
${\rm Twisted}_2$ & bottom & 
$6\times[\ft12({\bf 10},{\bf 1},{\bf 1},{\bf 2})
\oplus\ft12({\bf 1},{\bf 1},{\bf 6},{\bf 2})]$ 
& 96 \\[.1in]
& $M_7$ &  &\\[.1in]
\hline
\end{tabular} \\[.3in]
\parbox{4.5in}{Table 10: The hypermultiplet content of Model 7 in terms of 
$SO(10)\times SU(8)\times SU(4)\times SU(2)$ representations.}
\end{center}
\end{figure}

\begin{figure}
\begin{center}
\includegraphics[width=3.5in,angle=0]{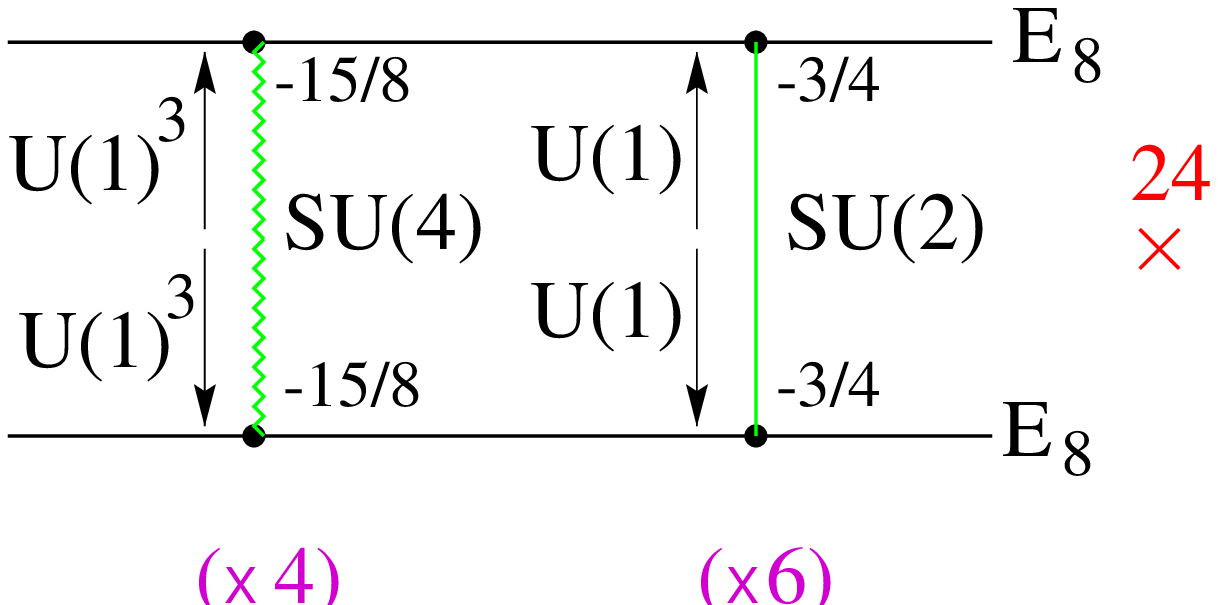} \\[.5in]
\parbox{5in}{Figure 14: Global picture of Model 8, which has 
gauge group
$E_8\times E_8\times U(1)^{18}$ and 24 fivebranes.
This model has
25 tensor multiplets (including 24 from the fivebranes),
514 vector multiplets transforming as 
$({\bf 248},{\bf 1})\oplus ({\bf 1},{\bf 248},{\bf 1},{\bf 1})
\oplus 18\,({\bf 1},{\bf 1}\,)$,
and 62 hypermultiplets transforming as shown in Table 11.}\\[.3in]
\begin{tabular}{|c||c||c||c|}
\hline
\multicolumn{4}{|c|}{} \\[-5mm]
\multicolumn{4}{|c|}{62 Hypers} \\[.1in]
\hline
\hline
&&&\\[-5mm]
&  Sugra & $2\,({\bf 1},{\bf 1},{\bf 1})$ & 2 \\[.1in]
Untwisted & $M_{10}^{\rm top}$      
&   &  \\[.1in]
& $M_{10}^{\rm bottom}$   & 
  &   \\[.1in]
\hline
&&&\\[-5mm]
& top &   &   \\[.1in]
${\rm Twisted}_1$ & bottom &  &  \\[.1in]
& $M_7$ & $4\times 6\,({\bf 1},{\bf 1},{\bf 1})$ & 24 \\[.1in]
\hline
&&&\\[-5mm]
& top &  & \\[.1in]
${\rm Twisted}_2$ & bottom &   
&   \\[.1in]
& $M_7$ & $6\times 2\,({\bf 1},{\bf 1},{\bf 1})$ & 12 \\[.1in]
\hline
&&&\\[-5mm]
Fivebranes & & $24\,({\bf 1},{\bf 1},{\bf 1})$ & 24 \\[.1in]
\hline
\end{tabular} \\[.3in]
\parbox{4.5in}{Table 11: The hypermultiplet content of Model 8 in terms of
$E_8\times E_8$ representations.}
\end{center}
\end{figure}

\begin{figure}
\begin{center}
\includegraphics[width=6in,angle=0]{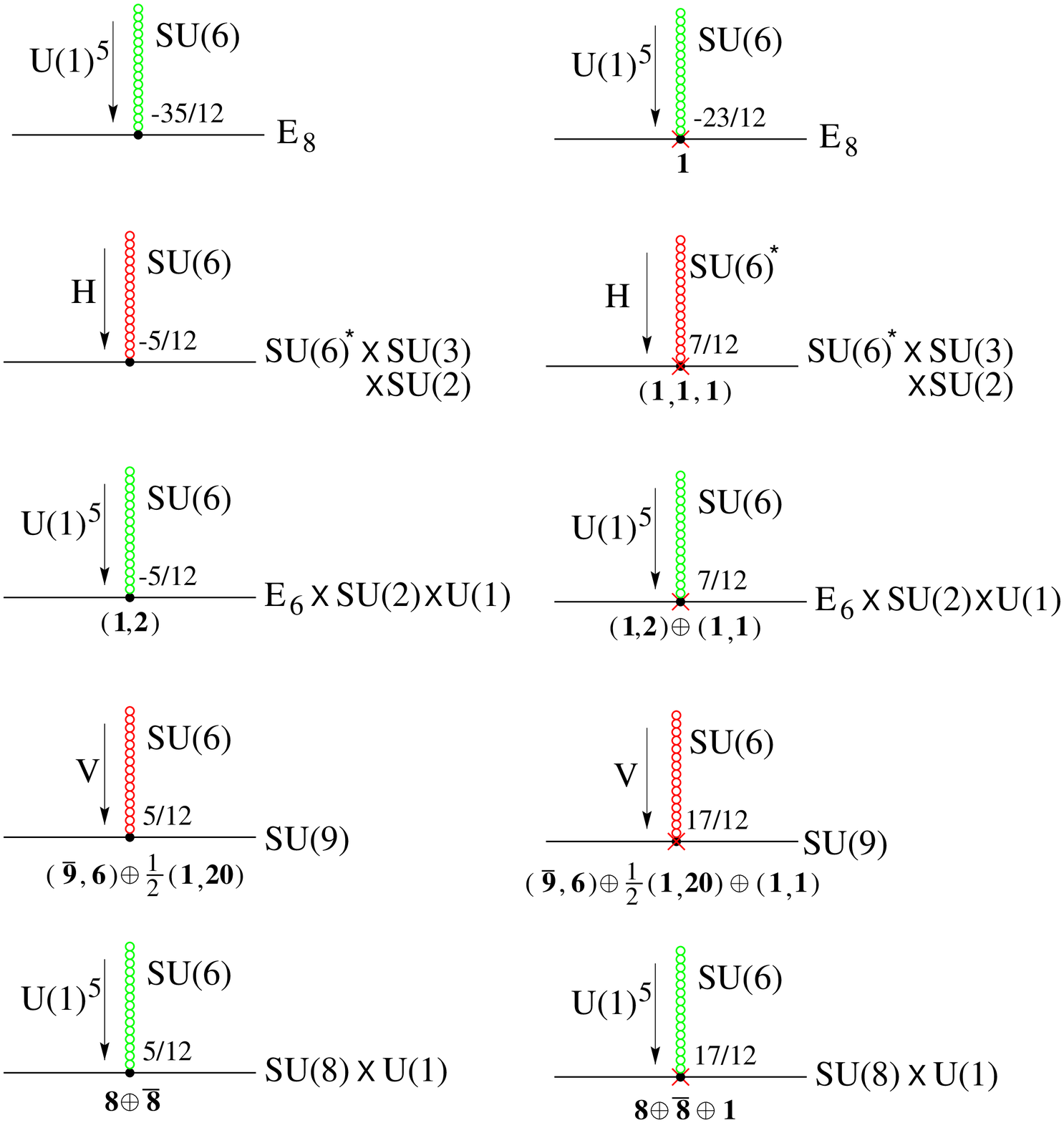}\\[.3in]
\parbox{5in}{Figure 15: Consistent primary vertices in the 
${\bf Z}_6$ orbifold}
\end{center} 
\end{figure}
 
\begin{figure}
\begin{center}
\includegraphics[width=4in,angle=0]{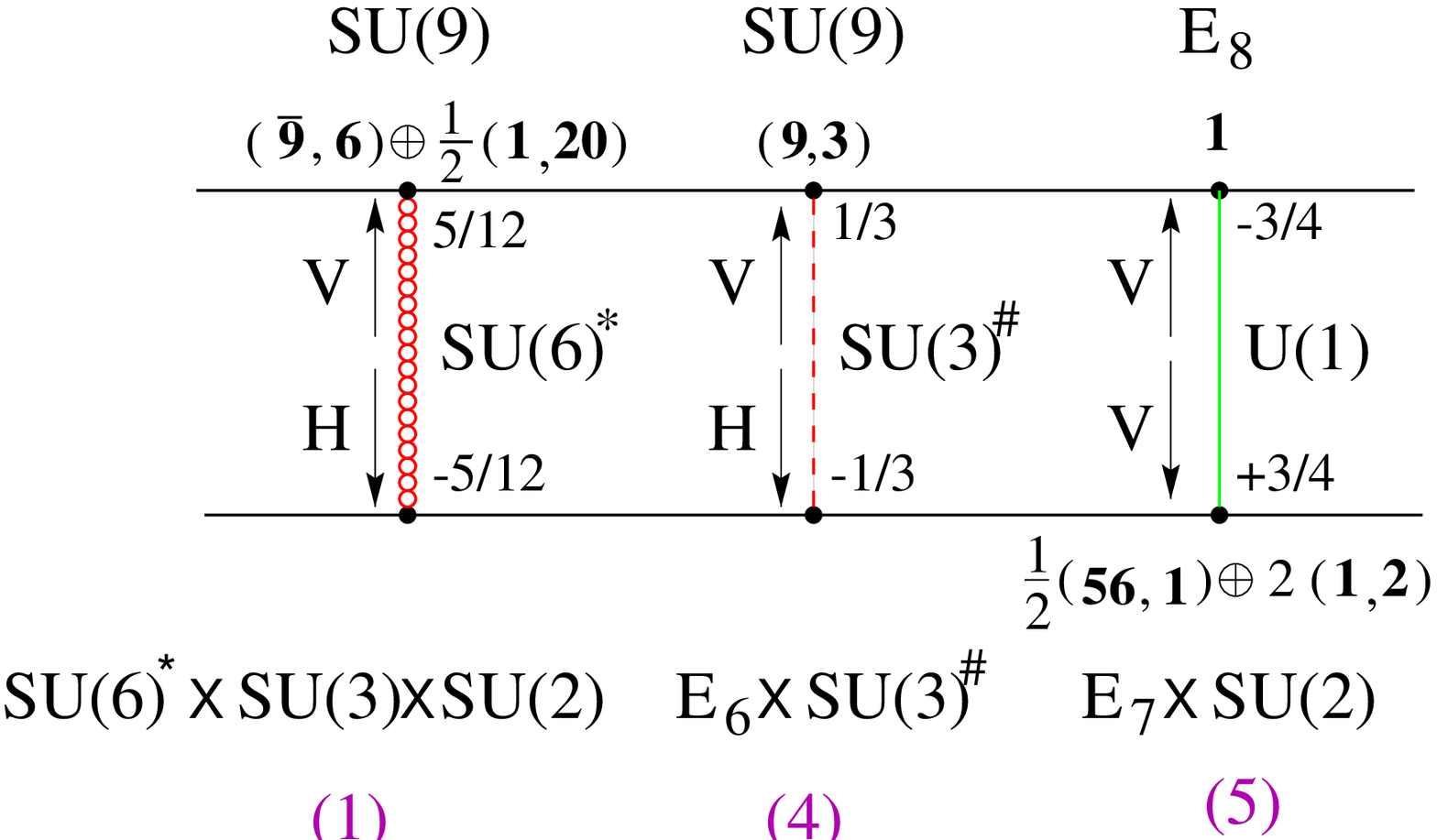} \\[.2in]
\parbox{6in}{Figure 16: Global picture of Model 9, which 
has gauge group
$SU(9)\times SU(6)\times SU(3)\times SU(2)\times U(1)^5$.
This model has one untwisted tensor multiplet, 
131 vector multiplets transforming as $({\bf 80},{\bf 1},{\bf 1},{\bf 1})\oplus ({\bf 1},{\bf 35},{\bf 1},{\bf 1})
\oplus ({\bf 1},{\bf 1},{\bf 8},{\bf 1})
\oplus ({\bf 1},{\bf 1},{\bf 1},{\bf 3})
\oplus 5\,({\bf 1},{\bf 1},{\bf 1},,{\bf 1})$,
and 375 hypermultiplets transforming as shown in Table 12.}\\[.1in]
\begin{tabular}{|c||c||c||c|}
\hline
\multicolumn{4}{|c|}{} \\[-5mm]
\multicolumn{4}{|c|}{375 Hypers} \\[.1in]
\hline
\hline
&&&\\[-5mm]
&  Sugra & $2({\bf 1},{\bf 1},{\bf 1},{\bf 1})$ & 2 \\[.1in]
Untwisted & $M_{10}^{\rm top}$      
&   &  \\[.1in]
& $M_{10}^{\rm bottom}$   & 
$({\bf 1},{\bf 6},{\bf 3},{\bf 2})$ & 36 \\[.1in]
\hline
& top & 
$({\bf 9},{\bf 6},{\bf 1},{\bf 1})
\oplus \ft12 ({\bf 1},{\bf 20},{\bf 1},{\bf 1})$ & 64  \\[.1in]
${\rm Twisted}_1$ & bottom &  &  \\[.1in]
& $M_7$ &  &  \\[.1in]
\hline
& top & $4\times ({\bf 9},{\bf 1},{\bf 3},{\bf 1})$ & 108 \\[.1in]
${\rm Twisted}_2$ & bottom &  &\\[.1in]
& $M_7$ &  &\\[.1in]
\hline
& top & $5\times ({\bf 1},{\bf 1},{\bf 1},{\bf 1})$ & 5  \\[.1in]
${\rm Twisted}_3$ & bottom & 
$5\times \,[\,({\bf 1},{\bf 6},{\bf 3},{\bf 1})
\oplus \ft12 ({\bf 1},{\bf 20},{\bf 1},{\bf 1})
\oplus 2 ({\bf 1},{\bf 1},{\bf 1},{\bf 2})\,]$ & 160 \\[.1in]
& $M_7$ & & \\[.1in]
\hline
\end{tabular} \\[.1in]
\parbox{4.5in}{Table 12: The hypermultiplet content of Model 9 in terms of
$SU(9)\times SU(6)\times SU(3)\times SU(2)$ representations.}
\end{center}
\end{figure}
 
\begin{figure}
\begin{center}
\includegraphics[width=4in,angle=0]{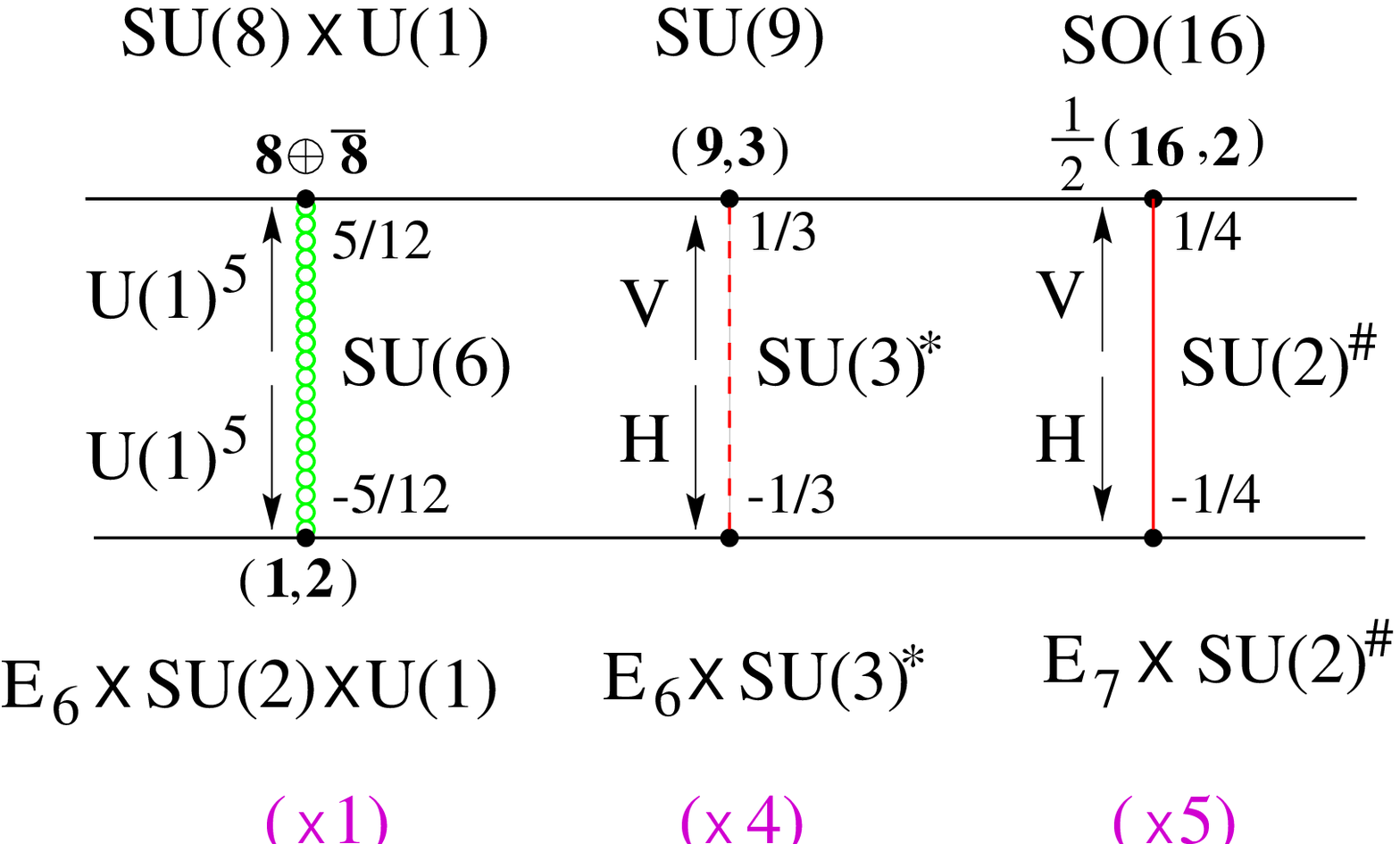} \\[.2in]
\parbox{6in}{Figure 17: Global picture of Model 10,
which has gauge group $E_6\times SU(8)\times SU(2)\times U(1)^7$.
This model has
one untwisted tensor multiplet, 
151 vector multiplets transforming as  
$({\bf 78},{\bf 1},{\bf 1})\oplus ({\bf 1},{\bf 63},{\bf 1})
\oplus ({\bf 1},{\bf 1},{\bf 3})\oplus 7\,({\bf 1},{\bf 1},{\bf 1})$,
and 395 hypermultiplets transforming as shown in Table 13.} \\[.1in]
\begin{tabular}{|c||c||c||c|}
\hline
\multicolumn{4}{|c|}{} \\[-5mm]
\multicolumn{4}{|c|}{395 Hypers} \\[.1in]
\hline
\hline
&&&\\[-5mm]
&  Sugra & $2({\bf 1},{\bf 1},{\bf 1})$ & 2 \\[.1in]
Untwisted & $M_{10}^{\rm top}$      
& $({\bf 1},{\bf 56},{\bf 1})
\oplus({\bf 1},{\bf 28},{\bf 1})
\oplus ({\bf 1},{\bf 8},{\bf 1})$ & 92 \\[.1in]
& $M_{10}^{\rm bottom}$   & 
$({\bf 27},{\bf 1},{\bf 1})
\oplus({\bf 27},{\bf 1},{\bf 2})
\oplus 2\,({\bf 1},{\bf 1},{\bf 2})$ & 85 \\[.1in]
\hline
& top & ${\bf 8}\oplus\bar{\bf 8}$ & 16 \\[.1in]
${\rm Twisted}_1$ & bottom & $({\bf 1},{\bf 1},{\bf 2})$ & 2 \\[.1in]
& $M_7$ & $10\,({\bf 1},{\bf 1},{\bf 1})$ & 10 \\[.1in]
\hline
& top & $4\times [\,({\bf 1},{\bf 8},{\bf 3})
\oplus ({\bf 1},{\bf 1},{\bf 3})\,]$ & 108\\[.1in]
${\rm Twisted}_2$ & bottom &  &\\[.1in]
& $M_7$ &  &\\[.1in]
\hline
& top & $5\times \ft12({\bf 1},{\bf 8},{\bf 2})$ & 80 \\[.1in]
${\rm Twisted}_3$ & bottom &  &\\[.1in]
& $M_7$ & & \\[.1in]
\hline
\end{tabular} \\[.1in]
\parbox{4.5in}{Table 13: The hypermultiplet content of Model 10 in terms of
$E_6\times SU(8)\times SU(2)$ representations.}
\end{center}
\end{figure}

\begin{figure}
\begin{center}
\includegraphics[width=4in,angle=0]{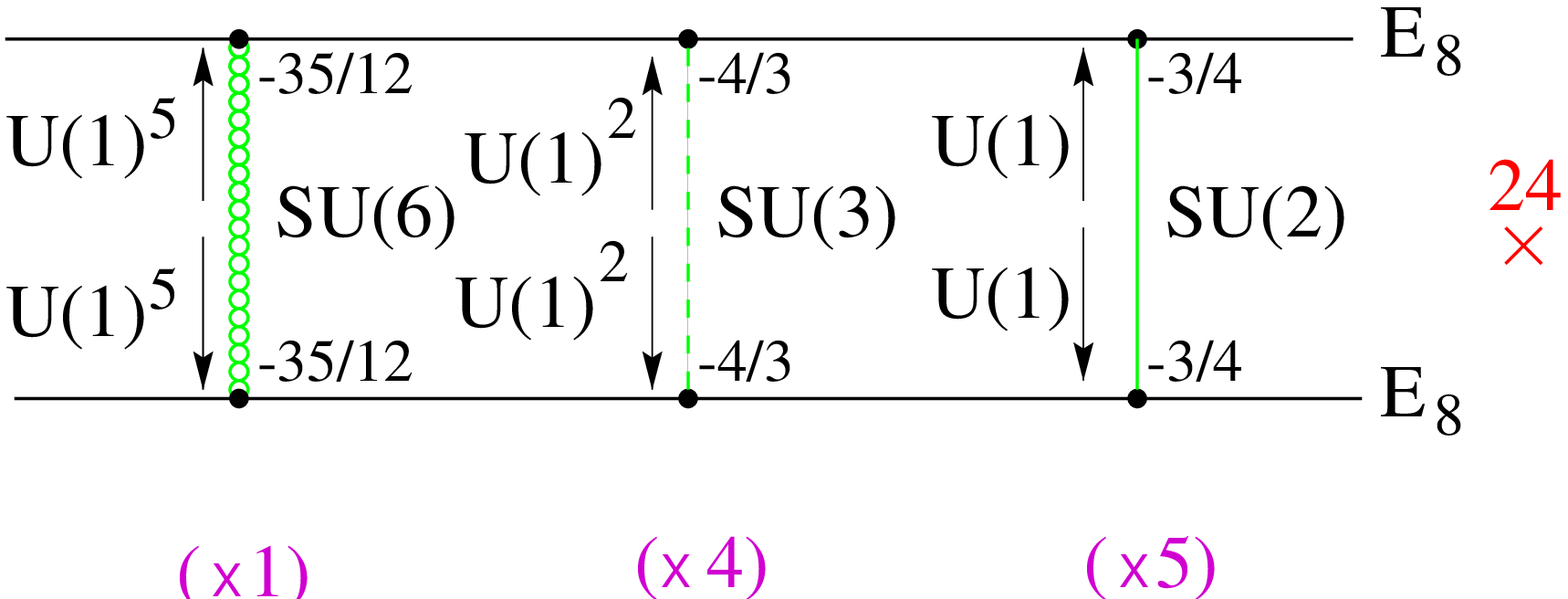} \\[.3in]
\parbox{5in}{Figure 18:  Global picture of Model 11,
which has gauge group $E_8\times E_8\times U(1)^{18}$ and 24 fivebranes.
This model has
25 tensor multiplets (including 24 from the fivebranes),
514 vector multiplets transforming as 
$({\bf 248},{\bf 1})\oplus ({\bf 1},{\bf 248})
\oplus 18\,({\bf 1},{\bf 1})$,
and 62 hypermultiplets transforming as shown in Table 14.}\\[.1in]
\begin{tabular}{|c||c||c||c|}
\hline
\multicolumn{4}{|c|}{} \\[-5mm]
\multicolumn{4}{|c|}{62 Hypers} \\[.1in]
\hline
\hline
&&&\\[-5mm]
&  Sugra & $2({\bf 1},{\bf 1})$ & 2 \\[.1in]
Untwisted & $M_{10}^{\rm top}$      
&   &   \\[.1in]
& $M_{10}^{\rm bottom}$   & 
  &   \\[.1in]
\hline
&&&\\[-5mm]
& top &   &   \\[.1in]
${\rm Twisted}_1$ & bottom &   &   \\[.1in]
& $M_7$ & $1\times 10\,({\bf 1},{\bf 1})$ & 10 \\[.1in]
\hline
& top &   &  \\[.1in]
${\rm Twisted}_2$ & bottom &  &\\[.1in]
& $M_7$ & $4\times 4\,({\bf 1},{\bf 1})$ & 16 \\[.1in]
\hline
&&&\\[-5mm]
& top &   &   \\[.1in]
${\rm Twisted}_3$ & bottom &  &\\[.1in]
& $M_7$ & $5\times 2\,({\bf 1},{\bf 1})$ & 10 \\[.1in]
\hline
&&&\\[-5mm]
Fivebranes & & $24\,({\bf 1},{\bf 1})$ & 24 \\[.1in]
\hline
\end{tabular} \\[.1in]
\parbox{4.5in}{Table 14: The hypermultiplet content of Model 11 in terms of
$E_8\times E_8$ representations.}
\end{center}
\end{figure}

\pagebreak
     
\section{Conclusions}
We have streamlined the technology developed in our previous papers 
\cite{mlo,phase} to enable us to determine 
{\it M}-theory orbifold anomalies soley in terms of
rational data. This permits an algorithmic search for consistent orbifold
vertices and allows us to complete twisted sector 
states, including those localized
on certain odd-dimensional planes. We have explained a comprehensive 
set of models describing portions of the low-energy moduli space 
corresponding to {\it M}-theory compactified on 
$S^1/{\bf Z}_2\times K3$ orbifolds, and discussed a number of technical and
conceptual issues associated with these.  This technology has been developed 
with the expressed intent of enabling a search for 
phenomenologically more interesting
compactifications to four dimensions.  We are currently investigating
a large class of $T^7/({\bf Z}_M\times {\bf Z}_N)$ and
$T^7/({\bf Z}_M\times {\bf Z}_N\times {\bf Z}_L)$ {\it M}-theory orbifolds,
where the technology described in this paper has a natural extension.
Our aim is to partially delineate how nonperturbative string effects,
particularly those involving fivebrane-mediated phase transitions can
modify phenomenological predictions based on perturbative orbifolds,
and facilitate the search for a more fundamental description of {\it M}-theory.
This will be described in forthcoming papers.

\renewcommand{\thesection}{Appendix:}
\section{Local Vertex Analyses}
\renewcommand{\thesection}{A}
\renewcommand{\theequation}{\arabic{equation}}

This appendix contains tables listing the rational parameters
associated with each of the consistent local vertices described in the 
main text.  The data contained in these tables are sufficient to
reconstruct the entire local anomaly for each vertex, 
including all gravitational, mixed
and pure gauge terms, and to verify consistency using equation
(\ref{zzzz}).

The method of analysis is as follows. For a specific vertex, we determine the 
$E_8$ branching rule to the desired subgroup. 
(These are tabulated in reference \cite{slansky}.)  
From this branching rule, it
is straightforward to determine $V^{(n)}$ and $H^{(n)}$.
The fixed input for a given
choice of breaking pattern include the 
orbifold geometric parameters $f_{(n)}$ (which determine $f$)
and the supergravity anomaly factor $P$, which is determined by
the untwisted hypermultiplet multiplicities $h_{(n)}$,
and the representation indices
${\cal I}_4$, ${\cal I}_{2,2}$ and ${\cal I}_2$ for each possible 
representation in the selected gauge group.  All of this data, except for the
representation indices, are also found in Table 1.  Representation indices
can be determined using the technology described in \cite{rsv}. 
In the tables, we include the
needed representation indices for easy
verification purposes.  These numbers represent the {\it keys} to the analysis,
while equation(\ref{zzzz}) represents the {\it lock}.
We encourage the reader to 
plug the data in the tables into (\ref{zzzz}) to show that each set
solves all of the requirements.  All of the parameters in the 
tables can be cross checked. 

As an example, we will describe the analysis ${\bf A.6}$
corresponding to the ${\bf Z}_3$ orbifold with branching
$E_8\to E_6\times SU(3)$.  The branching rule is
${\bf 248}\to (\,{\bf 78}\,,\,{\bf 1}\,)\oplus (\,{\bf 1}\,,\,{\bf 8}\,)
\oplus (\,{\bf 27}\,,\,{\bf 3}\,)\oplus (\,{\bf \bar{27}}\,,\,{\bf \bar{3}}\,)$.
The adjoint representation of the broken subgroup corresponds to
$78+8=86$ vector multiplets, so that $V^{(1)}=86$.  The remaining
terms in the decomposition correspond to $2\,(27\times 3)=162$ hypermultiplets.
However, half of these hypermultiplets are projected out by the ${\bf Z}_3$
projection, due to the nature of the ${\bf Z}_3$ action on the 
$E_8$ root lattice.  Because of this, 
$H^{(1)}=27\times 3=81$.  Although it is possible to resolve projections  
such as this by constructing projection matrices \footnote{We avoid
referring to ``shift vectors" since it is not known what analog {\it M}-theory
involves to replace vertex operators in the perturbative string.}, it is
relatively simple to resolve the effect of such projections by enforcing 
anomaly cancellation.  

The parameters $n_0({\cal R})$, for the four
representations ${\cal R}$ given by the ${\bf 78}$ and ${\bf 27}$ of
$E_6$ and the ${\bf 8}$ and ${\bf 3}$ of $SU(3)$, describe a straightforward
decomposition of $E_8$ which is insensitive to the ${\bf Z}_3$ projection.
As explained above, this is because the parameters $n_0({\cal R})$ appears in the inflow
anomaly, which involves only classical fields.  As a result of this,
the division by two reflected in $H^{(1)}$ is not reflected
in $n_0({\cal R})$. Thus, $n_0({\bf 27})=6$ rather than 3.   On the other hand,
the parameters $n_1({\cal R})$ appear in the quantum anomaly, which {\it is}
sensitive to the ${\bf Z}_3$ projection.  Thus, the division by two {\it is}
reflected in $n_1({\cal R})$.  (For instance $n_1({\bf 27})=-3$, since only
three ${\bf 27}$ hypermultiplets are ${\bf Z}_3$ invariant. Recall that  
hypermultiplets contribute with a minus sign in the quantum anomaly.) 
Analogous statements pertain to all of the tables included below.

\pagebreak

\subsection{The ${\bf Z}_2$ orbifold with $E_8\to E_8$}
\vspace{.3in}
\begin{flushleft} 
\begin{tabular}{|cc||c||cc|}
\hline
&&&&\\[-5mm]
\hspace{.1in} $n$ \hspace{.05in} &
\hspace{.05in} ${\cal G}_n$ \hspace{.1in} &
\hspace{.1in} $f_{(n)}$ \hspace{.1in} &
\hspace{.1in} $V^{(n)}$ \hspace{.05in} &
\hspace{.05in} $H^{(n)}$ \hspace{.1in} \\[.1in]
\hline
\hline
&&&&\\[-5mm]
1 & $E_8$ & 16 & 248 & 0 \\[.1in]
\hline
\end{tabular} \\[.3in]
\begin{tabular}{|c||cc||c|}
\hline
&&&\\[-5mm]
\hspace{.08in} $f$ \hspace{.05in} &
\hspace{.08in} $P$ \hspace{.05in} &
\hspace{.05in} $Q$ \hspace{.2in} &
\hspace{.2in} $P+Q$ \hspace{.15in}  \\[.1in]
\hline
\hline
&&&\\[-5mm]
16 & $\ft{15}{2}$ & $\ft{31}{2}$ & 23  \\[.1in]
\hline 
\end{tabular}\\[-1in]
\begin{tabular}{c}
\hspace{3in}
{\includegraphics[height=1in,angle=0]{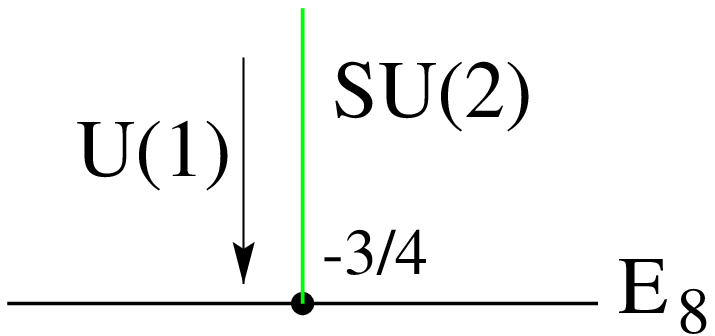}}
\end{tabular}
\vspace{.2in}
\begin{tabular}{|c||cc||ccc||ccc||c|}
\hline
&&&&&&&&&\\[-5mm]
\hspace{.1in} $I$ \hspace{.1in} &
\hspace{.1in} ${\cal G}_I$ \hspace{.05in} &
\hspace{.05in} ${\cal R}$ \hspace{.1in} &
\hspace{.1in}  ${\cal I}_4$ \hspace{.05in} & 
\hspace{.05in} ${\cal I}_{2,2}$ \hspace{.05in} & 
\hspace{.05in} ${\cal I}_2$ \hspace{.1in} &
\hspace{.05in} $n_0$ \hspace{.05in} &
\hspace{.05in} $n_1$ \hspace{.05in} &
\hspace{.05in} $\tilde{n}$ \hspace{.1in} &
\hspace{.1in} ${\rm z}$ \hspace{.1in} \\[.1in]
\hline
\hline
&&&&&&&&&\\[-5mm]
1 & $E_8$ & ${\bf 248}^*$ & 0 & 9 & 30 & 1 & 1 & 0 & $\ft{1}{16}$ \\[.1in]
\hline
\end{tabular} \\[.3in]
\begin{tabular}{|c||cccc||c|}
\hline
&&&&&\\[-5mm]
\hspace{.1in} $I$ \hspace{.1in} &
\hspace{.1in} $X_I$ \hspace{.05in} &
\hspace{.05in} $Z_I^{(4)}$ \hspace{.1in} &
\hspace{.05in} $Z_I^{(2)}$ \hspace{.05in} &
\hspace{.05in} $Z_I^{(2,2)}$ \hspace{.05in} & 
\hspace{.05in} $\rho_I$ \hspace{.05in} \\[.1in]
\hline
\hline
&&&&&\\[-5mm]
1 & 1 & 0 & $\ft{15}{8}$ & $\ft{9}{16}$ & 0  \\[.1in]
\hline
\end{tabular} \hspace{.3in}
\begin{tabular}{|c||c|}
\hline
&\\[-5mm]
\hspace{.1in} $g$ \hspace{.1in} & 
\hspace{.1in} $\rho$ \hspace{.1in} \\[.1in]
\hline
\hline
&\\[-5mm]
-$\ft34$ & 0 \\[.1in]
\hline
\end{tabular} \\[.2in]
\begin{tabular}{c}
\hspace{4.2in}
$n_H-n_V=\ft12$
\end{tabular}
   
\end{flushleft}
 
\pagebreak

\subsection{The ${\bf Z}_2$ orbifold with $E_8\to E_7\times SU(2)$}
\vspace{.3in}
\begin{flushleft} 
\begin{tabular}{|cc||c||cc|}
\hline
&&&&\\[-5mm]
\hspace{.1in} $n$ \hspace{.05in} &
\hspace{.05in} ${\cal G}_n$ \hspace{.1in} &
\hspace{.1in} $f_{(n)}$ \hspace{.1in} &
\hspace{.1in} $V^{(n)}$ \hspace{.05in} &
\hspace{.05in} $H^{(n)}$ \hspace{.1in} \\[.1in]
\hline
\hline
&&&&\\[-5mm]
1 & $E_7\times SU(2)$ & 16 & 136 & 112 \\[.1in]
\hline
\end{tabular} \\[.3in]
\begin{tabular}{|c||cc||c|}
\hline
&&&\\[-5mm]
\hspace{.08in} $f$ \hspace{.05in} &
\hspace{.08in} $P$ \hspace{.05in} &
\hspace{.05in} $Q$ \hspace{.2in} &
\hspace{.2in} $P+Q$ \hspace{.15in}  \\[.1in]
\hline
\hline
&&&\\[-5mm]
16 & $\ft{15}{2}$ & $\ft32$ & 9  \\[.1in]
\hline 
\end{tabular}\\[-1in]
\begin{tabular}{c}
\hspace{3in}
{\includegraphics[height=1in,angle=0]{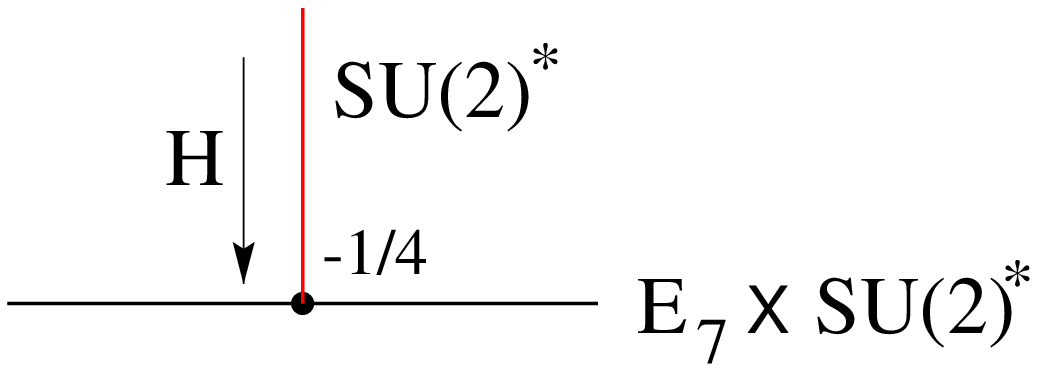}}
\end{tabular}
\vspace{.2in}
\begin{tabular}{|c||cc||ccc||ccc||c|}
\hline
&&&&&&&&&\\[-5mm]
\hspace{.1in}  $I$ \hspace{.1in} &
\hspace{.1in}  ${\cal G}_I$ \hspace{.05in} &
\hspace{.05in} ${\cal R}$ \hspace{.1in} &
\hspace{.1in}  ${\cal I}_4$ \hspace{.05in} & 
\hspace{.05in} ${\cal I}_{2,2}$ \hspace{.05in} & 
\hspace{.05in} ${\cal I}_2$ \hspace{.1in} &
\hspace{.05in} $n_0$ \hspace{.05in} &
\hspace{.05in} $n_1$ \hspace{.05in} &
\hspace{.05in} $\tilde{n}$ \hspace{.1in} &
\hspace{.1in}  ${\rm z}$ \hspace{.1in} \\[.1in]
\hline
\hline
&&&&&&&&&\\[-5mm]
1 & $E_7$ & ${\bf 133}^*$ & 0 & $\ft16$ & 3 & 1 & 1 & 0 & $\ft{1}{16}$ \\[.1in]
  &       & ${\bf 56}$  & 0 & $\ft{1}{24}$ & 1 & 2 & -2 & 0 & -$\ft18$ \\[.1in]
\hline
&&&&&&&&&\\[-5mm]
2 & $SU(2)$ & ${\bf 3}^*$ & 0 & 8 & 4 & 1 & 1 & -$\ft12$ & $\ft{1}{16}$ \\[.1in]
  &         & ${\bf 2}$ & 0 & $\ft12$ & 1 & 56 & -56 & 0 & -$\ft72$ \\[.1in]
\hline
\end{tabular} \\[.3in]
\begin{tabular}{|c||cccc||c|}
\hline
&&&&&\\[-5mm]
\hspace{.1in} $I$ \hspace{.1in} &
\hspace{.1in} $X_I$ \hspace{.05in} &
\hspace{.05in} $Z_I^{(4)}$ \hspace{.1in} &
\hspace{.05in} $Z_I^{(2)}$ \hspace{.05in} &
\hspace{.05in} $Z_I^{(2,2)}$ \hspace{.05in} &
\hspace{.05in} $\rho_I$ \hspace{.05in} \\[.1in]
\hline
\hline
&&&&&\\[-5mm]
1 & 1/6 & 0 & $\ft{1}{16}$ & $\ft{1}{192}$ & 0 \\[.1in]
2 & 2 & 0 & -$\ft{21}{4}$ & -$\ft{21}{4}$ & 1 \\[.1in]
\hline
\end{tabular} \hspace{.3in}
\begin{tabular}{|c||c|}
\hline
&\\[-5mm]
\hspace{.1in} $g$ \hspace{.1in} & 
\hspace{.1in} $\rho$ \hspace{.1in} \\[.1in]
\hline
\hline
&\\[-5mm]
-$\ft14$ & 1 \\[.1in]
\hline
\end{tabular} \\[-.15in]
\begin{tabular}{c}
\hspace{4.2in}
$n_H-n_V=\ft32$
\end{tabular}
   
\end{flushleft}

\pagebreak

\subsection{The ${\bf Z}_2$ orbifold with $E_8\to SO(16)$}
\vspace{.3in}
\begin{flushleft} 
\begin{tabular}{|cc||c||cc|}
\hline
&&&&\\[-5mm]
\hspace{.1in} $n$ \hspace{.05in} &
\hspace{.05in} ${\cal G}_n$ \hspace{.1in} &
\hspace{.1in} $f_{(n)}$ \hspace{.1in} &
\hspace{.1in} $V^{(n)}$ \hspace{.05in} &
\hspace{.05in} $H^{(n)}$ \hspace{.1in} \\[.1in]
\hline
\hline
&&&&\\[-5mm]
1 & $SO(16)$ & 16 & 120 & 128 \\[.1in]
\hline
\end{tabular} \\[.3in]
\begin{tabular}{|c||cc||c|}
\hline
&&&\\[-5mm]
\hspace{.08in} $f$ \hspace{.05in} &
\hspace{.08in} $P$ \hspace{.05in} &
\hspace{.05in} $Q$ \hspace{.2in} &
\hspace{.2in} $P+Q$ \hspace{.15in}  \\[.1in]
\hline
\hline
&&&\\[-5mm]
16 & $\ft{15}{2}$ & -$\ft12$ & 7 \\[.1in]
\hline 
\end{tabular}\\[-1in]
\begin{tabular}{c}
\hspace{3in}
{\includegraphics[height=1.3in,angle=0]{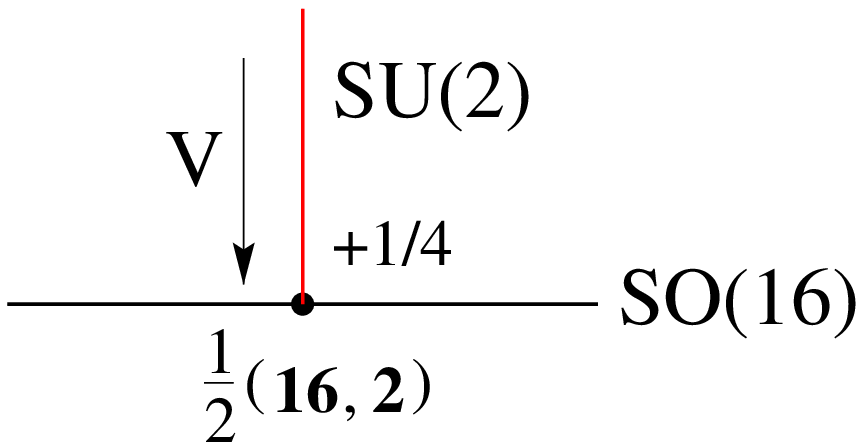}}
\end{tabular}
\vspace{.2in}
\begin{tabular}{|c||cc||ccc||ccc||c|}
\hline
&&&&&&&&&\\[-5mm]
\hspace{.1in}  $I$ \hspace{.1in} &
\hspace{.1in}  ${\cal G}_I$ \hspace{.05in} &
\hspace{.05in} ${\cal R}$ \hspace{.1in} &
\hspace{.1in}  ${\cal I}_4$ \hspace{.05in} & 
\hspace{.05in} ${\cal I}_{2,2}$ \hspace{.05in} & 
\hspace{.05in} ${\cal I}_2$ \hspace{.1in} &
\hspace{.05in} $n_0$ \hspace{.05in} &
\hspace{.05in} $n_1$ \hspace{.05in} &
\hspace{.05in} $\tilde{n}$ \hspace{.1in} &
\hspace{.1in}  ${\rm z}$ \hspace{.1in} \\[.1in]
\hline
\hline
&&&&&&&&&\\[-5mm]
1 & $SO(16)$ & ${\bf 128}$  & -8 & 6 & 16 & 1 & -1 & 0 & -$\ft{1}{16}$ \\[.1in]
  &          & ${\bf 120}^*$ & 8  & 3 & 14 & 1 & 1 & 0 & $\ft{1}{16}$ \\[.1in]
  &          & ${\bf 16}$    & 1 & 0 & 1 & 0 & 0 & -1 & 0 \\[.1in]
\hline
&&&&&&&&&\\[-5mm]
2 & $SU(2)$ & ${\bf 3}^*$ & 0 & 8 & 4 & 0 & 0 & $\ft12$ & 0 \\[.1in]
  &         & ${\bf 2}$   & 0 & $\ft12$ & 1 & 0 & 0 & -8 & 0 \\[.1in]
\hline
\end{tabular} \\[.3in]
\begin{tabular}{|c||cccc|c|}
\hline
&&&&&\\[-5mm]
\hspace{.1in} $I$ \hspace{.1in} &
\hspace{.1in} $X_I$ \hspace{.05in} &
\hspace{.05in} $Z_I^{(4)}$ \hspace{.1in} &
\hspace{.05in} $Z_I^{(2)}$ \hspace{.05in} &
\hspace{.05in} $Z_I^{(2,2)}$ \hspace{.05in} &
\hspace{.05in} $\rho_I$ \hspace{.05in} \\[.1in]
\hline
\hline
&&&&&\\[-5mm]
1 & 1 & 0 & -$\ft98$ & -$\ft{3}{16}$ & 0  \\[.1in]
2 & 0 & 0 & -6 & 0 & 1 \\[.1in]
\hline
\end{tabular} 
\hspace{.3in}
\begin{tabular}{|c||c|}
\hline
&\\[-5mm]
\hspace{.1in} $g$ \hspace{.1in} & 
\hspace{.1in} $\rho$ \hspace{.1in} \\[.1in]
\hline
\hline
&\\[-5mm]
$\ft14$ & 1 \\[.1in]
\hline
\end{tabular} \\[-.15in]
\begin{tabular}{c}
\hspace{4.2in}
$n_H-n_V=\ft{29}{2}$
\end{tabular}
  
\end{flushleft}

\pagebreak

\subsection{The ${\bf Z}_2$ orbifold with $E_8\to E_7\times SU(2)$}
\vspace{.3in}
\begin{flushleft} 
\begin{tabular}{|cc||c||cc|}
\hline
&&&&\\[-5mm]
\hspace{.1in} $n$ \hspace{.05in} &
\hspace{.05in} ${\cal G}_n$ \hspace{.1in} &
\hspace{.1in} $f_{(n)}$ \hspace{.1in} &
\hspace{.1in} $V^{(n)}$ \hspace{.05in} &
\hspace{.05in} $H^{(n)}$ \hspace{.1in} \\[.1in]
\hline
\hline
&&&&\\[-5mm]
1 & $E_7\times SU(2)$ & 16 & 136 & 112 \\[.1in]
\hline
\end{tabular} \\[.3in]
\begin{tabular}{|c||cc||c|}
\hline
&&&\\[-5mm]
\hspace{.08in} $f$ \hspace{.05in} &
\hspace{.08in} $P$ \hspace{.05in} &
\hspace{.05in} $Q$ \hspace{.2in} &
\hspace{.2in} $P+Q$ \hspace{.15in}  \\[.1in]
\hline
\hline
&&&\\[-5mm]
16 & $\ft{15}{2}$ & $\ft32$ & 9 \\[.1in]
\hline 
\end{tabular}\\[-1in]
\begin{tabular}{c}
\hspace{3in}
{\includegraphics[height=1.3in,angle=0]{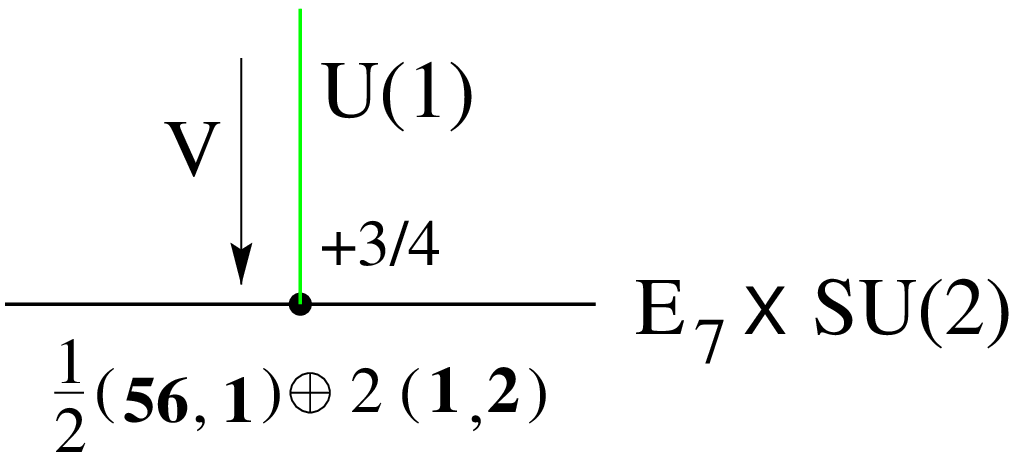}}
\end{tabular}
\vspace{.2in}
\begin{tabular}{|c||cc||ccc||ccc||c|}
\hline
&&&&&&&&&\\[-5mm]
\hspace{.1in}  $I$ \hspace{.1in} &
\hspace{.1in}  ${\cal G}_I$ \hspace{.05in} &
\hspace{.05in} ${\cal R}$ \hspace{.1in} &
\hspace{.1in}  ${\cal I}_4$ \hspace{.05in} & 
\hspace{.05in} ${\cal I}_{2,2}$ \hspace{.05in} & 
\hspace{.05in} ${\cal I}_2$ \hspace{.1in} &
\hspace{.05in} $n_0$ \hspace{.05in} &
\hspace{.05in} $n_1$ \hspace{.05in} &
\hspace{.05in} $\tilde{n}$ \hspace{.1in} &
\hspace{.1in}  ${\rm z}$ \hspace{.1in} \\[.1in]
\hline
\hline
&&&&&&&&&\\[-5mm]
1 & $E_7$ & ${\bf 133}^*$  & 0 & $\ft16$ & 3 & 1 & 1 & 0 & $\ft{1}{16}$ \\[.1in]
  &          & ${\bf 56}^*$ & 0 & $\ft{1}{24}$ &1&2&-2& -$\ft12$ & -$\ft{1}{8}$ \\[.1in]
\hline
&&&&&&&&&\\[-5mm]
2 & $SU(2)$ & ${\bf 3}^*$ & 0 & 8 & 4 & 1 & 1 & 0 & $\ft{1}{16}$ \\[.1in]
  &         & ${\bf 2}$   & 0 & $\ft12$ & 1 & 56 & -56 & -2 & -$\ft72$ \\[.1in]
\hline
\end{tabular} \\[.3in]
\begin{tabular}{|c||cccc|c|}
\hline
&&&&&\\[-5mm]
\hspace{.1in} $I$ \hspace{.1in} &
\hspace{.1in} $X_I$ \hspace{.05in} &
\hspace{.05in} $Z_I^{(4)}$ \hspace{.1in} &
\hspace{.05in} $Z_I^{(2)}$ \hspace{.05in} &
\hspace{.05in} $Z_I^{(2,2)}$ \hspace{.05in} &
\hspace{.05in} $\rho_I$ \hspace{.05in} \\[.1in]
\hline
\hline
&&&&&\\[-5mm]
1 & $\ft16$ & 0 & -$\ft{7}{16}$ & -$\ft{1}{64}$ & 0  \\[.1in]
2 & 2 & 0 & -$\ft{21}{4}$ & -$\ft94$ & 0 \\[.1in]
\hline
\end{tabular}
\hspace{.3in}
\begin{tabular}{|c||c|}
\hline
&\\[-5mm]
\hspace{.1in} $g$ \hspace{.1in} & 
\hspace{.1in} $\rho$ \hspace{.1in} \\[.1in]
\hline
\hline
&\\[-5mm]
$\ft34$ & 0 \\[.1in]
\hline
\end{tabular} \\[-.15in]
\begin{tabular}{c}
\hspace{4.2in}
$n_H-n_V=\ft{63}{2}$
\end{tabular}
  
\end{flushleft}

\pagebreak
 
\subsection{The ${\bf Z}_3$ orbifold with $E_8\to E_8$}
\vspace{.3in}
\begin{flushleft} 
\begin{tabular}{|cc||c||cc|}
\hline
&&&&\\[-5mm]
\hspace{.1in} $n$ \hspace{.05in} &
\hspace{.05in} ${\cal G}_n$ \hspace{.1in} &
\hspace{.1in} $f_{(n)}$ \hspace{.1in} &
\hspace{.1in} $V^{(n)}$ \hspace{.05in} &
\hspace{.05in} $H^{(n)}$ \hspace{.1in} \\[.1in]
\hline
\hline
&&&&\\[-5mm]
1 & $E_8$ & 9 & 248 & 0 \\[.1in]
\hline
\end{tabular} \\[.3in]
\begin{tabular}{|c||cc||c|}
\hline
&&&\\[-5mm]
\hspace{.08in} $f$ \hspace{.05in} &
\hspace{.08in} $P$ \hspace{.05in} &
\hspace{.05in} $Q$ \hspace{.2in} &
\hspace{.2in} $P+Q$ \hspace{.15in}  \\[.1in]
\hline
\hline
&&&\\[-5mm]
9 & $\ft{121}{9}$ & $\ft{248}{9}$ & 41  \\[.1in]
\hline 
\end{tabular}\\[-1in]
\begin{tabular}{c}
\hspace{3in}
{\includegraphics[height=1in,angle=0]{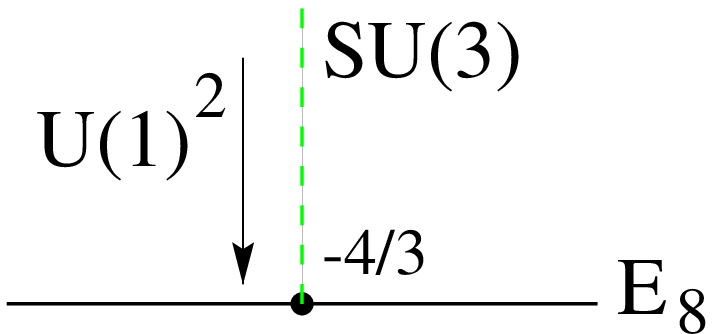}}
\end{tabular}
\vspace{.2in}
\begin{tabular}{|c||cc||ccc||ccc||c|}
\hline
&&&&&&&&&\\[-5mm]
\hspace{.1in}  $I$ \hspace{.1in} &
\hspace{.1in}  ${\cal G}_I$ \hspace{.05in} &
\hspace{.05in} ${\cal R}$ \hspace{.1in} &
\hspace{.1in}  ${\cal I}_4$ \hspace{.05in} & 
\hspace{.05in} ${\cal I}_{2,2}$ \hspace{.05in} & 
\hspace{.05in} ${\cal I}_2$ \hspace{.1in} &
\hspace{.05in} $n_0$ \hspace{.05in} &
\hspace{.05in} $n_1$ \hspace{.05in} &
\hspace{.05in} $\tilde{n}$ \hspace{.1in} &
\hspace{.1in}  ${\rm z}$ \hspace{.1in} \\[.1in]
\hline
\hline
&&&&&&&&&\\[-5mm]
1 & $E_8$ & ${\bf 248}^*$ & 0 & 9 & 30 & 1 & 1 & 0 & $\ft19$ \\[.1in]
 \hline
 \end{tabular} \\[.3in]
\begin{tabular}{|c||cccc||c|}
\hline
&&&&&\\[-5mm]
\hspace{.1in} $I$ \hspace{.1in} &
\hspace{.1in} $X_I$ \hspace{.05in} &
\hspace{.05in} $Z_I^{(4)}$ \hspace{.1in} &
\hspace{.05in} $Z_I^{(2)}$ \hspace{.05in} &
\hspace{.05in} $Z_I^{(2,2)}$ \hspace{.05in} &
\hspace{.05in} $\rho_I$ \hspace{.05in}  \\[.1in]
\hline
\hline
&&&&&\\[-5mm]
1 & 1 & 0 & $\ft{10}{3}$ & 1 & 0 \\[.1in]
\hline
\end{tabular}
\hspace{.3in}
\begin{tabular}{|c||c|}
\hline
&\\[-5mm]
\hspace{.1in} $g$ \hspace{.1in} & 
\hspace{.1in} $\rho$ \hspace{.1in} \\[.1in]
\hline
\hline
&\\[-5mm]
-$\ft43$ & 0 \\[.1in]
\hline
\end{tabular} \\[.05in]
\begin{tabular}{c}
\hspace{4.2in}
$n_H-n_V=1$
\end{tabular}

\end{flushleft}

\pagebreak

\subsection{The ${\bf Z}_3$ orbifold with $E_8\to E_6\times SU(3)$}
\vspace{.3in}
\begin{flushleft} 
\begin{tabular}{|cc||c||cc|}
\hline
&&&&\\[-5mm]
\hspace{.1in} $n$ \hspace{.05in} &
\hspace{.05in} ${\cal G}_n$ \hspace{.1in} &
\hspace{.1in} $f_{(n)}$ \hspace{.1in} &
\hspace{.1in} $V^{(n)}$ \hspace{.05in} &
\hspace{.05in} $H^{(n)}$ \hspace{.1in} \\[.1in]
\hline
\hline
&&&&\\[-5mm]
1 & $E_6\times SU(3)$ & 9 & 86 & 81  \\[.1in]
\hline
\end{tabular} \\[.3in]
\begin{tabular}{|c||cc||c|}
\hline
&&&\\[-5mm]
\hspace{.08in} $f$ \hspace{.05in} &
\hspace{.08in} $P$ \hspace{.05in} &
\hspace{.05in} $Q$ \hspace{.2in} &
\hspace{.2in} $P+Q$ \hspace{.15in}  \\[.1in]
\hline
\hline
&&&\\[-5mm]
9 & $\ft{121}{9}$ & $\ft59$ & 14  \\[.1in]
\hline 
\end{tabular}\\[-1in]
\begin{tabular}{c}
\hspace{3in}
{\includegraphics[height=1in,angle=0]{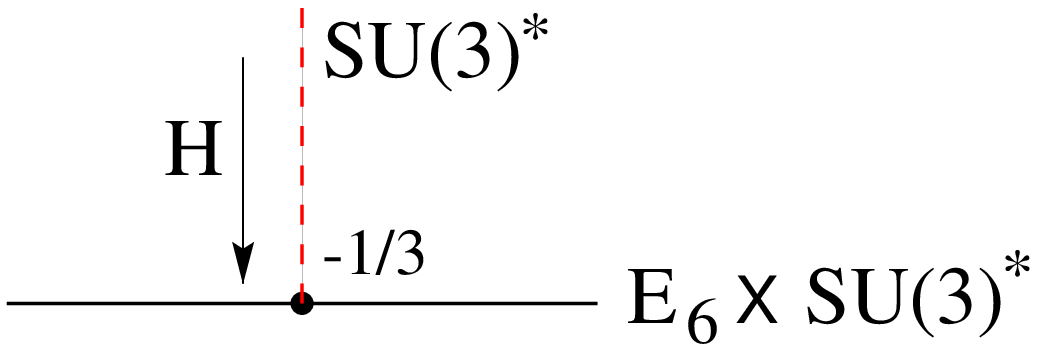}}
\end{tabular}
\vspace{.2in}
\begin{tabular}{|c||cc||ccc||ccc||c|}
\hline
&&&&&&&&&\\[-5mm]
\hspace{.1in}  $I$ \hspace{.1in} &
\hspace{.1in}  ${\cal G}_I$ \hspace{.05in} &
\hspace{.05in} ${\cal R}$ \hspace{.1in} &
\hspace{.1in}  ${\cal I}_4$ \hspace{.05in} & 
\hspace{.05in} ${\cal I}_{2,2}$ \hspace{.05in} & 
\hspace{.05in} ${\cal I}_2$ \hspace{.1in} &
\hspace{.05in} $n_0$ \hspace{.05in} &
\hspace{.05in} $n_1$ \hspace{.05in} &
\hspace{.05in} $\tilde{n}$ \hspace{.1in} &
\hspace{.1in}  ${\rm z}$ \hspace{.1in} \\[.1in]
\hline
\hline
&&&&&&&&&\\[-5mm]
1 & $E_6$ & ${\bf 78}^*$ & 0 & $\ft12$ & 4 & 1 & 1 & 0 & $\ft19$ \\[.1in]
  &       & ${\bf 27}$ & 0 & $\ft{1}{12}$ & 1 & 6 & -3 & 0 & -$\ft13$ \\[.1in]
\hline
&&&&&&&&&\\[-5mm]
2 & $SU(3)$ & ${\bf 8}^*$ & 0 & 9 & 6 & 1 & 1 & -$\ft12$ & $\ft19$ \\[.1in]
  &         & ${\bf 3}$   & 0 & $\ft12$ & 1 & 54 & -27 & 0 & -3 \\[.1in]
\hline
\end{tabular} \\[.3in]
\begin{tabular}{|c||cccc||c|}
\hline
&&&&&\\[-5mm]
\hspace{.1in} $I$ \hspace{.1in} &
\hspace{.1in} $X_I$ \hspace{.05in} &
\hspace{.05in} $Z_I^{(4)}$ \hspace{.1in} &
\hspace{.05in} $Z_I^{(2)}$ \hspace{.05in} &
\hspace{.05in} $Z_I^{(2,2)}$ \hspace{.05in} &
\hspace{.05in} $\rho_I$ \hspace{.05in} \\[.1in]
\hline
\hline
&&&&&\\[-5mm]
1 & $\ft13$ & 0 & $\ft19$ & $\ft{1}{36}$ & 0 \\[.1in]
2 & 2 & 0 & -$\ft{16}{3}$ & -5 & 1 \\[.1in]
\hline
\end{tabular}
\hspace{.3in}
\begin{tabular}{|c||c|}
\hline
&\\[-5mm]
\hspace{.1in} $g$ \hspace{.1in} & 
\hspace{.1in} $\rho$ \hspace{.1in} \\[.1in]
\hline
\hline
& \\[-5mm]
-$\ft13$ & 1 \\[.1in]
\hline
\end{tabular} \\[-.15in]
\begin{tabular}{c}
\hspace{4.2in}
$n_H-n_V=4$
\end{tabular}
  
\end{flushleft}

\pagebreak

\subsection{The ${\bf Z}_3$ orbifold with $E_8\to SU(9)$}
\vspace{.3in}
\begin{flushleft} 
\begin{tabular}{|cc||c||cc|}
\hline
&&&&\\[-5mm]
\hspace{.1in} $n$ \hspace{.05in} &
\hspace{.05in} ${\cal G}_n$ \hspace{.1in} &
\hspace{.1in} $f_{(n)}$ \hspace{.1in} &
\hspace{.1in} $V^{(n)}$ \hspace{.05in} &
\hspace{.05in} $H^{(n)}$ \hspace{.1in} \\[.1in]
\hline
\hline
&&&&\\[-5mm]
1 & $SU(9)$ & 9 & 80 & 84 \\[.1in]
\hline
\end{tabular} \\[.3in]
\begin{tabular}{|c||cc||c|}
\hline
&&&\\[-5mm]
\hspace{.08in} $f$ \hspace{.05in} &
\hspace{.08in} $P$ \hspace{.05in} &
\hspace{.05in} $Q$ \hspace{.2in} &
\hspace{.2in} $P+Q$ \hspace{.15in}  \\[.1in]
\hline
\hline
&&&\\[-5mm]
9 & $\ft{121}{9}$ & -$\ft49$ & 13  \\[.1in]
\hline 
\end{tabular}\\[-1in]
\begin{tabular}{c}
\hspace{3in}
{\includegraphics[height=1.3in,angle=0]{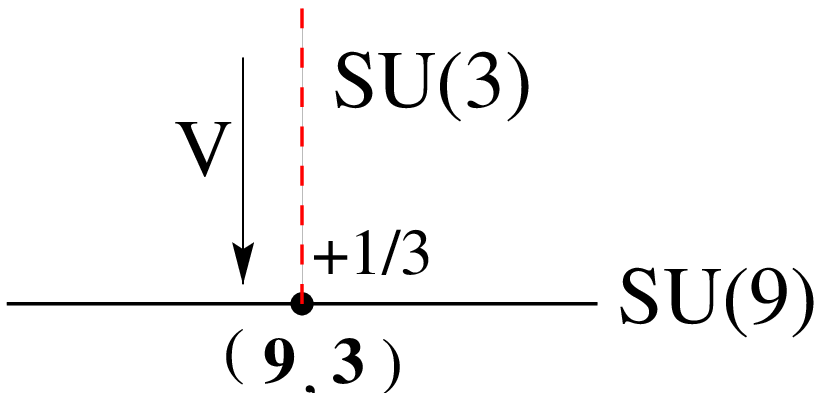}}
\end{tabular}
\vspace{.2in}
\begin{tabular}{|c||cc||ccc||ccc||c|}
\hline
&&&&&&&&&\\[-5mm]
\hspace{.1in}  $I$ \hspace{.1in} &
\hspace{.1in}  ${\cal G}_I$ \hspace{.05in} &
\hspace{.05in} ${\cal R}$ \hspace{.1in} &
\hspace{.1in}  ${\cal I}_4$ \hspace{.05in} & 
\hspace{.05in} ${\cal I}_{2,2}$ \hspace{.05in} & 
\hspace{.05in} ${\cal I}_2$ \hspace{.1in} &
\hspace{.05in} $n_0$ \hspace{.05in} &
\hspace{.05in} $n_1$ \hspace{.05in} &
\hspace{.05in} $\tilde{n}$ \hspace{.1in} &
\hspace{.1in}  ${\rm z}$ \hspace{.1in} \\[.1in]
\hline
\hline
&&&&&&&&&\\[-5mm]
1 & $SU(9)$  & ${\bf 84}$   & -9  & 15 & 21 & 2 & -1 & 0 & -$\ft19$ \\[.1in]
  &          & ${\bf 80}^*$ & 18  &  6 & 18 & 1 & 1 & 0 & $\ft19$ \\[.1in]
  &          & ${\bf 9}$    &  1  &  0 &  1 & 0 & 0 & -3 & 0 \\[.1in]
\hline
&&&&&&&&&\\[-5mm]
2 & $SU(3)$  & ${\bf 8}^*$   & 0 & 9 & 6 & 0 & 0 & $\ft12$ & 0 \\[.1in]
  &          & ${\bf 3}$ & 0  & $\ft12$ & 1 & 0 & 0 & -9 & 0 \\[.1in]
\hline
\end{tabular} \\[.3in]
\begin{tabular}{|c||cccc||c|}
\hline
&&&&&\\[-5mm]
\hspace{.1in} $I$ \hspace{.1in} &
\hspace{.1in} $X_I$ \hspace{.05in} &
\hspace{.05in} $Z_I^{(4)}$ \hspace{.1in} &
\hspace{.05in} $Z_I^{(2)}$ \hspace{.05in} &
\hspace{.05in} $Z_I^{(2,2)}$ \hspace{.05in} &
\hspace{.05in} $\rho_I$ \hspace{.05in} \\[.1in]
\hline
\hline
&&&&&\\[-5mm]
1 & 2 & 0 & -$\ft{10}{3}$ & -1 & 0  \\[.1in]
2 & 0 & 0 & -6 & 0 & 1  \\[.1in]
\hline
\end{tabular}
\hspace{.3in}
\begin{tabular}{|c||c|}
\hline
&\\[-5mm]
\hspace{.1in} $g$ \hspace{.1in} & 
\hspace{.1in} $\rho$ \hspace{.1in} \\[.1in]
\hline
\hline
&\\[-5mm]
$\ft13$ & 1 \\[.1in]
\hline
\end{tabular} \\[-.15in]
\begin{tabular}{c}
\hspace{4.2in}
$n_H-n_V=23$
\end{tabular}
  
\end{flushleft}

\pagebreak
 
\subsection{The ${\bf Z}_4$ orbifold with $E_8\to E_8$}
\vspace{.3in}
\begin{flushleft} 
\begin{tabular}{|cc||c||cc|}
\hline
&&&&\\[-5mm]
\hspace{.1in} $n$ \hspace{.05in} &
\hspace{.05in} ${\cal G}_n$ \hspace{.1in} &
\hspace{.1in} $f_{(n)}$ \hspace{.1in} &
\hspace{.1in} $V^{(n)}$ \hspace{.05in} &
\hspace{.05in} $H^{(n)}$ \hspace{.1in} \\[.1in]
\hline
\hline
&&&&\\[-5mm]
1 & $E_8$ & 4 & 248 & 0 \\[.1in]
2 & $E_8$ & $\ft{32}{3}$ & 248 & 0 \\[.1in]
\hline
\end{tabular} \\[.2in]
\begin{tabular}{|c||cc||c|}
\hline
&&&\\[-5mm]
\hspace{.08in} $f$ \hspace{.05in} &
\hspace{.08in} $P$ \hspace{.05in} &
\hspace{.05in} $Q$ \hspace{.2in} &
\hspace{.2in} $P+Q$ \hspace{.15in}  \\[.1in]
\hline
\hline
&&&\\[-5mm]
$\ft{32}{5}$ & 19 & $\ft{155}{4}$ & $\ft{231}{4}$  \\[.1in]
\hline 
\end{tabular}\\[-1in]
\begin{tabular}{c}
\hspace{3in}
{\includegraphics[height=1in,angle=0]{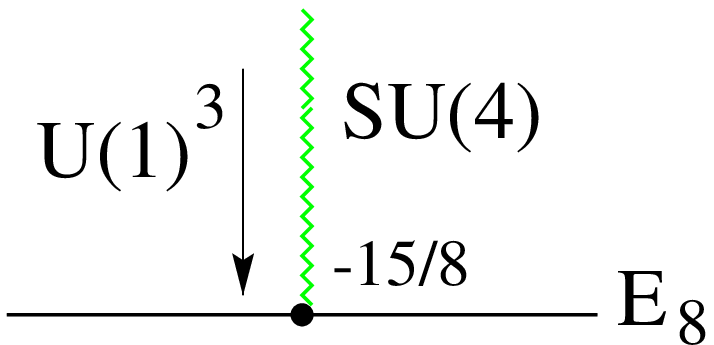}}
\end{tabular}
\vspace{.2in}
\begin{tabular}{|c||cc||ccc||cccc||c|}
\hline
&&&&&&&&&&\\[-5mm]
\hspace{.1in}  $I$ \hspace{.1in} &
\hspace{.1in}  ${\cal G}_I$ \hspace{.05in} &
\hspace{.05in} ${\cal R}$ \hspace{.08in} &
\hspace{.08in}  ${\cal I}_4$ \hspace{.03in} & 
\hspace{.03in} ${\cal I}_{2,2}$ \hspace{.03in} & 
\hspace{.03in} ${\cal I}_2$ \hspace{.08in} &
\hspace{.05in} $n_0$ \hspace{.05in} &
\hspace{.05in} $n_1$ \hspace{.05in} &
\hspace{.05in} $n_2$ \hspace{.05in} &
\hspace{.05in} $\tilde{n}$ \hspace{.1in} &
\hspace{.1in}  ${\rm z}$ \hspace{.1in} \\[.1in]
\hline
\hline
&&&&&&&&&&\\[-5mm]
1 & $E_8$ & ${\bf 248}^*$ & 0 & 9 & 30 & 1 & 1 & 1  & 0 & $\ft{5}{32}$ \\[.1in]
\hline
\end{tabular} \\[.2in]
\begin{tabular}{|c||cccc||c|}
\hline
&&&&&\\[-5mm]
\hspace{.1in} $I$ \hspace{.1in} &
\hspace{.1in} $X_I$ \hspace{.05in} &
\hspace{.05in} $Z_I^{(4)}$ \hspace{.1in} &
\hspace{.05in} $Z_I^{(2)}$ \hspace{.05in} &
\hspace{.05in} $Z_I^{(2,2)}$ \hspace{.05in} & 
\hspace{.05in} $\rho_I$ \hspace{.05in} \\[.1in]
\hline
\hline
&&&&&\\[-5mm]
1 & 1 & 0 & $\ft{75}{16}$ & $\ft{45}{32}$ & 0 \\[.1in]
\hline
\end{tabular}
\hspace{.3in}
\begin{tabular}{|c||c|}
\hline
&\\[-5mm]
\hspace{.1in} $g$ \hspace{.1in} & 
\hspace{.1in} $\rho$ \hspace{.1in} \\[.1in]
\hline
\hline
&\\[-5mm]
-$\ft{15}{8}$ & 0 \\[.1in]
\hline
\end{tabular} \\[.1in]
\begin{tabular}{c}
\hspace{4.2in}
$n_H-n_V=\ft32$
\end{tabular}
   
\end{flushleft}

\pagebreak

\subsection{The ${\bf Z}_4$ orbifold with $E_8\to SO(10)\times SU(4)$}
\vspace{.3in}
\begin{flushleft} 
\begin{tabular}{|cc||c||cc|}
\hline
&&&&\\[-5mm]
\hspace{.1in} $n$ \hspace{.05in} &
\hspace{.05in} ${\cal G}_n$ \hspace{.1in} &
\hspace{.1in} $f_{(n)}$ \hspace{.1in} &
\hspace{.1in} $V^{(n)}$ \hspace{.05in} &
\hspace{.05in} $H^{(n)}$ \hspace{.1in} \\[.1in]
\hline
\hline
&&&&\\[-5mm]
1 & $SO(10)\times SU(4)$ & 4 & 60 & 64 \\[.1in]
2 & $SO(16)$ & $\ft{32}{3}$ & 120 & 128 \\[.1in]
\hline
\end{tabular} \\[.2in]
\begin{tabular}{|c||cc||c|}
\hline
&&&\\[-5mm]
\hspace{.08in} $f$ \hspace{.05in} &
\hspace{.08in} $P$ \hspace{.05in} &
\hspace{.05in} $Q$ \hspace{.2in} &
\hspace{.2in} $P+Q$ \hspace{.15in}  \\[.1in]
\hline
\hline
&&&\\[-5mm]
$\ft{32}{5}$ & 19 & -$\ft14$ & $\ft{75}{4}$  \\[.1in]
\hline 
\end{tabular}\\[-1in]
\begin{tabular}{c}
\hspace{3in}
{\includegraphics[height=.8in,angle=0]{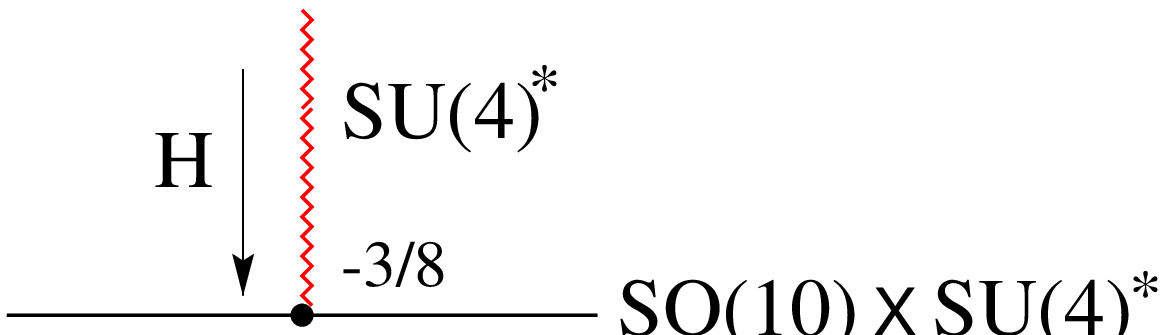}} \\[.2in]
\end{tabular}
\begin{tabular}{|c||cc||ccc||cccc||c|}
\hline
&&&&&&&&&&\\[-5mm]
\hspace{.1in}  $I$ \hspace{.1in} &
\hspace{.1in}  ${\cal G}_I$ \hspace{.05in} &
\hspace{.05in} ${\cal R}$ \hspace{.08in} &
\hspace{.08in}  ${\cal I}_4$ \hspace{.03in} & 
\hspace{.03in} ${\cal I}_{2,2}$ \hspace{.03in} & 
\hspace{.03in} ${\cal I}_2$ \hspace{.08in} &
\hspace{.05in} $n_0$ \hspace{.05in} &
\hspace{.05in} $n_1$ \hspace{.05in} &
\hspace{.05in} $n_2$ \hspace{.05in} &
\hspace{.05in} $\tilde{n}$ \hspace{.1in} &
\hspace{.1in}  ${\rm z}$ \hspace{.1in} \\[.1in]
\hline
\hline
&&&&&&&&&&\\[-5mm]
1 & $SO(10)$ & ${\bf 45}^*$ & 2  & 3 & 8 & 1  & 1  & 1  & 0 & $\ft{5}{32}$ \\[.1in]
  &  & ${\bf 16}$ & -1 & $\ft34$ & 2 & 8 & -4 & -8 & 0 & -$\ft14$ \\[.1in]
  &  & ${\bf 10}$ & 1 & 0 & 1 & 6 & 0 & 6 & 0 & -$\ft{9}{16}$ \\[.1in]
\hline
&&&&&&&&&&\\[-5mm]
2 & $SU(4)$ & ${\bf 15}^*$ & 8&6&8&1&1&1& -$\ft12$ & $\ft{5}{32}$ \\[.1in]
  & & ${\bf 6}$ & -4 & 3 & 2 & 10 & 0 & 10 & 0 & -$\ft{15}{16}$ \\[.1in]
  & & ${\bf 4}$ & 1 & 0 & 1 & 32 & -16 & -32 & 0 & -1 \\[.1in]
\hline
\end{tabular} \\[.2in]
\begin{tabular}{|c||cccc||c|}
\hline
&&&&&\\[-5mm]
\hspace{.1in} $I$ \hspace{.1in} &
\hspace{.1in} $X_I$ \hspace{.05in} &
\hspace{.05in} $Z_I^{(4)}$ \hspace{.1in} &
\hspace{.05in} $Z_I^{(2)}$ \hspace{.05in} &
\hspace{.05in} $Z_I^{(2,2)}$ \hspace{.05in} &
\hspace{.05in} $\rho_I$ \hspace{.05in}\\[.1in]
\hline
\hline
&&&&&\\[-5mm]
1 & 1 & 0 & $\ft{3}{16}$ & $\ft{9}{32}$ & 0 \\[.1in]
2 & 2 & 0 & -$\ft{45}{8}$ & -$\ft{39}{8}$ & 1  \\[.1in]
\hline
\end{tabular}
\hspace{.3in}
\begin{tabular}{|c||c|}
\hline
&\\[-5mm]
\hspace{.1in} $g$ \hspace{.1in} & 
\hspace{.1in} $\rho$ \hspace{.1in} \\[.1in]
\hline
\hline
&\\[-5mm]
-$\ft{3}{8}$ & 1 \\[.1in]
\hline
\end{tabular} \\[-.15in]
\begin{tabular}{c}
\hspace{4.2in}
$n_H-n_V=\ft{15}{2}$
\end{tabular}
  
\end{flushleft}

\pagebreak

\subsection{The ${\bf Z}_4$ orbifold with $E_8\to SU(8)\times SU(2)$}
\vspace{.3in}
\begin{flushleft} 
\begin{tabular}{|cc||c||cc|}
\hline
&&&&\\[-5mm]
\hspace{.1in} $n$ \hspace{.05in} &
\hspace{.05in} ${\cal G}_n$ \hspace{.1in} &
\hspace{.1in} $f_{(n)}$ \hspace{.1in} &
\hspace{.1in} $V^{(n)}$ \hspace{.05in} &
\hspace{.05in} $H^{(n)}$ \hspace{.1in} \\[.1in]
\hline
\hline
&&&&\\[-5mm]
1 & $SU(8)\times SU(2)$ & 4 & 66 & 56 \\[.1in]
2 & $E_7\times SU(2)$ & $\ft{32}{3}$ & 136 & 112 \\[.1in]
\hline
\end{tabular} \\[.2in]
\begin{tabular}{|c||cc||c|}
\hline
&&&\\[-5mm]
\hspace{.08in} $f$ \hspace{.05in} &
\hspace{.08in} $P$ \hspace{.05in} &
\hspace{.05in} $Q$ \hspace{.2in} &
\hspace{.2in} $P+Q$ \hspace{.15in}  \\[.1in]
\hline
\hline
&&&\\[-5mm]
$\ft{32}{5}$ & 19 & $\ft14$ & $\ft{77}{4}$  \\[.1in]
\hline 
\end{tabular}\\[-1in]
\begin{tabular}{c}
\hspace{3in}
{\includegraphics[width=3in,angle=0]{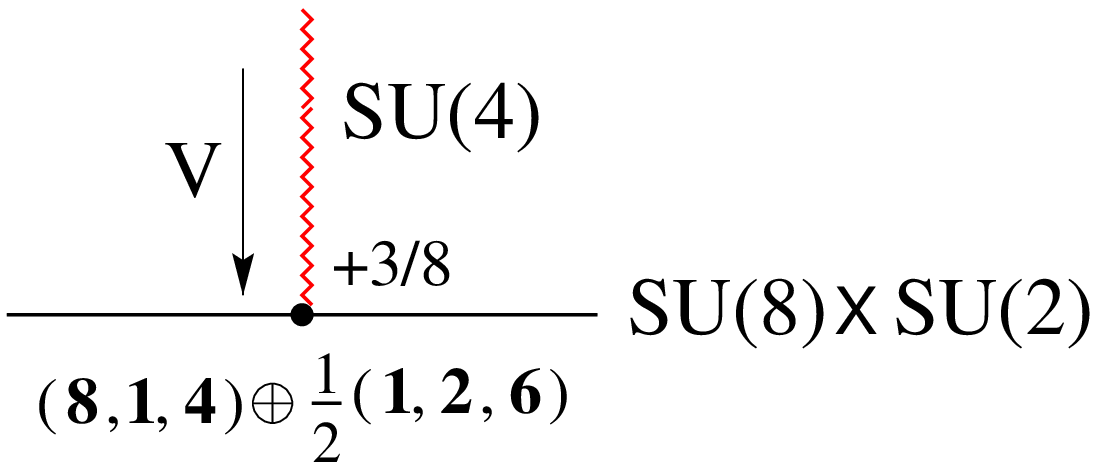}}
\end{tabular}
\vspace{.2in}
\begin{tabular}{|c||cc||ccc||cccc||c|}
\hline
&&&&&&&&&&\\[-5mm]
\hspace{.1in}  $I$ \hspace{.1in} &
\hspace{.1in}  ${\cal G}_I$ \hspace{.05in} &
\hspace{.05in} ${\cal R}$ \hspace{.08in} &
\hspace{.08in}  ${\cal I}_4$ \hspace{.03in} & 
\hspace{.03in} ${\cal I}_{2,2}$ \hspace{.03in} & 
\hspace{.03in} ${\cal I}_2$ \hspace{.08in} &
\hspace{.05in} $n_0$ \hspace{.05in} &
\hspace{.05in} $n_1$ \hspace{.05in} &
\hspace{.05in} $n_2$ \hspace{.05in} &
\hspace{.05in} $\tilde{n}$ \hspace{.1in} &
\hspace{.1in}  ${\rm z}$ \hspace{.1in} \\[.1in]
\hline
\hline
&&&&&&&&&&\\[-5mm]
1 & $SU(8)$ & ${\bf 70}$ & -16 & 18 & 20 &1& 0 & 1 & 0 & -$\ft{3}{32}$ \\[.1in]
 & & ${\bf 63}^*$ & 16 & 6 & 16 & 1 & 1 & 1 & 0 & $\ft{5}{32}$ \\[.1in]
  &  & ${\bf 28}$ & 0 & 3 & 6 & 4 & -2 & -4 & 0 & -$\ft18$ \\[.1in]
  &  & ${\bf 8}$ & 1 & 0 & 1 & 0 & 0 & 0 & -4 & 0 \\[.1in]
\hline
&&&&&&&&&&\\[-5mm]
2 & $SU(2)$ & ${\bf 3}^*$ & 0&8&4&1&1&1& 0 & $\ft{5}{32}$ \\[.1in]
  & & ${\bf 2}$ & 0 & $\ft12$ & 1 & 56 & -28 & -56 & -3 & -$\ft74$ \\[.1in]
\hline
&&&&&&&&&&\\[-5mm]
3 & $SU(4)$ & ${\bf 15}^*$ & 8&6&8& 0&0&0& $\ft12$ & 0 \\[.1in]
  & & ${\bf 6}$ & -4 & 3 & 2 & 0 & 0 & 0 & $-1$ & 0 \\[.1in]
  & & ${\bf 4}$ & 1 & 0 & 1 & 0 & 0 & 0 & $-8$ & 0 \\[.1in]
\hline
\end{tabular} \\[.2in]
\begin{tabular}{|c||cccc||c|}
\hline
&&&&&\\[-5mm]
\hspace{.1in} $I$ \hspace{.1in} &
\hspace{.1in} $X_I$ \hspace{.05in} &
\hspace{.05in} $Z_I^{(4)}$ \hspace{.1in} &
\hspace{.05in} $Z_I^{(2)}$ \hspace{.05in} &
\hspace{.05in} $Z_I^{(2,2)}$ \hspace{.05in} &
\hspace{.05in} $\rho_I$ \hspace{.05in}\\[.1in]
\hline
\hline
&&&&&\\[-5mm]
1 & 2 & 0 & -$\ft{33}{8}$ & -$\ft{9}{8}$ & 0 \\[.1in]
2 & 2 & 0 & -$\ft{33}{8}$ & -$\ft{9}{8}$ & 0 \\[.1in]
3 & 0 & 0 & $-6$ & 0 & 1  \\[.1in]
\hline
\end{tabular} \hspace{.3in}
\begin{tabular}{|c||c|}
\hline
&\\[-5mm]
\hspace{.1in} $g$ \hspace{.1in} & 
\hspace{.1in} $\rho$ \hspace{.1in} \\[.1in]
\hline
\hline
&\\[-5mm]
$\ft{3}{8}$ & 1 \\[.1in]
\hline
\end{tabular} \\[-.21in]
\begin{tabular}{c}
\hspace{4.2in}
$n_H-n_V=\ft{61}{2}$
\end{tabular}

\end{flushleft}

\pagebreak

\subsection{The ${\bf Z}_6$ orbifold with $E_8\to E_8$}
\vspace{.3in}
\begin{flushleft} 
\begin{tabular}{|cc||c||cc|}
\hline
&&&&\\[-5mm]
\hspace{.1in} $n$ \hspace{.05in} &
\hspace{.05in} ${\cal G}_n$ \hspace{.1in} &
\hspace{.1in} $f_{(n)}$ \hspace{.1in} &
\hspace{.1in} $V^{(n)}$ \hspace{.05in} &
\hspace{.05in} $H^{(n)}$ \hspace{.1in} \\[.1in]
\hline
\hline
&&&&\\[-5mm]
1 & $E_8$ & 1 & 248 & 0 \\[.1in]
2 & $E_8$ & $\ft94$ & 248 & 0 \\[.1in]
3 & $E_8$ & $\ft{16}{5}$ & 248 & 0 \\[.1in]
\hline
\end{tabular} \\[.2in]
\begin{tabular}{|c||cc||c|}
\hline
&&&\\[-5mm]
\hspace{.08in} $f$ \hspace{.05in} &
\hspace{.08in} $P$ \hspace{.05in} &
\hspace{.05in} $Q$ \hspace{.2in} &
\hspace{.2in} $P+Q$ \hspace{.15in}  \\[.1in]
\hline
\hline
&&&\\[-5mm]
$\ft{144}{35}$ & $\ft{535}{18}$ & $\ft{1085}{18}$ & 90  \\[.1in]
\hline 
\end{tabular}\\[-1in]
\begin{tabular}{c}
\hspace{3in}
{\includegraphics[height=1in,angle=0]{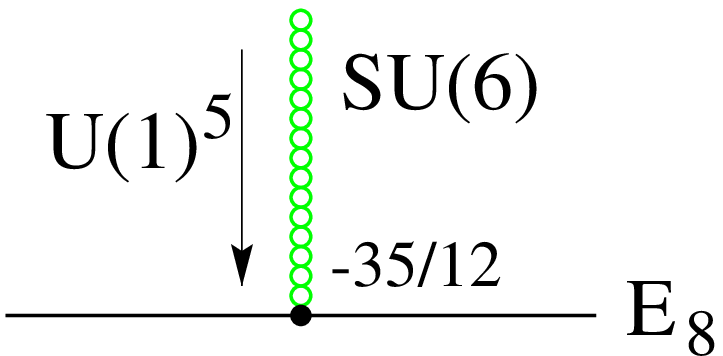}}
\end{tabular}
\vspace{.2in}
\begin{tabular}{|c||cc||ccc||ccccc||c|}
\hline
&&&&&&&&&&&\\[-5mm]
\hspace{.08in}  $I$ \hspace{.1in} &
\hspace{.08in}  ${\cal G}_I$ \hspace{.03in} &
\hspace{.03in} ${\cal R}$ \hspace{.07in} &
\hspace{.07in}  ${\cal I}_4$ \hspace{.02in} & 
\hspace{.02in} ${\cal I}_{2,2}$ \hspace{.02in} & 
\hspace{.02in} ${\cal I}_2$ \hspace{.07in} &
\hspace{.03in} $n_0$ \hspace{.02in} &
\hspace{.02in} $n_1$ \hspace{.02in} &
\hspace{.02in} $n_2$ \hspace{.02in} &
\hspace{.02in} $n_3$ \hspace{.02in} &
\hspace{.03in} $\tilde{n}$ \hspace{.1in} &
\hspace{.08in}  ${\rm z}$ \hspace{.1in} \\[.1in]
\hline
\hline
&&&&&&&&&&&\\[-5mm]
1 & $E_8$ & ${\bf 248}^*$ & 0 & 9 & 30 & 1 & 1 & 1 & 1 & 0 & $\ft{35}{144}$ \\[.1in]
\hline
\end{tabular} \\[.2in]
\begin{tabular}{|c||cccc||c|}
\hline
&&&&&\\[-5mm]
\hspace{.1in} $I$ \hspace{.1in} &
\hspace{.1in} $X_I$ \hspace{.05in} &
\hspace{.05in} $Z_I^{(4)}$ \hspace{.1in} &
\hspace{.05in} $Z_I^{(2)}$ \hspace{.05in} &
\hspace{.05in} $Z_I^{(2,2)}$ \hspace{.05in} &
\hspace{.05in} $\rho_I$ \hspace{.05in} \\[.1in]
\hline
\hline
&&&&&\\[-5mm]
1 & 1 & 0 & $\ft{175}{24}$ & $\ft{35}{16}$ & 0  \\[.1in]
\hline
\end{tabular}
\hspace{.3in}
\begin{tabular}{|c||c|}
\hline
&\\[-5mm]
\hspace{.1in} $g$ \hspace{.1in} & 
\hspace{.1in} $\rho$ \hspace{.1in} \\[.1in]
\hline
\hline
&\\[-5mm]
-$\ft{35}{12}$ & 0 \\[.1in]
\hline
\end{tabular} \\[.1in]
\begin{tabular}{c}
\hspace{4.2in}
$n_H-n_V=\ft52$
\end{tabular}
  
\end{flushleft}

\pagebreak

\subsection{The ${\bf Z}_6$ orbifold with $E_8\to SU(8)\times U(1)$}
\vspace{.3in}
\begin{flushleft} 
\begin{tabular}{|cc||c||cc|}
\hline
&&&&\\[-5mm]
\hspace{.1in} $n$ \hspace{.05in} &
\hspace{.05in} ${\cal G}_n$ \hspace{.1in} &
\hspace{.1in} $f_{(n)}$ \hspace{.1in} &
\hspace{.1in} $V^{(n)}$ \hspace{.05in} &
\hspace{.05in} $H^{(n)}$ \hspace{.1in} \\[.1in]
\hline
\hline
&&&&\\[-5mm]
1 & $SU(8)\times U(1)$ & 1 & 64 & 92 \\[.1in]
2 & $SU(9)$ & $\ft94$ & 80 & 84 \\[.1in]
3 & $SO(16)$ & $\ft{16}{5}$ & 120 & 128 \\[.1in]
\hline
\end{tabular} \\[.2in]
\begin{tabular}{|c||cc||c|}
\hline
&&&\\[-5mm]
\hspace{.08in} $f$ \hspace{.05in} &
\hspace{.08in} $P$ \hspace{.05in} &
\hspace{.05in} $Q$ \hspace{.2in} &
\hspace{.2in} $P+Q$ \hspace{.15in}  \\[.1in]
\hline
\hline
&&&\\[-5mm]
$\ft{144}{35}$ & $\ft{535}{18}$ & -$\ft{427}{18}$ & 6  \\[.1in]
\hline 
\end{tabular}\\[-1in]
\begin{tabular}{c}
\hspace{3in}
{\includegraphics[height=1in,angle=0]{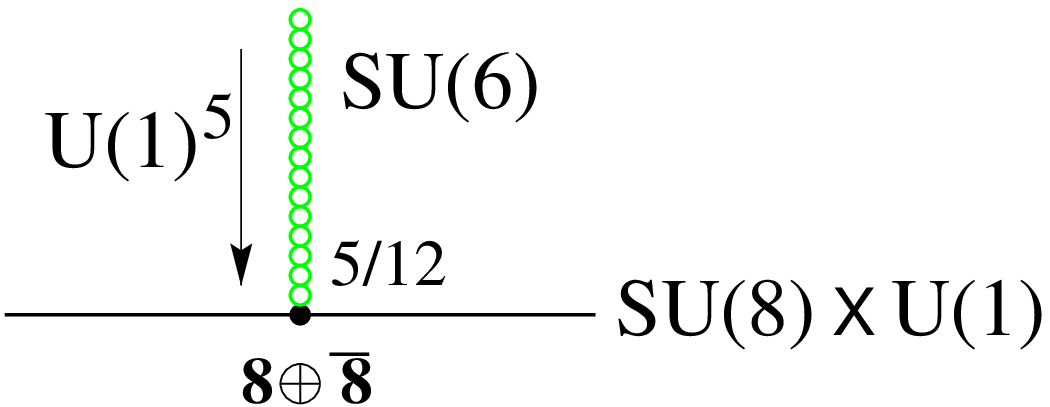}}
\end{tabular}
\vspace{.2in}
\begin{tabular}{|c||cc||ccc||ccccc||c|}
\hline
&&&&&&&&&&&\\[-5mm]
\hspace{.07in}  $I$ \hspace{.1in} &
\hspace{.01in}  ${\cal G}_I$ \hspace{.02in} &
\hspace{.03in} ${\cal R}$ \hspace{.07in} &
\hspace{.07in}  ${\cal I}_4$ \hspace{.02in} & 
\hspace{.02in} ${\cal I}_{2,2}$ \hspace{.02in} & 
\hspace{.02in} ${\cal I}_2$ \hspace{.07in} &
\hspace{.03in} $n_0$ \hspace{.02in} &
\hspace{.02in} $n_1$ \hspace{.02in} &
\hspace{.02in} $n_2$ \hspace{.02in} &
\hspace{.02in} $n_3$ \hspace{.02in} &
\hspace{.03in} $\tilde{n}$ \hspace{.1in} &
\hspace{.06in}  ${\rm z}$ \hspace{.1in} \\[.1in]
\hline
\hline
&&&&&&&&&&&\\[-5mm]
1 & $SU(8)$ & ${\bf 63}^*$ & 16 & 6 & 16 & 1&1&1&1&0& $\ft{35}{144}$ \\[.1in]
&& ${\bf 56}$ & -9 & 12 & 15 & 2 & -1 & -1 & -2 & 0 & $\ft{5}{72}$ \\[.1in]
&& ${\bf 28}$ & 0 & 3 & 6 & 2 & -1 & -1 & +2 & 0 & -$\ft{85}{72}$ \\[.1in]
&& ${\bf 8}$ & 1 & 0 & 1 & 2 & -1 & +2 & -2 & -2 & -$\ft{91}{72}$ \\[.1in]
\hline
\end{tabular} \\[.2in]
\begin{tabular}{|c||cccc||c|}
\hline
&&&&&\\[-5mm]
\hspace{.1in} $I$ \hspace{.1in} &
\hspace{.1in} $X_I$ \hspace{.05in} &
\hspace{.05in} $Z_I^{(4)}$ \hspace{.1in} &
\hspace{.05in} $Z_I^{(2)}$ \hspace{.05in} &
\hspace{.05in} $Z_I^{(2,2)}$ \hspace{.05in} &
\hspace{.05in} $\rho_I$ \hspace{.05in} \\[.1in]
\hline
\hline
&&&&&\\[-5mm]
1 & 2 & 0 & -$\ft{65}{12}$ & -$\ft54$ & 0  \\[.1in]
\hline
\end{tabular} \hspace{.2in}
\begin{tabular}{|c||c|}
\hline
&\\[-5mm]
\hspace{.1in} $g$ \hspace{.1in} & 
\hspace{.1in} $\rho$ \hspace{.1in} \\[.1in]
\hline
\hline
&\\[-5mm]
$\ft{5}{12}$ & 0 \\[.1in]
\hline
\end{tabular} \\[.2in]
\begin{tabular}{c}
\hspace{4.2in}
$n_H-n_V=\ft{37}{2}$
\end{tabular}
 
\end{flushleft}
 
\pagebreak

\subsection{The ${\bf Z}_6$ orbifold with $E_8\to E_6\times SU(2)\times U(1)$}
\vspace{.3in}
\begin{flushleft} 
\begin{tabular}{|cc||c||cc|}
\hline
&&&&\\[-5mm]
\hspace{.1in} $n$ \hspace{.05in} &
\hspace{.05in} ${\cal G}_n$ \hspace{.1in} &
\hspace{.1in} $f_{(n)}$ \hspace{.1in} &
\hspace{.1in} $V^{(n)}$ \hspace{.05in} &
\hspace{.05in} $H^{(n)}$ \hspace{.1in} \\[.1in]
\hline
\hline
&&&&\\[-5mm]
1 & $E_6\times SU(2)\times U(1)$ & 1 & 82 & 85 \\[.1in]
2 & $E_6\times SU(3)$ & $\ft94$ & 86 & 81 \\[.1in]
3 & $E_7\times SU(2)$ & $\ft{16}{5}$ & 136 & 112 \\[.1in]
\hline
\end{tabular} \\[.2in]
\begin{tabular}{|c||cc||c|}
\hline
&&&\\[-5mm]
\hspace{.08in} $f$ \hspace{.05in} &
\hspace{.08in} $P$ \hspace{.05in} &
\hspace{.05in} $Q$ \hspace{.2in} &
\hspace{.2in} $P+Q$ \hspace{.15in}  \\[.1in]
\hline
\hline
&&&\\[-5mm]
$\ft{144}{35}$ & $\ft{535}{18}$ & -$\ft{229}{18}$ & 17  \\[.1in]
\hline 
\end{tabular}\\[-1in]
\begin{tabular}{c}
\hspace{3in}
{\includegraphics[height=.9in,angle=0]{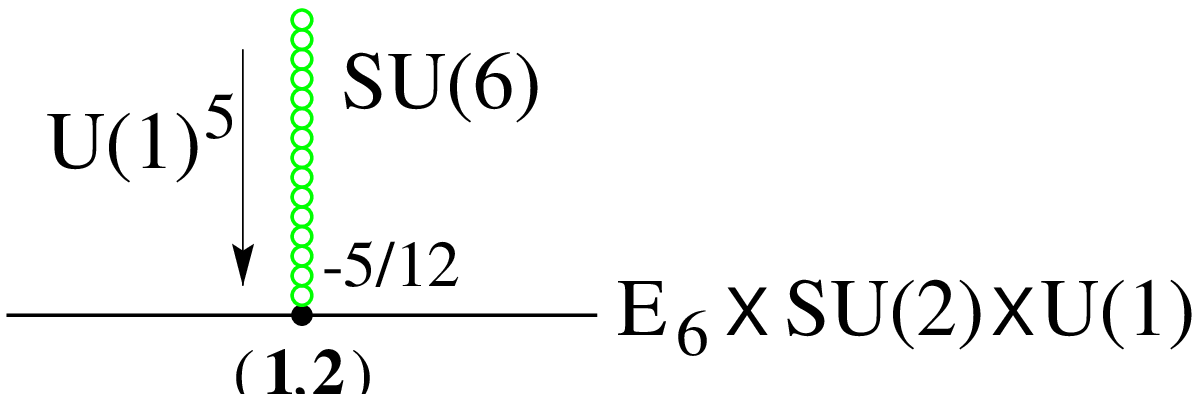}} \\[.2in]
\end{tabular}
\begin{tabular}{|c||cc||ccc||ccccc||c|}
\hline
&&&&&&&&&&&\\[-5mm]
\hspace{.07in}  $I$ \hspace{.1in} &
\hspace{.01in}  ${\cal G}_I$ \hspace{.02in} &
\hspace{.03in} ${\cal R}$ \hspace{.07in} &
\hspace{.07in}  ${\cal I}_4$ \hspace{.02in} & 
\hspace{.02in} ${\cal I}_{2,2}$ \hspace{.02in} & 
\hspace{.02in} ${\cal I}_2$ \hspace{.07in} &
\hspace{.03in} $n_0$ \hspace{.02in} &
\hspace{.02in} $n_1$ \hspace{.02in} &
\hspace{.02in} $n_2$ \hspace{.02in} &
\hspace{.02in} $n_3$ \hspace{.02in} &
\hspace{.03in} $\tilde{n}$ \hspace{.1in} &
\hspace{.06in}  ${\rm z}$ \hspace{.1in} \\[.1in]
\hline
\hline
&&&&&&&&&&&\\[-5mm]
1 & $E_6$ & ${\bf 78}^*$ & 0 & $\ft12$ & 4 & 1&1&1&1&0& $\ft{35}{144}$ \\[.1in]
&& ${\bf 27}$ & 0 & $\ft{1}{12}$ & 1 & 6 & -3 & -3 & -2 & 0 & -$\ft{25}{24}$ \\[.1in]
\hline
&&&&&&&&&&&\\[-5mm]
2 & $SU(2)$ & ${\bf 3}^*$ & 0 & 8 & 4 & 1&1&1&1&0& $\ft{35}{144}$ \\[.1in]
&& ${\bf 2}$ & 0 & $\ft{1}{2}$ &1&56&-29&-25&-56& -1 & -$\ft{7}{18}$ \\[.1in]
\hline
\end{tabular} \\[.2in]
\begin{tabular}{|c||cccc||c|}
\hline
&&&&&\\[-5mm]
\hspace{.1in} $I$ \hspace{.1in} &
\hspace{.1in} $X_I$ \hspace{.05in} &
\hspace{.05in} $Z_I^{(4)}$ \hspace{.1in} &
\hspace{.05in} $Z_I^{(2)}$ \hspace{.05in} &
\hspace{.05in} $Z_I^{(2,2)}$ \hspace{.05in} &
\hspace{.05in} $\rho_I$ \hspace{.05in} \\[.1in]
\hline
\hline
&&&&&\\[-5mm]
1 & $\ft13$ & 0 & -$\ft{5}{72}$ & $\ft{5}{144}$ & 0 \\[.1in]
&&&&&\\[-5mm]
2 & 2 & 0 & -$\ft{5}{12}$ & $\ft54$ & 0  \\[.1in]
\hline
\end{tabular} \hspace{.2in}
\begin{tabular}{|c||c|}
\hline
&\\[-5mm]
\hspace{.1in} $g$ \hspace{.1in} & 
\hspace{.1in} $\rho$ \hspace{.1in} \\[.1in]
\hline
\hline
&\\[-5mm]
-$\ft{5}{12}$ & 0 \\[.1in]
\hline
\end{tabular} \\[-.15in]
\begin{tabular}{c}
\hspace{4.2in}
$n_H-n_V=\ft92$
\end{tabular}
  
\end{flushleft}

\pagebreak

\subsection{The ${\bf Z}_6$ orbifold with $E_8\to SU(9)$}
\vspace{.3in}
\begin{flushleft} 
\begin{tabular}{|cc||c||cc|}
\hline
&&&\\[-5mm]
\hspace{.1in} $n$ \hspace{.05in} &
\hspace{.05in} ${\cal G}_n$ \hspace{.1in} &
\hspace{.1in} $f_{(n)}$ \hspace{.1in} &
\hspace{.1in} $V^{(n)}$ \hspace{.05in} &
\hspace{.05in} $H^{(n)}$ \hspace{.1in} \\[.1in]
\hline
\hline
&&&&\\[-5mm]
1 & $SU(9)$ & 1 & 80 & 0 \\[.1in]
2 & $SU(9)$ & $\ft94$ & 80 & 84 \\[.1in]
3 & $E_8$ & $\ft{16}{5}$ & 248 & 0 \\[.1in]
\hline
\end{tabular} \\[.2in]
\begin{tabular}{|c||cc||c|}
\hline
&&&\\[-5mm]
\hspace{.08in} $f$ \hspace{.05in} &
\hspace{.08in} $P$ \hspace{.05in} &
\hspace{.05in} $Q$ \hspace{.2in} &
\hspace{.2in} $P+Q$ \hspace{.15in}  \\[.1in]
\hline
\hline
&&&\\[-5mm]
$\ft{144}{35}$ & $\ft{535}{18}$ & $\ft{77}{18}$ & 34 \\[.1in]
\hline 
\end{tabular}\\[-1in]
\begin{tabular}{c}
\hspace{3in}
{\includegraphics[height=1.3in,angle=0]{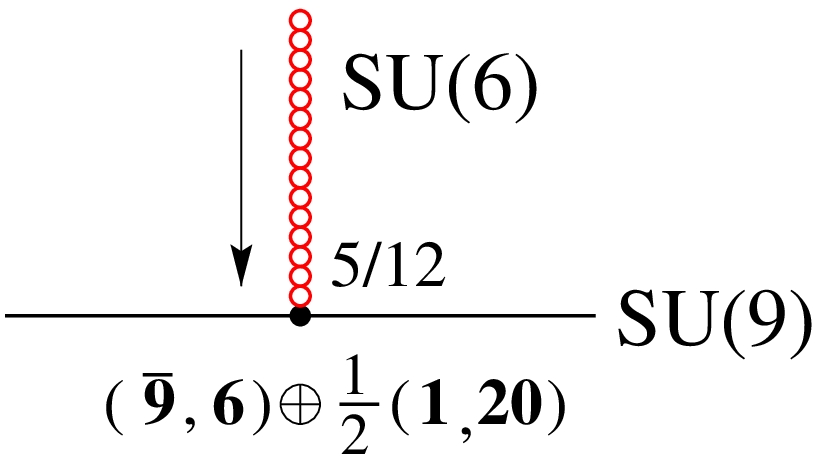}}
\end{tabular}
\vspace{.2in}
\begin{tabular}{|c||cc||ccc||ccccc||c|}
\hline
&&&&&&&&&&&\\[-5mm]
\hspace{.07in}  $I$ \hspace{.1in} &
\hspace{.01in}  ${\cal G}_I$ \hspace{.02in} &
\hspace{.03in} ${\cal R}$ \hspace{.07in} &
\hspace{.07in}  ${\cal I}_4$ \hspace{.02in} & 
\hspace{.02in} ${\cal I}_{2,2}$ \hspace{.02in} & 
\hspace{.02in} ${\cal I}_2$ \hspace{.07in} &
\hspace{.03in} $n_0$ \hspace{.02in} &
\hspace{.02in} $n_1$ \hspace{.02in} &
\hspace{.02in} $n_2$ \hspace{.02in} &
\hspace{.02in} $n_3$ \hspace{.02in} &
\hspace{.03in} $\tilde{n}$ \hspace{.1in} &
\hspace{.06in}  ${\rm z}$ \hspace{.1in} \\[.1in]
\hline
\hline
&&&&&&&&&&&\\[-5mm]
1 & $SU(9)$ & ${\bf 84}$ & -9 & 15 & 21 & 2&0&-1&2&0& -$\ft{13}{72}$ \\[.1in]
&& ${\bf 80}^*$ & 18 & 6 & 18 & 1 & 1 & 1 & 1 & 0 & $\ft{35}{144}$ \\[.1in]
&& ${\bf 9}$ & 1 & 0 & 1 & 0&0&0&0&-6& 0 \\[.1in]
\hline
&&&&&&&&&&&\\[-5mm]
1 & $SU(6)$ & ${\bf 35}^*$ &12&6&12&0&0&0&0&$\ft12$& 0 \\[.1in]
&& ${\bf 20}$ & -6 & 6 & 6 & 0 & 0 & 0 & 0 & -$\ft12$ & 0 \\[.1in]
&& ${\bf 6}$ & 1 & 0 & 1 & 0&0&0&0&-9& 0 \\[.1in]
\hline
\end{tabular} \\[.2in]
\begin{tabular}{|c||cccc||c|}
\hline
&&&&&\\[-5mm]
\hspace{.1in} $I$ \hspace{.1in} &
\hspace{.1in} $X_I$ \hspace{.05in} &
\hspace{.05in} $Z_I^{(4)}$ \hspace{.1in} &
\hspace{.05in} $Z_I^{(2)}$ \hspace{.05in} &
\hspace{.05in} $Z_I^{(2,2)}$ \hspace{.05in} &
\hspace{.05in} $\rho_I$ \hspace{.05in} \\[.1in]
\hline
\hline
&&&&&\\[-5mm]
1 & 2 & 0 & -$\ft{65}{12}$ & -$\ft{5}{4}$ & 0 \\[.1in]
&&&&&\\[-5mm]
2 & 0 & 0 & -6 & 0 & 1  \\[.1in]
\hline
\end{tabular} \hspace{.2in}
\begin{tabular}{|c||c|}
\hline
&\\[-5mm]
\hspace{.1in} $g$ \hspace{.1in} & 
\hspace{.1in} $\rho$ \hspace{.1in} \\[.1in]
\hline
\hline
&\\[-5mm]
$\ft{5}{12}$ & 1 \\[.1in]
\hline
\end{tabular} \\[-.15in]
\begin{tabular}{c}
\hspace{4.1in}
$n_H-n_V=\ft{93}{2}$
\end{tabular}
  
\end{flushleft}

\pagebreak

\subsection{The ${\bf Z}_6$ orbifold with $E_8\to SU(6)\times 
SU(3)\times SU(2)$}
\vspace{.3in}
\begin{flushleft} 
\begin{tabular}{|cc||c||cc|}
\hline
&&&&\\[-5mm]
\hspace{.1in} $n$ \hspace{.05in} &
\hspace{.05in} ${\cal G}_n$ \hspace{.1in} &
\hspace{.1in} $f_{(n)}$ \hspace{.1in} &
\hspace{.1in} $V^{(n)}$ \hspace{.05in} &
\hspace{.05in} $H^{(n)}$ \hspace{.1in} \\[.1in]
\hline
\hline
&&&&\\[-5mm]
1 & $SU(6)\times SU(3)\times SU(2)$ & 1 & 46 & 36 \\[.1in]
2 & $E_6\times SU(3)$ & $\ft94$ & 86 & 81 \\[.1in]
3 & $E_7\times SU(2)$ & $\ft{16}{5}$ & 136 & 112 \\[.1in]
\hline
\end{tabular} \\[.2in]
\begin{tabular}{|c||cc||c|}
\hline
&&&\\[-5mm]
\hspace{.08in} $f$ \hspace{.05in} &
\hspace{.08in} $P$ \hspace{.05in} &
\hspace{.05in} $Q$ \hspace{.2in} &
\hspace{.2in} $P+Q$ \hspace{.15in}  \\[.1in]
\hline
\hline
&&&\\[-5mm]
$\ft{144}{35}$ & $\ft{535}{18}$ & $\ft58$ & 30  \\[.1in]
\hline 
\end{tabular}\\[-1in]
\begin{tabular}{c}
\hspace{3in}
{\includegraphics[height=1in,angle=0]{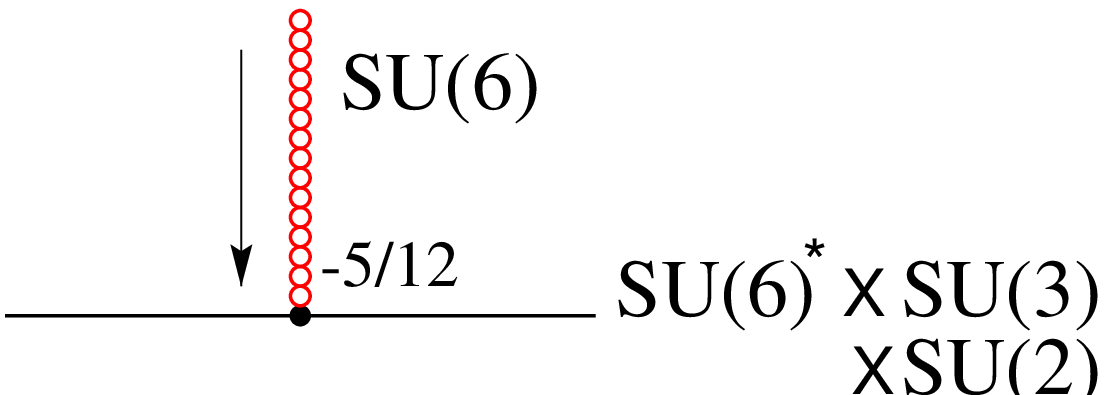}}
\end{tabular}
\vspace{.2in}
\begin{tabular}{|c||cc||ccc||ccccc||c|}
\hline
&&&&&&&&&&&\\[-5mm]
\hspace{.07in}  $I$ \hspace{.1in} &
\hspace{.01in}  ${\cal G}_I$ \hspace{.02in} &
\hspace{.03in} ${\cal R}$ \hspace{.07in} &
\hspace{.07in}  ${\cal I}_4$ \hspace{.02in} & 
\hspace{.02in} ${\cal I}_{2,2}$ \hspace{.02in} & 
\hspace{.02in} ${\cal I}_2$ \hspace{.07in} &
\hspace{.03in} $n_0$ \hspace{.02in} &
\hspace{.02in} $n_1$ \hspace{.02in} &
\hspace{.02in} $n_2$ \hspace{.02in} &
\hspace{.02in} $n_3$ \hspace{.02in} &
\hspace{.03in} $\tilde{n}$ \hspace{.1in} &
\hspace{.06in}  ${\rm z}$ \hspace{.1in} \\[.1in]
\hline
\hline
&&&&&&&&&&&\\[-5mm]
1 & $SU(6)$ & ${\bf 35}^*$ & 12&6&12&1&1&1&1&-$\ft12$&$\ft{35}{144}$ \\[.1in]
&& ${\bf 20}$ & -6 & 6 & 6 & 2 & 0 & 2 & -2 & 0 & -$\ft{19}{72}$ \\[.1in]
&& ${\bf 15}$ & -2 & 3 & 4 & 6 & 0 & -3 & 6 & 0 & -$\ft{13}{24}$ \\[.1in]
&& ${\bf 6}$ & 1 & 0 & 1 & 12 & -6 & -6 & -12 & 0 & $\ft{5}{72}$ \\[.1in]
\hline
&&&&&&&&&&&\\[-5mm]
2 & $SU(3)$ & ${\bf 8}^*$ & 0 & 9 & 6 & 1&1&1&1&0& $\ft{35}{144}$ \\[.1in]
&& ${\bf 3}$ & 0 & $\ft12$ &1&54&-12&-27&6& 0 & -$\ft{15}{8}$ \\[.1in]
\hline
&&&&&&&&&&&\\[-5mm]
2 & $SU(2)$ & ${\bf 3}^*$ & 0 & 8 & 4 & 1&1&1&1&0& $\ft{35}{144}$ \\[.1in]
&& ${\bf 2}$ & 0 & $\ft{1}{2}$ &1&56&-18&2&-56& 0 & -$\ft{25}{18}$ \\[.1in]
\hline
\end{tabular} \\[.2in]
\begin{tabular}{|c||cccc||c|}
\hline
&&&&&\\[-5mm]
\hspace{.1in} $I$ \hspace{.1in} &
\hspace{.1in} $X_I$ \hspace{.05in} &
\hspace{.05in} $Z_I^{(4)}$ \hspace{.1in} &
\hspace{.05in} $Z_I^{(2)}$ \hspace{.05in} &
\hspace{.05in} $Z_I^{(2,2)}$ \hspace{.05in} &
\hspace{.05in} $\rho_I$ \hspace{.05in} \\[.1in]
\hline
\hline
&&&&&\\[-5mm]
1 & 2 & 0 & -$\ft{77}{12}$ & -$\ft{19}{4}$ & 1 \\[.1in]
&&&&&\\[-5mm]
2 & 2 & 0 & -$\ft{5}{12}$ & $\ft54$ & 0  \\[.1in]
&&&&&\\[-5mm]
3 & 2 & 0 & -$\ft{5}{12}$ & $\ft54$ & 0  \\[.1in]
\hline
\end{tabular} \hspace{.2in}
\begin{tabular}{|c||c|}
\hline
&\\[-5mm]
\hspace{.1in} $g$ \hspace{.1in} & 
\hspace{.1in} $\rho$ \hspace{.1in} \\[.1in]
\hline
\hline
&\\[-5mm]
-$\ft{5}{12}$ & 1 \\[.1in]
\hline
\end{tabular} \\[-.3in]
\begin{tabular}{c}
\hspace{4.2in}
$n_H-n_V=\ft{35}{2}$
\end{tabular}
  
\end{flushleft}

\pagebreak

\end{document}